\documentclass[twoside,twocolumn,9pt]{article}
\usepackage{extsizes}
\usepackage[super,sort&compress,comma]{natbib} 
\usepackage[version=3]{mhchem}
\usepackage[left=1.5cm, right=1.5cm, top=1.785cm, bottom=2.0cm]{geometry}
\usepackage{balance}
\usepackage{mathptmx}
\usepackage{sectsty}
\usepackage{graphicx} 
\usepackage{lastpage}
\usepackage[format=plain,justification=justified,singlelinecheck=false,font={stretch=1.125,small,sf},labelfont=bf,labelsep=space]{caption}
\usepackage{float}
\usepackage{fancyhdr}
\usepackage{fnpos}
\usepackage[english]{babel}
\addto{\captionsenglish}{%
  
}
\usepackage{array}
\usepackage{droidsans}
\usepackage{charter}
\usepackage[T1]{fontenc}
\usepackage[usenames,dvipsnames]{xcolor}
\usepackage{setspace}
\usepackage[compact]{titlesec}
\usepackage{hyperref}
\usepackage{pdfpages}

\usepackage{subfigure}
\usepackage{dblfloatfix} 

\usepackage[normalem]{ulem}

\usepackage{epstopdf}

\definecolor{cream}{RGB}{222,217,201}

\begin{document}

\pagestyle{fancy}
\thispagestyle{plain}
\fancypagestyle{plain}{
\renewcommand{\headrulewidth}{0pt}
}

\makeFNbottom
\makeatletter
\renewcommand\LARGE{\@setfontsize\LARGE{15pt}{17}}
\renewcommand\Large{\@setfontsize\Large{12pt}{14}}
\renewcommand\large{\@setfontsize\large{10pt}{12}}
\renewcommand\footnotesize{\@setfontsize\footnotesize{7pt}{10}}
\makeatother

\renewcommand{\thefootnote}{\fnsymbol{footnote}}
\renewcommand\footnoterule{\vspace*{1pt}%
\color{cream}\hrule width 3.5in height 0.4pt \color{black}\vspace*{5pt}} 
\setcounter{secnumdepth}{5}

\makeatletter 
\renewcommand\@biblabel[1]{#1}            
\renewcommand\@makefntext[1]%
{\noindent\makebox[0pt][r]{\@thefnmark\,}#1}
\makeatother 
\renewcommand{\figurename}{\small{Fig.}~}
\sectionfont{\sffamily\Large}
\subsectionfont{\normalsize}
\subsubsectionfont{\bf}
\setstretch{1.125} 
\setlength{\skip\footins}{0.8cm}
\setlength{\footnotesep}{0.25cm}
\setlength{\jot}{10pt}
\titlespacing*{\section}{0pt}{4pt}{4pt}
\titlespacing*{\subsection}{0pt}{15pt}{1pt}

\fancyfoot{}
\fancyfoot[LO,RE]{\vspace{-7.1pt}\includegraphics[height=9pt]{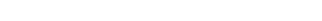}}
\fancyfoot[CO]{\vspace{-7.1pt}\hspace{13.2cm}\includegraphics{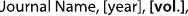}}
\fancyfoot[CE]{\vspace{-7.2pt}\hspace{-14.2cm}\includegraphics{head_foot/RF}}
\fancyfoot[RO]{\footnotesize{\sffamily{1--\pageref{LastPage} ~\textbar  \hspace{2pt}\thepage}}}
\fancyfoot[LE]{\footnotesize{\sffamily{\thepage~\textbar\hspace{3.45cm} 1--\pageref{LastPage}}}}
\fancyhead{}
\renewcommand{\headrulewidth}{0pt} 
\renewcommand{\footrulewidth}{0pt}
\setlength{\arrayrulewidth}{1pt}
\setlength{\columnsep}{6.5mm}
\setlength\bibsep{1pt}

\makeatletter 
\newlength{\figrulesep} 
\setlength{\figrulesep}{0.5\textfloatsep} 

\newcommand{\topfigrule}{\vspace*{-1pt}%
\noindent{\color{cream}\rule[-\figrulesep]{\columnwidth}{1.5pt}} }

\newcommand{\botfigrule}{\vspace*{-2pt}%
\noindent{\color{cream}\rule[\figrulesep]{\columnwidth}{1.5pt}} }

\newcommand{\dblfigrule}{\vspace*{-1pt}%
\noindent{\color{cream}\rule[-\figrulesep]{\textwidth}{1.5pt}} }

\makeatother

\twocolumn[
  \begin{@twocolumnfalse}
{\includegraphics[height=30pt]{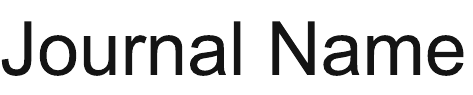}\hfill\raisebox{0pt}[0pt][0pt]{\includegraphics[height=55pt]{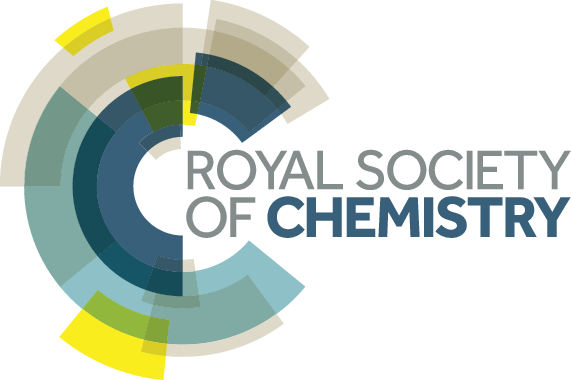}}\\[1ex]
\includegraphics[width=18.5cm]{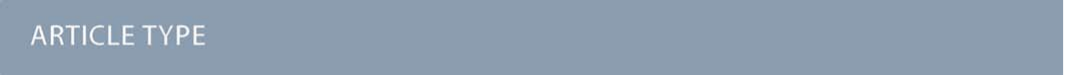}}\par
\vspace{1em}
\sffamily
\begin{tabular}{m{4.5cm} p{13.5cm} }

\includegraphics{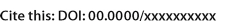} & \noindent\LARGE{\textbf{Mapping the Configuration Space of Half-Heusler Compounds via Subspace Identification for Thermoelectric Materials Discovery $^\dag$}} \\
\vspace{0.3cm} & \vspace{0.3cm} \\

 & \noindent\large{Angela Pak,\textit{$^{a}$} Kamil Ciesielski,\textit{$^{b}$} Maria Wroblewska\textit{$^{b}$}, Eric S. Toberer\textit{$^{b}$}, and Elif Ertekin\textit{$^{a}$}} \\

\includegraphics{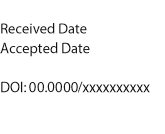} & \noindent\normalsize{
Half-Heuslers are a promising family for thermoelectric (TE) applications, yet only a small fraction of their potential chemistries has been experimentally explored. 
In this work, we introduce a distinct computational high-throughput screening approach designed to identify underexplored yet promising material subspaces, and apply it to half-Heusler thermoelectrics. 
We analyze 1,126 half-Heuslers satisfying the “18 valence electron rule,” including 332 predicted to be semiconductors, using electronic structure calculations, semi-empirical transport models, and thermoelectric quality factor $\beta$. 
Unlike conventional filtering workflows, our approach employs statistical analysis of candidate material groups to uncover trends in their collective behavior, providing robust insights and minimizing reliance on uncertain predictions for individual compounds.
Our findings link $n$-type performance to ultra-high mobility at conduction band edges and $p$-type performance to high band degeneracy. 
Statistical correlations reveal elemental subspaces associated with high $\beta$.
We identify two primary (Y- and Zr-containing) and two secondary (Au- and Ir-containing) subspaces that reinforce key physical design principles, making them promising candidates for further exploration.
These recommendations align with previous experimental results on yttrium pnictides.
Inspired by these insights, we synthesize and characterize rare-earth gold stannides (REAuSn), finding Sc$_{0.5}$Lu$_{0.5}$AuSn to exhibit low thermal conductivity (0.9–2.3 Wm$^{-1}$K$^{-1}$ at 650 K). 
This work demonstrates alternative strategies for high throughput screening when using approximate but unbiased models, and offers predictive tools and design strategies for optimizing half-Heusler chemistries for TE performance.}

\end{tabular}

 \end{@twocolumnfalse} \vspace{0.6cm}

  ]

\renewcommand*\rmdefault{bch}\normalfont\upshape
\rmfamily
\section*{}
\vspace{-1cm}


\footnotetext{\textit{$^{a}$~University
of Illinois at Urbana-Champaign, Urbana, IL 61801, USA; E-mail: ertekin@illinois.edu}}
\footnotetext{\textit{$^{b}$~Colorado School of Mines, Golden, CO 80401, USA. }}

\footnotetext{\dag~Electronic Supplementary Information (ESI) available: See DOI: 00.0000/00000000.}



\section{Introduction}
Half-Heuslers are a family of face-centered-cubic crystals with chemical formula ABC where A and B are transition metals and C is a p-block element. 
These elements crystallize in space group F$\overline{4}3m$, where the more electronegative of the two transition metals, B, occupies the FCC sublattice and A and C species form a rock salt structure.\cite{graf2011simple} The 18 valence electron half-Heuslers are particularly of interest in semiconductor applications due to their anticipated band gaps and stability,\cite{lim2021systematic,nesper2014zintl} but only 188 of these chemistries have been realized experimentally as listed in the Inorganic Crystal Structure Database.\cite{bergerhoff1983inorganic} Such realized compounds have shown promise for applications such as topological insulators and thermoelectrics (TEs).\cite{gautier2015prediction,  ciesielski2021mobility, ciesielski2020high, carrete2014nanograined, chadov2010tunable, lin2010half} 
Yet, the 188 experimentally realized half-Heuslers represent only a small subset of all possible 18 valence electron chemistries.
Thus, the half-Heusler space offers a vast range of unique intermetallic chemistries and transport properties that have yet to be characterized. 
To fill this gap, we introduce a modified high-throughput materials screening workflow that is designed to identify subspaces, or families of related compounds, that are promising search spaces for new thermoelectrics.  

The disparate properties required for good performance, as described by TE figure of merit $zT$, make TEs a challenging case for computational materials discovery. 
Figure of merit $zT$ is a composite parameter that involves both intrinsic (e.g., intrinsic mobility and thermal conductivity\cite{chasmar1959thermoelectric,bipasha2022intrinsic}  and extrinsic (e.g., temperature and carrier concentration \cite{snyder2008complex, qu2020doping} ) properties.
Prediction of intrinsic properties relies on models for electron and phonon transport and scattering.\cite{gorai2017computationally}
The need to evaluate transport coefficients, and other relevant material properties has motivated the design of ``funnel'' workflows, where candidate materials are passed through computational filters one by one.\cite{chen2016understanding,gorai2015computational,madsen2006automated,miller2017capturing} 

Computational approaches to predicting transport properties, however, inherently involve trade-offs between accuracy and cost, ranging from semi-empirical transport descriptions,\cite{yan2015material} to constant relaxation time Boltzmann transport equations,\cite{madsen2006boltztrap} to fully first-principles mode-resolved calculations of scattering rates \cite{ganose2021efficient}. 
While fully first-principles approaches show the highest predictive capacity, they are not amenable to screening large material sets. 
Semi-empirical models, by contrast, are most suited for evaluating large numbers of materials and have been successfully applied to TE materials discovery,\cite{qu2020doping, bipasha2022intrinsic, qu2021controlling} but their simplified and approximate nature limit quantitative precision. 

\begin{figure}[]
  \centering
  \includegraphics[width=3in]{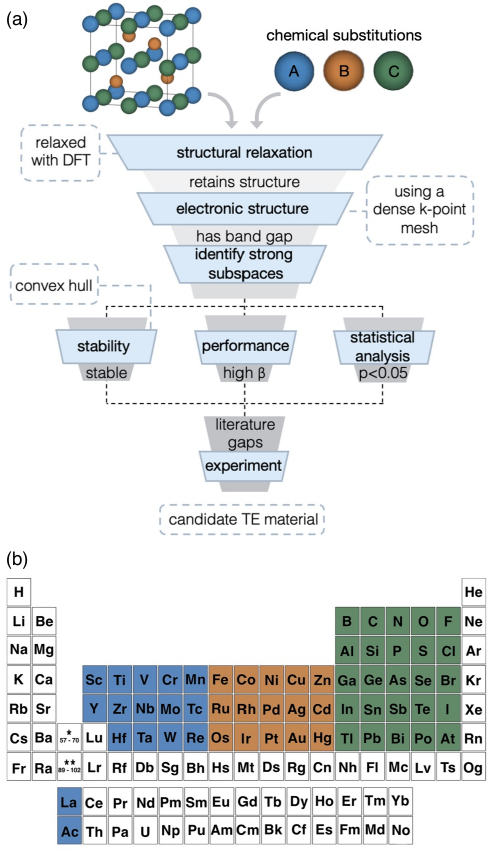}
  \caption{ Overview of the approach for identifying promising elemental subspaces in Half-Heusler thermoelectric materials.
(a) The method resembles a funnel approach, but integrates parallel assessments of stability, performance, and statistical significance to identify optimal subspaces.
(b) Elements evaluated for potential chemical substitution.
}
  \label{fig1:funnel-ptable}
\end{figure}



These trade-offs present challenges for high-throughput screening workflows that rely on traditional ``funnel'' approaches. 
In standard funnel workflows, a large pool of candidate materials is gradually narrowed through a sequence of computational filters  
by passing only candidates that meet predefined criteria to the next stage. 
Materials that make it through the entire funnel are recommended for synthesis and experimental validation.
While this approach can be effective, it has notable pitfalls. 
Early stage reliance on computationally inexpensive but approximate filters introduces uncertainty that propagates through later stages.
Furthermore, these workflows often rely on rank-ordered lists of performance descriptors \cite{bhattacharya2015high,gorai2015computational,jia2020screening,becke2006simple,qu2020doping,wang2011assessing,xi2018discovery}, placing a heavy emphasis on the accuracy of individual predictions, over identifying broader patterns or trends in the material space. 
Approximate but unbiased models are prone to variability and noise, making individual property predictions less reliable. 
Trends, when analyzed collectively, serve as more reliable indicators of performance. 
Thus, strict rank-ordering risks overlooking candidates with valuable traits visible only when viewed across the broader material space. 

To address these challenges, we present an alternative high-throughput screening approach centered on identifying promising subspaces within a broad materials space. 
We employ the dimensionless TE quality factor, $\beta$,\cite{yan2015material} alongside evaluations of electronic structure, thermodynamic stability, and transport properties to target key subspaces within the half-Heusler family (Figure \ref{fig1:funnel-ptable}a). Instead of sequential filtering, our method applies statistical analysis across the dataset to reveal robust, chemistry-based patterns, leveraging the collective behavior of groups over less reliable individual predictions.
Unlike standard workflows, this methodology actively promotes integration of physical insights with computational screening, considering material stability, novelty, and future synthesis compatibility. 
Our approach focuses on element-based regions of half-Heusler chemical space where high TE performance aligns with stable chemistries, providing actionable insights for experimental synthesis. 
This intersection of performance and stability is visually represented in Figure \ref{fig:subspace_visual}. 
Such element-based subspaces provide advantages in their practicality, particularly for strategies like alloying to enhance properties.\cite{joshi2011enhancement,li2019n,luo2023tafesb,rogl2017v,zhu2018discovery,sagar2024substantial} 

Specifically, we analyze an extensive set of 1,126 candidate half-Heuslers with the chemical formula ABC, satisfying the 18-valence electron rule. 
These candidates were generated by selecting elements 
from the periodic table (Figure \ref{fig1:funnel-ptable}b), of which 332 are predicted to be semiconductors.
To date, the vast majority of half-Heusler workflows start from considering only known chemistries or those that exist in online databases \cite{carrete2014nanograined}, or limit the search to a narrow set of elements \cite{gautier2015prediction}, and often downselect through filters. 
In contrast, our study leverages the full dataset of 332 semiconductors to identify trends, distinguish desirable $n$-type and $p$-type characteristics for half-Heuslers, and establish statistical links to promising elemental subspaces. Our analysis reveals that excellent $n$-type performance is associated with compounds featuring ultra-high mobility at conduction band edges, while strong $p$-type behavior stems from high band degeneracy. We further demonstrate that elemental subspaces statistically linked to high $\beta$ reinforce underlying physical design principles, leading us to identify four promising, yet underexplored subspaces for thermoelectric materials discovery. These include two primary subspaces of (i) yttrium- and (ii) zirconium-containing chemistries, along with two secondary subspaces for (iii) gold- and (iv) iridium-containing chemistries. Guided by these element-based recommendations, we present experimental work on rare earth gold stannides which yields ultra-low thermal conductivity half-Heusler alloys.

\section{Methods}

\subsection{Chemical Replacements}

In order to methodically consider an expansive variety of theoretical half-Heusler compounds, all elements with known electronegativities from groups 3 to 17 were utilized as possible inputs to the half-Heusler structure (Figure \ref{fig1:funnel-ptable}b). Transition metals were split between groups 3-7 and groups 8-12, with the former as chemical replacements for half-Heusler atom A and the latter groups' elements as atom B. An important constraint applied to this work was the requirement of a net 18 valence electrons per half-Heusler system. In such systems, atom A acts as a cation and donates all of its valence electrons to the covalently bonded and anionic BC pair. When these elements share and transfer electrons in this manner and with a net 18 valence electron count, the half-Heusler \begin{math} A^{n+}(BC)^{n-} \end{math}
has a completely filled d-shell. This gap in energies before the next available orbital contributes to the likelihood of 18 valence electron half-Heuslers being semiconductors - a necessity for thermoelectric applications.\cite{lim2021systematic,nesper2014zintl, zeier2016engineering} This 18 electron constraint in combination with the groups of elements considered led to an initial pool 1,126 half-Heuslers for this work.

\subsection{Structural Relaxation and Electronic Properties}
First-principles calculations were performed with density functional theory (DFT) using the Vienna Ab Initio Simulation Package (VASP) \cite{kresse1996efficient} with Projector Augmented Wave (PAW) pseudopotentials.\cite{blochl1994projector} The Perdew-Burke-Ernzerhof (PBE) \cite{perdew1996generalized} exchange correlation functional was used for structural relaxation. A plane wave cut off of 400 eV was utilized on a 8x8x8 Monkhorst-Pack sampled k-grid for relaxation. The total energy and total force convergence criteria for structural relaxations were \begin{math} 10^{-5} \end{math} eV and 0.01 eV/\r{A} respectively.

The PBE exchange-correlation functional as well as the Modified-Becke-Johnson (mBJ) meta-GGA functional \cite{becke2006simple} were utilized to evaluate band structure along a high-symmetry k-points path. Band gaps were analyzed from the results of both of these functionals, and the results of the mBJ band structure were kept in cases where only the PBE band gap closed but the mBJ gap did not. A high density k-grid, 14x14x14 \begin{math} \Gamma \end{math}-centered, was utilized in order to converge electronic properties for trends analysis across all half-Heuslers calculated to have a band gap. Such calculations also utilized total energy criteria of \begin{math} 10^{-5} \end{math} eV and total force convergence criteria of 0.01 eV/\r{A}.

\begin{figure}
  \centering
  \includegraphics[width=3.25in]{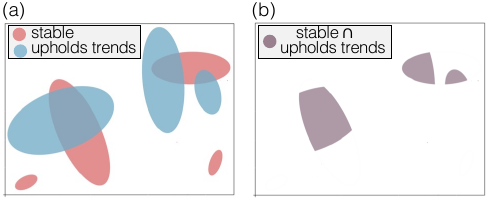}
  \caption{ Half-Heusler subspaces of interest are are identified by (a) identifying chemical spaces associated with physical, interpretable trends (blue) and chemical spaces that harbor stable compounds (pink). 
 (b) The intersection of these regions defines a unique and promising elemental subspace.  
  }
  \label{fig:subspace_visual}
\end{figure}

\subsection{Thermoelectric Quality Factor}

\begin{equation}
\beta \propto \frac{\mu_0 {m_{DOS}^*}^{3/2}}{\kappa_L}T^{5/2} 
\label{eq:beta}
\end{equation}
Thermoelectric quality factor, $\beta$, is a strong indicator of intrinsic thermoelectric performance.\cite{yan2015material} As seen in Equation (\ref{eq:beta}), $\beta$ is directly evaluated from the ratio of electronic transport parameters to lattice thermal conductivity, where $\mu_0$ is the intrinsic carrier charge mobility, $m_{DOS}^*$ is the density of states (DOS) effective mass, $\kappa_L$ is the lattice thermal conductivity and $T$ is temperature. Intrinsic carrier charge mobility is evaluated from a semi-empirical model utilizing $m_{DOS}$, band degeneracy, and bulk modulus.\cite{yan2015material} $m_{DOS}$ is evaluated utilizing the density of states from a high k-point density DFT calculation as described in Section 2.2 and using a single parabolic band approximation. Bulk modulus is evaluated by fitting the Birch-Murnaghan equation of state. \cite{birch1988elasticity}. $\kappa_L$ is also evaluated from a semi-empirical model, and uses bulk modulus, number of atoms in the primitive unit cell, density, volume, mass, and average coordination number as inputs.\cite{miller2017capturing}


\subsection{Phase Stability}
The thermodynamic stabilities of 332 half-Heuslers with band gaps were evaluated against all competing ternaries and binaries available on the Inorganic Crystal Structure Database (ICSD)\cite{bergerhoff1983inorganic} through convex hull analysis. Due to the half-Heusler space being largely unexplored experimentally, some compounds had no known competing binaries. In these cases lowest energy competing phases from Materials Project (MP) were used to construct a convex hull. The lowest-energy elemental phase structures from MP were also retrieved for each compound. 

Total energy calculations were performed for all competing phases. The majority of parameters for these calculations were generated from Pymatgen's MPRelaxSet,\cite{ong2013python} but the energy convergence criteria was lowered to $10^{-6}$ eV and a total force convergence criteria of 0.01 eV/\r{A} (as used in our geometry optimizations described in Section 2.2) was added. Compounds directly on the hull (distance = 0 eV) were marked as stable and compounds off the hull were marked to be unstable. Ranges of chemical potentials for which the predicted stable compounds remain stable were found by visualizing each stable system in chemical potential space.

\begin{figure*}[t]
  \centering
   \includegraphics[width=6.25in]{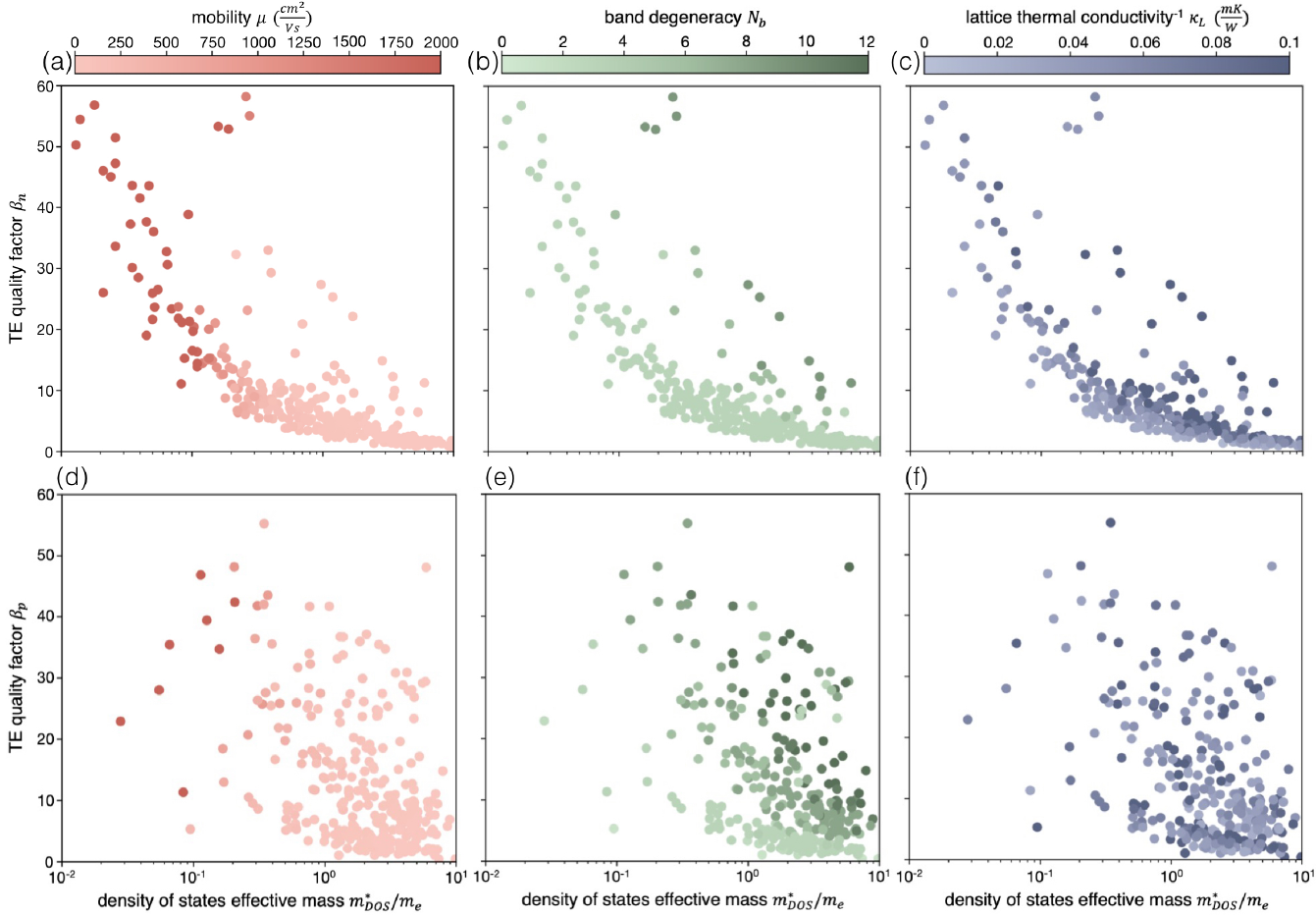}
  \caption{When thermoelectric figure of merit, $\beta$, is visualized with respect to its components, distinct trends emerge for $n$-type and $p$-type performance. For $n$-type materials, a) high $\beta$ values occur exclusively at low $m_{DOS}$ and high electron mobility. b) As their band edge degeneracy is almost always fixed at $N_b = 3$, high $\beta$ compounds must consistently exhibit low band mass $m^*b$. c) For a given $m_{DOS}$, compounds with lower lattice thermal conductivity achieve higher $\beta$. For $p$-type materials, d) the highest $\beta$ values are less concentrated at lower $m_{DOS}$. e) Valence band degeneracy, particularly combined with low $m^*_b$ at the valence band edge, correlates strongly with high $\beta_p$. e) While the lowest performing compounds tend to have the highest $\kappa_L$, the p-type association between $\kappa_L$ and $\beta_p$ is also less pronounced than in n-type. }
  \label{fig2:pairwise-trends}
\end{figure*}

\subsection{Synthesis}

The samples of ScAuSn, LuAuSn and Sc$_{0.5}$Lu$_{0.5}$AuSn were prepared first using arc-melting of constituent elements (Sc 99.8\%, Lu 99.9\%, Au 99.99\%, Sn 99.99\%)
under an ultra-pure argon atmosphere. The ingots  were remelted and flipped several times to ensure homogeneity. The arc-melted materials were subsequently ground using an agate mortar and sintered in the induction hot press with conditions 700$^{\mathrm{o}}$C, 40 MPa, 15 min for ScAuSn
and 625 $^{\mathrm{o}}$C, 40 MPa, 60-90 min for LuAuSn and Sc$_{0.5}$Lu$_{0.5}$AuSn.

\subsection{Measurement}

Crystal structure was studied with X-ray diffraction using a Bruker D2 Phaser device with a Cu radiation source. Rietveld refinement was performed with FullProf software. Density was measured with geometrical methods; for all the samples the values were $\sim$90\% or higher. 
The resistivity and Hall effect were studied on a custom-built apparatus\cite{borup2012measurement} using the Van der Pauw technique. The current supplied was 150 mA, while the magnetic field amounted to 1 T. Seebeck coefficient was also studied on a home-built apparatus\cite{iwanaga2011high}. Thermal conductivity was calculated from relation $\kappa =d C_p D $, there $d$ stands for the density, $C_p$ denotes heat capacity approximated from the Dulong-Petit law, while $D$ is the diffusivity coefficient measured with the commercial Netzsch 467 apparatus. Throughout the analysis, lattice thermal conductivity was obtained using the Franz-Wiedemann law, where the Lorenz number was calculated as shown in Ref.\cite{kim2015characterization}.

\section{Results and Discussion}

\subsection{Trends in Transport and Thermoelectric Quality Factor}
To identify the features of half-Heuslers associated with promising performance, Figure \ref{fig2:pairwise-trends} highlights several key trends observed in our computational dataset for both $n$-type and $p$-type performance. 
These trends are based on the 332 half-Heuslers (out of a total of 1,126) that exhibit a non-zero band gap (see Section 2.2).
As described by Equation \ref{eq:beta}, high intrinsic mobilities ($\mu_o$), large density of states (DOS) effective masses ($m_{DOS}^\ast$), and low lattice thermal conductivities ($\kappa_L$) are desirable for achieving a high thermoelectric (TE) quality factor. 
However, the expression for $\beta$ conceals additional complexities. DOS effective masses and mobilities are interconnected and influenced by band effective masses and band degeneracies. 
Optimally, a large DOS effective mass is achieved through multiple bands (high degeneracy) with low effective masses, which helps preserve high carrier mobilities, rather than by a single, heavy band.
To explore these interdependencies, Figure \ref{fig2:pairwise-trends} systematically examines pairwise relationships, offering insights that can guide the identification of promising subspaces. Correlation matrices of both $n$- and $p$-type data corresponding to Figure 3 is available in Figure S2 of the ESI.

Figure \ref{fig2:pairwise-trends}a illustrates the variation of $\beta_n$ with the DOS effective mass. 
The data reveal an inverse relationship, where high DOS effective masses correspond to lower thermoelectric (TE) quality factors. 
Notably, the highest $\beta$ values are observed in compounds with the lowest DOS effective masses.
Each point in Figure \ref{fig2:pairwise-trends}a is shaded based on the calculated electron mobility ($\mu_e$), highlighting that compounds with low $m_{DOS}^\ast$ but high $\beta_n$ achieve their superior predicted performance due to exceptionally high electron mobilities. Figure \ref{fig2:pairwise-trends}b also plots $\beta_n$ against the DOS effective mass, here with points colored by the band degeneracy ($N_b$) at the conduction band edge, with nearly the entire dataset sharing a degeneracy of 3. This visualization reveals that the high mobilities are linked to small band effective masses, as lower  $m_{DOS}^\ast$ must come from a small ($m_b^\ast$) given fixed $N_b$.

These trends can be further understood through the common structural features of half-Heusler conduction bands: the placement of the conduction band minimum (CBM) at the point $X$ of the fcc lattice Brillouin zone, coupled with strongly curved bands.\cite{graf2011simple,guo2022conduction}
The CBM placement limits the $n$-type band degeneracy to 3, the multiplicity of the $X$ point in the first Brillouin zone, or a theoretical maximum of 9 if each ion contributes an equivalent band at that point.\cite{guo2022conduction,zeier2016engineering} As seen in Figure \ref{fig2:pairwise-trends}b, 286 of the 332 semiconductors considered here exhibit a total conduction band degeneracy of 3 (see Table S1 of the ESI). This is much lower than the degeneracies seen in $p$-type half Heuslers (Figure \ref{fig2:pairwise-trends}e). This limitation implies that achieving a high DOS effective mass -- often regarded as beneficial for maximizing the thermoelectric quality factor $\beta$ -- in practice necessitates a large band effective mass and hinders mobility.  
Consequently, low DOS and band effective masses emerge as the primary, intertwined factors correlating with higher $\beta$ for $n$-type performance. 

To examine the influence of lattice thermal conductivity $\kappa_L$ on predicted quality factor $\beta_n$, Figure \ref{fig2:pairwise-trends}c once again plots $\beta_n$ against $m_{DOS}^\ast$, this time with compounds color-coded by the inverse of their predicted lattice thermal conductivity, with darker colors corresponding to lower $\kappa_L$.
Achieving low lattice thermal conductivity in half-Heuslers is a well-documented challenge,\cite{xie2012recent,yan2011enhanced,sakurada2005effect} and our computational results reflect this difficulty. 
The predicted $\kappa_L$ values span a wide range, from 0.2 to 98 W/mK, with a median of 18 W/mK and first and third quartiles at 12 and 26 W/mK, respectively. 
These values contrast starkly with the < 1 W/mK achieved by state-of-the-art thermoelectric materials above room temperature.
It is worthwhile to explore how this range of $\kappa_L$ impacts the observed trends. 
Figure \ref{fig2:pairwise-trends}c shows that, for a given DOS effective mass, higher $\beta_n$ values are generally associated with lower lattice thermal conductivities. The lighter color band, corresponding to compounds with relatively high $\kappa_L$, forms an envelope beneath the data points.
As $\kappa_L$ decreases (indicated by darker colors), the data points shift away from this envelope toward higher $\beta_n$. 

\begin{figure}[ht!]
    \centering
    \includegraphics[width=3in]{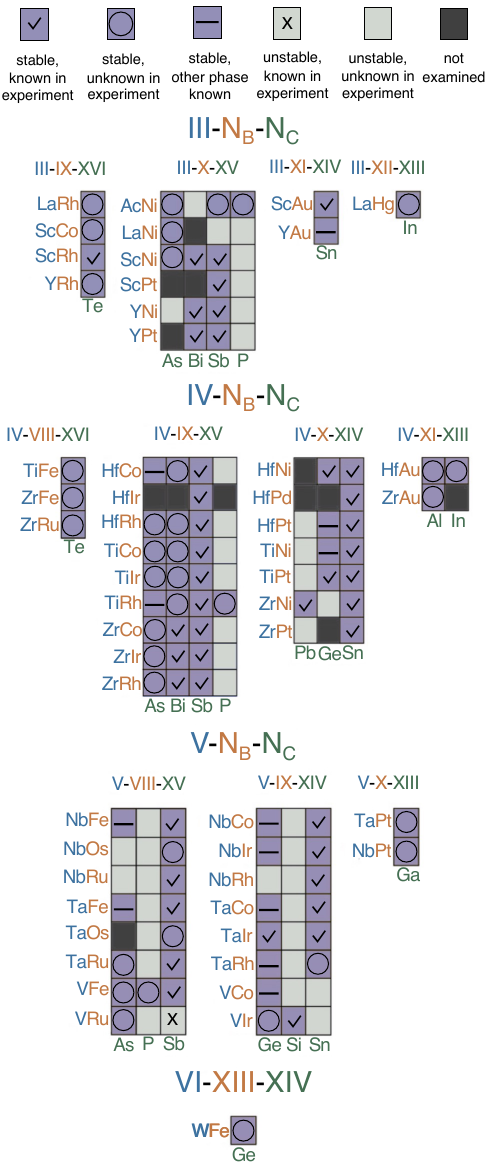}
  \caption{Subset of 332 semiconductors analyzed for phase stability, including all 93 compounds predicted to be thermodynamically stable. Purple blocks represent compounds calculated to lie on the convex hull, those with checkmarks are experimentally known and those with circles are unreported in experiment. Purple blocks with horizontal lines indicate half-Heuslers predicted stable in this work but experimentally realized in another structure. Light grey blocks correspond to compounds that lie off the calculated convex hull. Dark grey blocks indicate compounds that were not considered for stability analysis.}
  \label{fig3:stability-chart}
\end{figure}

Unlike $n$-type thermoelectric (TE) quality factors, $p$-type performance reveals much weaker correlations between $\beta_p$ and the valence band DOS effective mass ($m_{DOS}^\ast$).  
Hole mobility (Figure \ref{fig2:pairwise-trends}d) is much more limited than that of electrons in $n$-type systems. Instead, Figure  \ref{fig2:pairwise-trends}e indicates that high $\beta$ can be achieved by higher DOS masses arising from high band degeneracy. Additionally, the highest lattice thermal conductivities (Figure \ref{fig2:pairwise-trends}f) tend to be paired with low $\beta_n$, but the distinct layers of higher TE quality factor seen with lower $\kappa_L$ in $n$-type are not carried over in $p$-type. Rather, Figure \ref{fig2:pairwise-trends}d reveals that high $\beta_p$ values are most strongly associated with small band effective masses combined with large valence band degeneracies as predictors of good performance.

The significance of band degeneracy for $p$-type performance can also be traced to the characteristics of half-Heusler band structures. 
While the conduction band minimum (CBM) is consistently located at the point $X$, the valence band maximum (VBM) can occur at $\Gamma$, $W$, or $L$ within the first Brillouin zone.\cite{graf2011simple,dylla2020machine} 
This diversity in band edge placement, combined with the higher multiplicities of $W$ and $L$ (6 and 4, respectively) compared to $X$ (3), allows for a broader range of band degeneracies in the valence band (see Table S2 in the ESI), especially when multiple of these extrema lie close in energy (see Figure S1). 
As a result, the $p$-type data encompass  several categories: compounds with high DOS effective masses enabled by high valence band degeneracies but still maintaining low band effective masses, compounds with low DOS effective masses due to low valence band degeneracies but similarly low band effective masses, and a spectrum of intermediate DOS effective mass compounds. Figure \ref{fig2:pairwise-trends}e illustrates increasingly higher layers of $\beta_p$ that can be achieved for a given DOS effective mass as valence band degeneracy increases.

Overall, this analysis yields distinct design rules for identifying promising chemical subspaces for $n$- and $p$-type performance. 
For $n$-type, we seek materials with sharp conduction band edges that yield low band effective mass and therefore high mobility. 
For $p$-type, we looks for materials with high valence band degeneracy characterized by degenerate energy eigenstates across any of the $\Gamma$, $W$, or $L$ points of the Brillouin zone. 
Such degeneracy, when coupled with steeper band edges when feasible, achieves high $\beta_p$.  
For both carrier types, we further seek compounds with lower $\kappa_L$ where possible, as this can only increase TE quality factor $\beta$ more.
Below, we utilize these design rules across all calculated half-Heusler semiconductors to inform subspace identification. 

\subsection{Phase Stability}

With an understanding of desirable transport properties for $n$- and $p$-type established, we next carry out phase stability analysis across the full space of 332 semiconducting half-Heuslers. 
We use convex hull analyses of all 332 semiconductors (as described in Section 2.4) from computed formation energies, and visualize the predicted compounds' stability in chemical potential space. 
The convex hull represents the lower envelope of phases that appear when phase formation energies are plotted as a function of composition.  
Phases that lie on the hull are thermodynamically stable, while those that lie above it are unstable or metastable. 
Analyses of hulls help identify which phases are stable within a chemical composition space, and the nearby competing phases.  

\begin{figure}[ht!]
    \centering
    \includegraphics[width=3in]{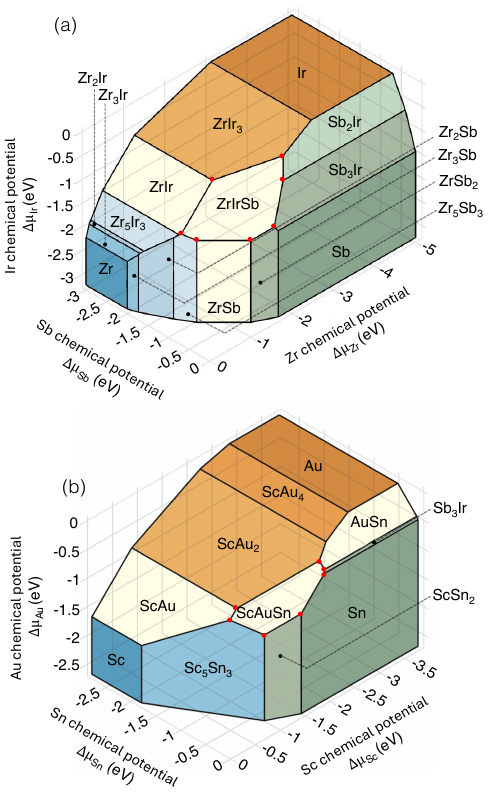}
  \caption{Phase stability diagram of a) ZrIrSb and b) ScAuSn. Blue faces correspond to competing phases closest in composition to A-element, orange faces correspond to competing phases closest to B-element, and green faces correspond to competing phases closest in composition to C-element Bi. The red dots mark the invariant points at which the Half-Heusler is in equilibrium with competing phases.}
  \label{fig4:ZrIrSb}
\end{figure} 

A summary of all half-Heusler chemistries that lie on the convex hull is presented in Figure \ref{fig3:stability-chart}. 
Figure \ref{fig3:stability-chart} is organized into 4 groups of $N_A - N_B - N_C$, where N refers to the group number of the periodic table shared between elements A, B, or C. 
The chemistries are first organized by the group number of element A, then B, and finally C. 
Compounds not included in Figure \ref{fig3:stability-chart} are predicted to be unstable. 
Of the 332 half-Heuslers considered, 93 (marked in gold blocks) are predicted to be stable. We note that the 12 compounds marked as ``stable, other phase known" were predicted to be thermodynamically stable here as half-Heuslers, but are only experimentally known in other phases (several are predicted stable in other studies as well \cite{bergerhoff1983inorganic}). 
These compounds may be false positives; alternatively targetted synthesis strategies may be effective for ultimately realizing them. 
Additionally, we note one false negative result: VRuSb, calculated as unstable but experimentally known as a half-Heusler. 

There are 38 chemistries that are not experimentally known but predicted here to be stable labeled as ``stable, unknown in experiment''. While computational analyses of their properties exist, the majority of their thermodynamic stabilities had not been previously assessed; here, they are predicted to be stable and our findings align with existing computational literature where available. This label also includes 6 chemistries not reported in the literature either experimentally or computationally: YRhTe, AcNiAs, AcNiSb, AcNiP, HfAuIn, and LaHgIn. 
These predicted compounds are sparsely distributed throughout the chemical space, but reflect comparatively less explored chemical trends, such as actinide containing, B-site gold containing, or atypical p-block elements such as Te, In, and P.

\begin{figure*}[h!]
  \centering
  \includegraphics[width=6.25in]{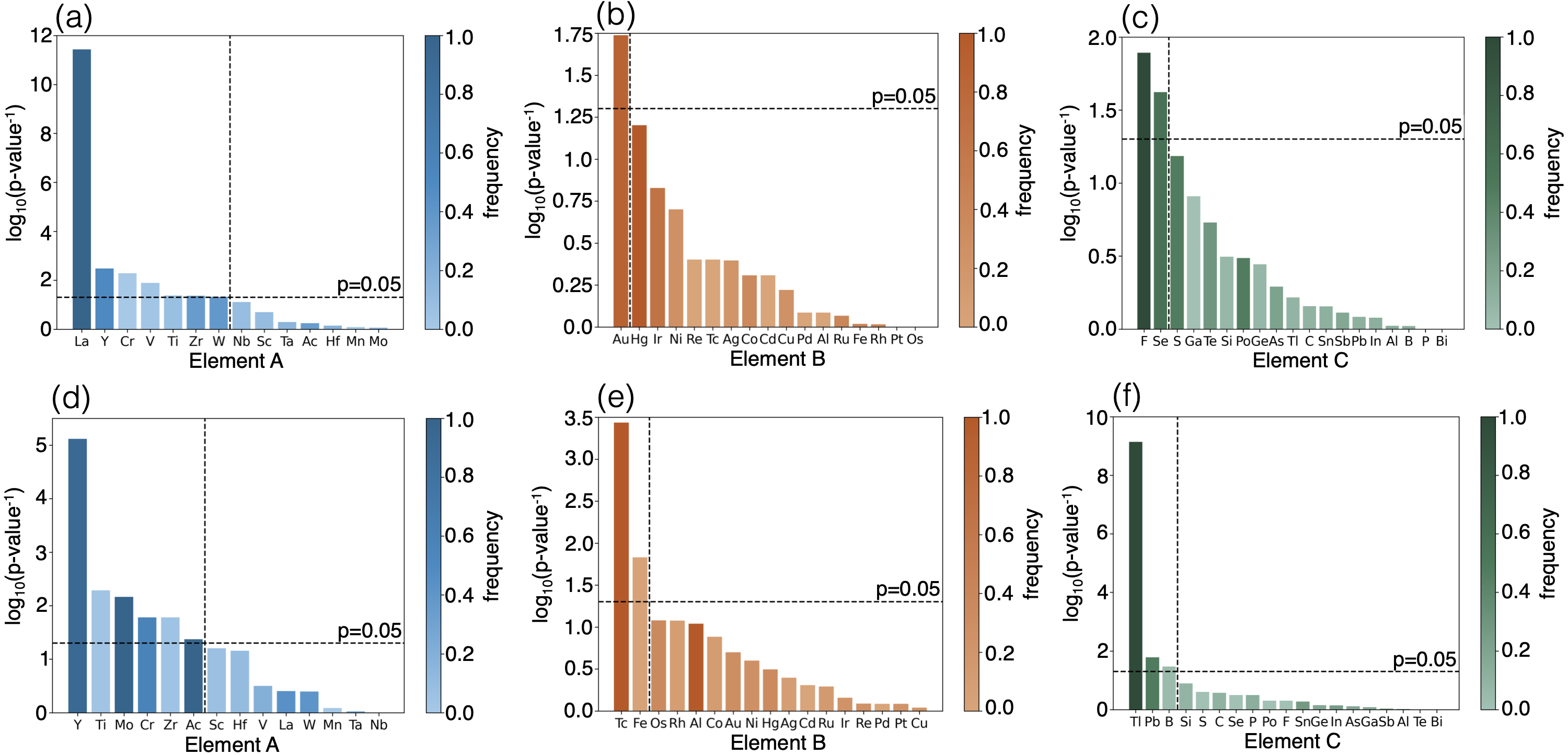}
  \caption{Statistical analysis results using Barnard's exact test. Frequency coloring represents the proportion of times an element was associated with high $\beta$ values.The y-axis shows $\log_{10} (1/p)$, so that the statistically significant elements yield the largest bars. Subfigures (a), (b), and (c) show $n$-type data for A, B, and C elements, respectively, while subfigures (d), (e), and (f) present $p$-type data for the same groups. Horizontal lines denote the significance threshold (0.05), and vertical lines mark the cutoff beyond which elements are not statistically significant.}
  \label{fig5:stats}
\end{figure*}

It is interesting to analyze trends across the stability analysis, to provide an understanding of the chemistries that tend to be stable and those that do not. 
For example, in Figure \ref{fig3:stability-chart} the block IV-IX-XV appears to be heavily explored and largely composed of stable compounds including well-known thermoelectric half-Heuslers HfCoSb and ZrIrSb. 
Across multiple blocks, we see that nearly all Sb-containing half-Heuslers are stable, and also have been previously reported in the literature.


It is also interesting to understand the compounds that are predicted to be unstable, given the goal of identifying comparatively underexplored but promising subspaces that contain realizable chemistries. For instance, Figure \ref{fig3:stability-chart} shows that chemistries with C-elements P, Pb, or Si often lie off the convex hull. This instability frequently arises from competing 1:1:1 stoichiometric phases, typically in the orthorhombic $Pnma$ structure. Examples include ScNiP and TaCoSi, which have higher formation energies than their experimentally known $Pnma$ counterparts \cite{bergerhoff1983inorganic}. Additionally, LaNiP and YNiP remain off the convex hull due to hexagonal $P63/Mmc$ phases, while YPtP has a more favorable ternary in hexagonal $P6m2$

As for the unstable group of Pb-based compounds seen in Figure \ref{fig3:stability-chart}, here we often observe a competing AB  binary. 
Favorable compounds HfPt, TiNi, TiPt, and ZrPt effectively exclude HfPtPb, TiNiPb, TiPtPb, and ZrPtPb respectively from the convex hull. 
In comparison, these competing AB-element binaries exhibit higher formation energies for largely stable groups of half-Heuslers such as the IV-IX-XV group.

Figure \ref{fig4:ZrIrSb} shows two example stable chemistries, ZrIrSb and ScAuSn, visualized in chemical potential space. In chemical potential space, a chemical potential of element i, $\Delta \mu_i$, that is closer to 0 indicates element i-rich conditions while a more negative $\Delta \mu_i$ value indicates element i-poor conditions. This visualization gives further insight to the stability of each predicted-stable compound, including the bounding phases and the thermodynamic conditions under which it may be likely to form. 

\begin{figure*}[ht!]
  \centering
   \includegraphics[width=6.5in]{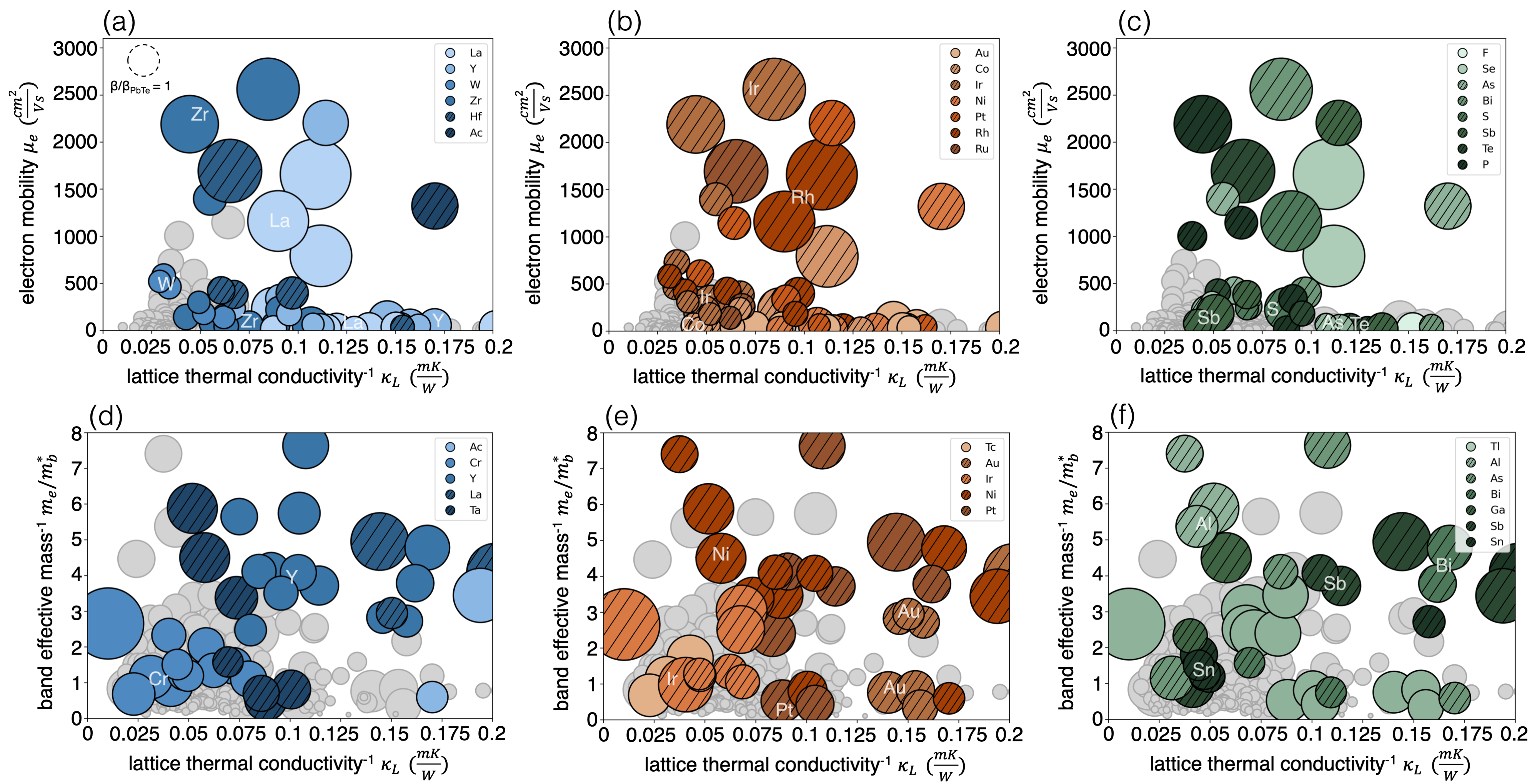}
  \caption{Visualization of the distribution of transport properties among all 332 half-Heusler semiconductors.    Half heusler $n$-type performance is visualized by (a) A-element, (b) B-element, and (c) C-element, and $p$-type performance is also visualized by (d) A-element (e) B-element and (f) C-element. Marker size corresponds to value of TE quality factor $\beta$, a reference to classic TE material PbTe is supplied in (a). Compounds with predicted values of  $\beta$ in the top 20\% are marked in color, while lower $\beta$ value compounds are marked in gray. The top 20\% compounds contain many of the elements that show statistical significance (shaded) as well as additional elements (shaded and cross-hatched).} 
  \label{fig6:bubble}
\end{figure*}
\subsection{Subspace Identification}

\subsubsection{Statistical Testing.}


To begin identifying promising subspaces within half-Heusler chemistries, Barnard's exact test was applied to evaluate correlations between specific elements in the A, B, or C sublattices and TE quality factor $\beta$. 
Known for its precision in small-sample comparisons, Barnard's test provides robust insights into specific elements most strongly associated with high or low $\beta$ and therefore TE behavior. Alternative statistical methods were attempted but ultimately deemed unsuitable due to specific limitations. 
For example, chi-squared and Mutual Information tests require larger sample sizes (e.g., n > 1000, compared to our dataset of 332 semiconductors), and ANOVA assumes linear trends and homogeneous variance between groups, which do not align with our data. 
Fisher's exact test, though designed for small sample sizes, was also excluded because it conditions only on contingency tables with fixed marginal totals, unlike Barnard's test, which considers all possible tables.

Barnard's test was performed on all A, B, and C  elements from the list of half-Heuslers predicted to exhibit a non-zero band gap. 
As the test requires categorical data, compounds were classified as ``high'' or ``low'' $\beta$ for $n$-type and $p$-type materials. 
High $\beta$ values were defined as those in the top 20\%, with the remainder categorized as low. 
The p-values for each element are shown in Figure \ref{fig5:stats}, where bar shading indicates the frequency with which compounds containing each element are classified as high $\beta$. 
The significance threshold for associations was set at p$=$0.05.

The analysis reveals several elements that achieve statistical significance in their associations with $\beta$. 
For $n$-type thermoelectrics, the A-elements La, Y, Cr, V, Ti, Zr, and W show significant p-values (Figure \ref{fig5:stats}a), while Au among the B-elements (Figure \ref{fig5:stats}b) and F and Se among the C-elements (Figure \ref{fig5:stats}c) also stand out. 
In $p$-type materials, significant A-elements include Y, Ti, Mo, Cr, Zr, and Ac (Figure \ref{fig5:stats}d); significant B-elements are Tc and Fe (Figure \ref{fig5:stats}e); and significant C-elements are Tl, Pb, and B (Figure \ref{fig5:stats}f).

It is essential to note that Barnard's exact test identifies significant associations but does not inherently indicate the direction of these correlations (i.e., whether an element is associated with high or low $\beta$). 
To address this, the proportion of high-$\beta$ versus low-$\beta$ compounds within each element’s contingency table was examined, and only elements with a stronger association to high $\beta$ were retained for further analysis. 
These elements are represented by darker bars in Figure \ref{fig5:stats}.
For $n$-type performance, the most significant A-elements then become La, Y, Zr, and W; Au is the significant B-element; and F and Se are significant C-elements. 
For $p$-type performance, significant A-elements are Y, Mo, Cr, and Ac; Tc is the significant B-element; and Tl and Pb emerge as significant C-elements. 
By contrast, elements such as Cr, V, and Ti for $n$-type, and Ti, Zr, Fe, and B for $p$-type, are excluded as their light shading indicates stronger correlations with low $\beta$.

We note that Figure \ref{fig5:stats} shows the presence of elements frequently associated with high $\beta$ values but without statistically significant p-values. 
For example, Hg and Ir as B elements, and S and Po as C element for $n$-type (Figure \ref{fig5:stats}b,c), 
and Al as B element for $p$-type (Figure \ref{fig5:stats}e) exhibit this behavior. 
The most common explanation for this behavior is that, although these chemistries correlate with high $\beta$, they are typically associated with small sample sizes. Their infrequent occurrence within the dataset results in greater variability and reduced statistical power. 
Consequently, their associations with high $\beta$ are less likely to achieve significance in Figure \ref{fig5:stats} despite their apparent relevance.

Lastly, we note that subspaces defined by descriptors other than elemental species are also viable in principle. 
For example, we examined subspaces based on elemental group numbers and group number differences (Figure S3 in the ESI). 
However, these groupings do not exhibit the strong correlations observed in the element-specific perspective. 
Although alternative descriptors may exist, the pronounced trends in elemental chemistry -- and their connection to underlying physical insights (see next subsection) -- support the focus on element-based subspaces here. 
A key advantage of these subspaces is the direct identification of promising elements, which can then be targeted for alloying strategies, as alloyed half-Heuslers have demonstrated the highest thermoelectric potential to date. 
However, it is important to acknowledge that not all physical properties or performance factors can be fully captured through elemental subspaces alone, highlighting the potential need for additional refinement or complementary descriptors in future studies.

\subsubsection{Applying Transport Trends to Significant Elements.}

Next, we analyze the elements frequently found in high-$\beta$ compounds and their overlap with the statistically significant elements identified by Barnard's exact test. While Barnard's test evaluates whether the presence of a specific element increases the likelihood of high performance, this analysis identifies the elements most likely to be present in a high-performing compound. Additionally, we assess whether these elements align with the transport trends established in Section 3.1. This approach  ensures the observed correlations reflect reproducible and interpretable physical mechanisms that offer meaningful insights into transport behavior.

Figure \ref{fig6:bubble} provides a detailed visualization of the transport trends identified in Figure \ref{fig2:pairwise-trends}.
Each marker represents a compound, with marker size corresponding to the magnitude of quality factor $\beta$. 
Colored markers denote elements frequently appearing in the top 20\%, while gray markers represent elements that are either infrequent in the top 20\% or associated with lower-performing compounds outside this range. 
Solid colored markers denote frequent elements that were also identified as statistically significant with Barnard's test, while cross-hatched markers highlight frequent elements not identified as statistically significant. 
Notably, the elements identified as frequently occurring in the top 20\% overlap substantially with those found using Barnard's exact test. 

In Figure \ref{fig6:bubble}, 
panels (a-c) illustrate the predicted $n$-type trend, where high electron mobility and low lattice thermal conductivity drive superior thermoelectric (TE) performance.  As expected, the highest-performing compounds are clustered in the upper-right corner. 
Panels (d-f) depict the $p$-type trend, highlighting the roles of low band mass, low lattice thermal conductivity, and high band degeneracy.
The broader distribution of the largest markers in Figure \ref{fig6:bubble}(d-f) reflects the influence of band degeneracy, which allows high $\beta$ values to arise across the parameter space rather than being confined to the upper-right corner.
Even so, Figure \ref{fig6:bubble}d contains a series of yttrium-based compounds, all sharing the same degeneracy, that consistently achieve higher $\beta$ values as they progress from the lower left to the upper right. 


An important takeaway from Figure \ref{fig6:bubble} is that the statistically significant A elements identified in Figure \ref{fig5:stats} dominate the top 20\% of compounds for both $n$-type and $p$-type half-Heuslers (Figure \ref{fig6:bubble}(a,d)), 
as indicated by the prevalence of solid colored markers over cross-hatched ones. Combined with the higher prevalence of significant A elements compared to B and C elements (Figure \ref{fig5:stats}), these  findings suggest that TE performance is most strongly driven by the A element. 
In contrast, the top-performing compounds exhibit greater variability in B and C elements, suggesting a lesser role. Further, the significant B and C elements from Figure \ref{fig5:stats} account for a smaller fraction of the top-performing compounds (Figure \ref{fig6:bubble}(b,c,e,f)). 
The prominence of the A element likely arises from its unique role in orbital hybridization. 
While B and C atoms form a covalently bonded zinc blende sublattice acting as an effective anion (see Section 2.1), the A element retains much of its elemental character, allowing it to exert greater influence. 

Considering $n$-type semiconductors in detail, all elements that were identified as statistically significant and associated with high $\beta$ are found in Figure \ref{fig6:bubble}(a-c): A-elements La, Y, W, and Zr; B-element Au, and the C-elements F and Se.
Additional elements (marked with cross-hatching) also appear frequently in high $\beta$ compounds. 
These additional elements fall into two main categories: (i) elements present in a significant proportion of top 20\% $\beta$ compounds but also frequently found in low $\beta$ compounds, and (ii) elements that were frequently associated with high $\beta$ values in statistical testing but lacked sufficient data to achieve statistically significant p-values.
For most cross-hatched elements, their presence in the top-performing compounds seems incidental, as many poor-performing compounds also contain these elements (case (i)). 
An exception we identified is Ir, which, as a B-element, consistently contributes to high $\beta$ values due to its exceptionally high electron mobilities. 
However, Ir was not identified as statistically significant in Barnard's exact test due to its small representation in the dataset (case (ii)), despite its strong association with high $\beta$ compounds.

Considering $p$-type semiconductors in detail, again elements that were identified as statistically significant and associated with high $\beta$ are found in Figure \ref{fig6:bubble}(d-f).
Many additional elements appear frequently in the top performing $p$-type compounds as well. 
As in $n$-type, the prevalence of these elements in poor TEs made these non-statistically significant with Barnard's test. 
Here Au and again Ir are exceptions, achieving uniquely high $\beta_p$ in within the dataset. 
From Figure \ref{fig6:bubble}(e), Au-containing chemistries conform to the expected trends of low $m^*_b$ and $\kappa$ for optimizing $p$-type performance, while Ir-based half-Heuslers exhibit exceptionally high band degeneracy at the valence band edge. 

The frequent appearance of Au and Ir in the top 20\% of p-type compounds, despite their lack of statistical significance, is more complex to interpret.
For Ir, the smaller sample size likely again contributes to this outcome.
However, as shown by the shading of the Au and Ir bars in Figure 6e, these elements are not consistently associated with a high TE quality factor. 
Instead, a subset of Au and Ir compounds exhibits exceptionally high $\beta$ values -- Au due to low thermal conductivity combined with decent electronic transport, and Ir due to exceptionally high band degeneracies.
These findings reflect that elemental subspaces alone do not fully capture all physical properties relevant to thermoelectric performance. The combination of Ir or Au with more nuanced features -- such as degree of orbital hybridization, bonding characteristic (as characterized by COHP for instance), or other aspects of electronic structure -- could yield insights into compounds with high $\beta$.
Future studies may benefit from exploring hybrid subspaces that incorporate both elemental and other descriptors, to better understand the exceptional cases. 

In summary, all elements identified as strongly associated with high $\beta$ using Barnard's test align with the physical trends driving high performance in both $n$-type and $p$-type semiconductors and frequently appear in the top 20\% of compounds.
Specifically, for $n$-type, these include the A-elements La, Y, W, and Zr; the B-element Au; and the C-elements F and Se.
For $p$-type, they are the A-elements Ac, Cr, and Y; the B-element Tc; and the C-element Tl.
Among these, the A-element plays a particularly critical role in determining transport behavior, with its presence serving as a strong predictor of good thermoelectric performance.
Additionally, Ir (for both $p$-type and $n$-type) and Au (for $p$-type) are noteworthy, as they frequently appear in the top 20\% of performers, even though their presence alone does not consistently guarantee high predicted TE performance.

\begin{figure*}[ht!]
  \centering
  \includegraphics[width=6.25in]{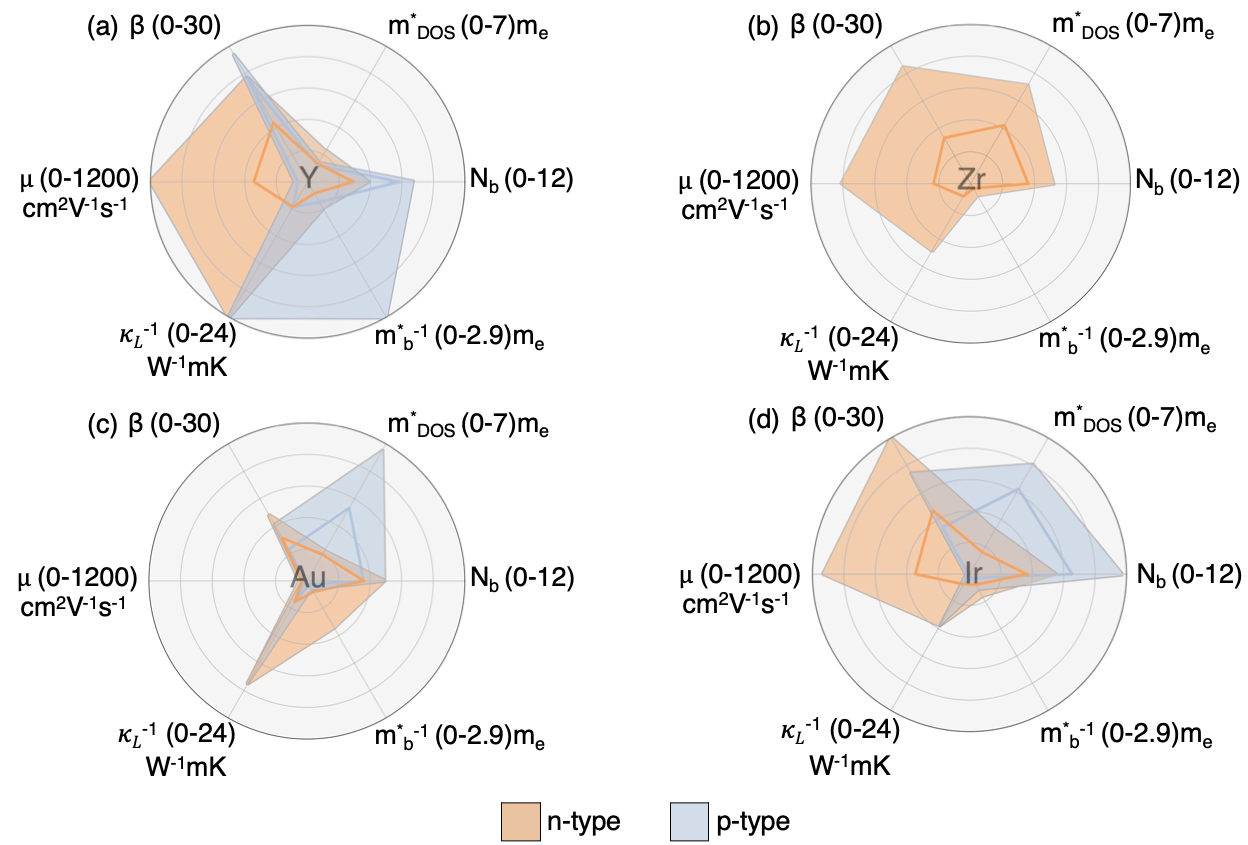}
  \caption{Spider plot visualization of elemental subspaces: Primary subspaces are a) Yttrium containing chemistries, and b) Zirconium containing chemistries. Secondary subspaces are c) Gold, and d) Iridium containing chemistries. Different properties are plotted with respect to their relationship with $\beta$, so that maximum shading indicates higher performance. Blue corresponds to $p$-type performance and orange to $n$-type. Darker shaded lines are plotted along the average values and lighter shading indicates the standard deviation for all stable compounds considered containing that element.}
  \label{fig7:spider}
\end{figure*}

\subsubsection{Filtering Elements by Stability.}

The considerations above identify elemental spaces of interest from which our promising subspaces are to be selected. 
As the goal is to identify promising, element-based subspaces for utilizable half-Heuslers, we now incorporate the phase stability analysis provided in Section 3.2. 
As shown in Figure \ref{fig:subspace_visual}, it is practically of interest to identify the overlap between areas that are promising in terms of their physical trends, and likely to yield stable compounds that are, when possible, in comparatively less explored regions of phase space. 


In $n$-type, the consideration of physical trends (Figure \ref{fig6:bubble}(a-c)) in conjunction with stability analyses identifies a more refined set of elements.
More specifically, La, W, F, and Se are removed from consideration due to their being part of very few stable chemistries -- especially those that are not already extensively explored. 
This leaves A-elements Y and Zr along with Au as a B-element.  
In $p$-type (Figure \ref{fig6:bubble}(d-f), Ac, Cr, Mo, Tc, Tl, and Pb are also removed due to yielding extremely limited stable chemistries (Figure \ref{fig3:stability-chart}); Ac, Tc, and Tl are also particularly less desirable by default given their toxicity and/or radioactivity. Thus, A-element Y demonstrates strong potential for superior TE performance in $p$-type as well. 
Finally, Au as a $p$-type element and Ir as both a $p$-type and $n$-type element are retained due to their presence in numerous stable chemistries.
This refined selection highlights the balance between statistical significance, transport properties, and material stability.

\subsubsection{Selection and Analysis of Promising Subspaces.} 
With clear insights into transport trends, key elements, and stability, we finalize the selection of promising subspaces for exploration.
The chosen subspaces are illustrated in Figure \ref{fig7:spider}.
\textit{Primary subspaces} include A-elements Y and Zr (Figures \ref{fig7:spider}a,b). Yttrium is featured for both $n$- and $p$-type performance due to its statistical significance in both categories, while zirconium focuses on $n$-type performance.
In these primary subspaces, the presence of the element strongly correlates with high $\beta$, making them the most promising areas for investigation.
\textit{Secondary subspaces} include B-elements Au and Ir (Figures \ref{fig7:spider}c,d) for both $n$- and $p$-type performance. 
Unlike primary subspaces, the presence of these elements does not always strongly correlate with high $\beta$ (with $n$-type Au being a partial exception). 
Additionally, as B-elements, their influence is generally weaker compared to A-elements. 
However, they are noteworthy due to the existence of many Au- or Ir-containing compounds that demonstrate exceptionally high $\beta$ values. 

The visualization of these subspaces in Figure \ref{fig7:spider} embody key components of $\beta$ that drive high TE performance in both $n$-type and $p$-type materials. 
Each subspace is summarized using data from all associated chemistries predicted to be thermodynamically stable. 
In the Y-based subspace, high $\beta$ is driven by lower lattice thermal conductivity, a rare trait in half-Heusler desirable for TE efficiency (Figure \ref{fig7:spider}a). 
As predicted, Y achieves superior electronic transport properties as a result of exceptionally high electron mobilities despite low DOS effective mass ($n$-type), and a combination of low band mass and high degeneracy for $p$-type. 
An advantage of Y is that it is predicted to be a high performance subspace, independent of whether the material turns out to be natively n or p -type. 
For Zr, $n$-type $\beta$ is again enhanced by particularly high electron mobility 
(Figure \ref{fig7:spider}c). 
These primary subspaces form the basis for a range of high-performing thermoelectric materials. 

The Au-based subspace, as anticipated for a secondary B-element subspace, demonstrates more modest overall statistics but includes notable standout properties. Half-Heuslers in the Au-based subspace, for example, show a low lattice thermal conductivity which benefits both carrier types. 
The interplay between a high DOS effective mass and low lattice thermal conductivity benefits $p$-type TE quality factor specifically, but the high DOS effective mass limits carrier mobility 
and adversely affects electronic transport 
(Figure \ref{fig7:spider}c). 
Even so, its chemistries with particularly low lattice thermal conductivity exhibit some of the highest $\beta$ values across the whole dataset. 
The Ir-based subspace stands out for its exceptionally high electron mobility in $n$-type applications, while $p$-type performance is once again driven by high band degeneracy. This nuanced understanding of how different elements influence TE performance enables a targeted approach to optimize chemistries for both $n$-type and $p$-type materials, while also guiding the intentional exploration of the most promising groups within the largely uncharacterized half-Heusler space.

\subsection{Validation}

 \begin{figure*}[t]
    \centering
    \includegraphics[width=6.5in]{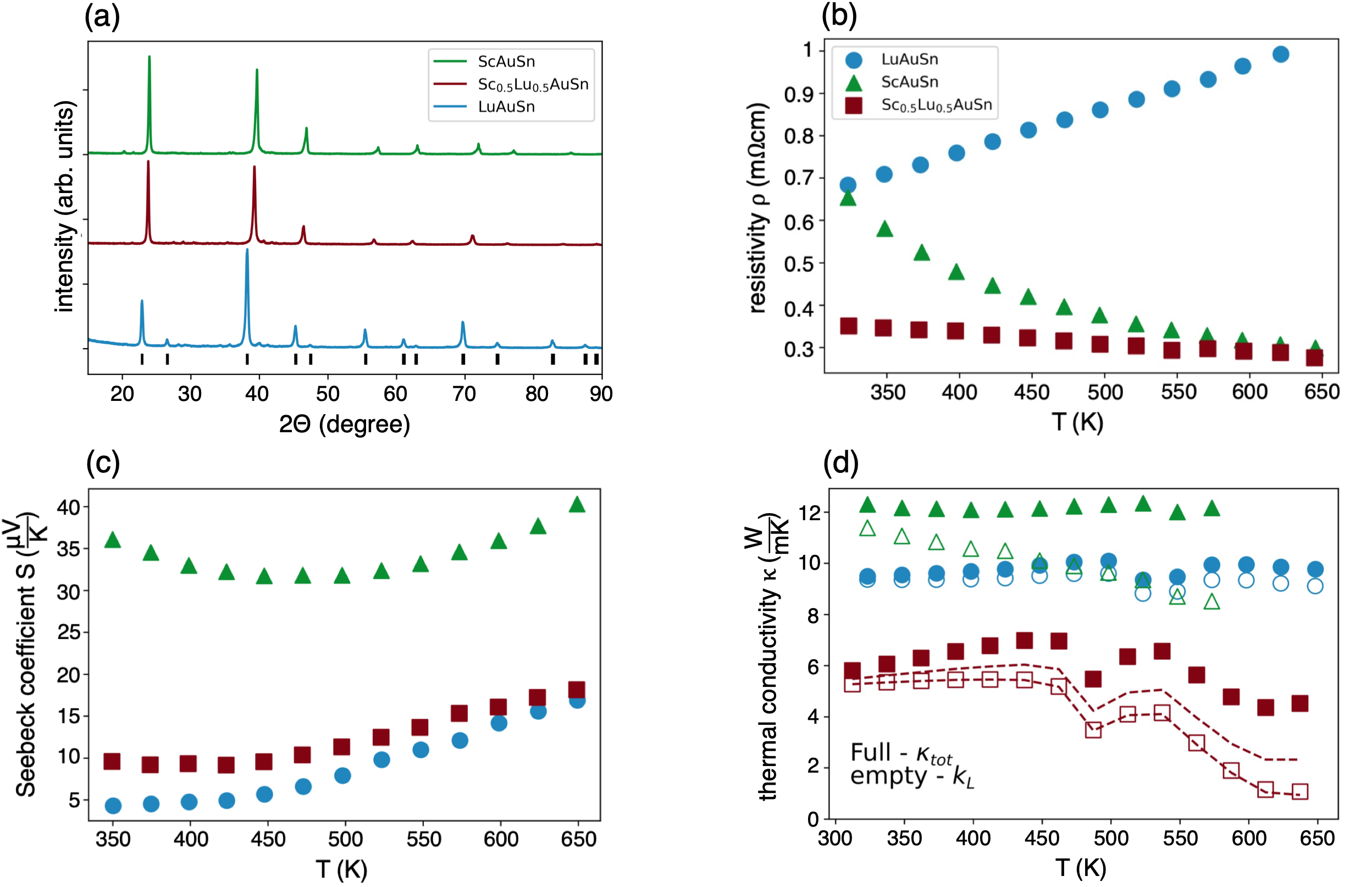}
  \caption{Experimental studies of rare-earth gold stannides revealed ultra-low thermal conductivity. Panel (a) displays almost single-phase XRD; panel (b) demonstrates similar semimetallic resistivity for all the samples. Sc$_{0.5}$Lu$_{0.5}$AuSn has the lowest $\rho$ due to decent mobility and rather high carrier concentration, see Tab. \ref{tbl:gold-stannides}. Panel (c) shows Seebeck coefficient, while (d) presents thermal conductivity - $\kappa_{tot}$ with full symbols and $\kappa_L$ with open symbols. Two dashed lines represent upper and lower limit of $\kappa_L$ depending on Lorenz number (see text for details).}
  \label{fig8:experiment}
\end{figure*}

In this section we gather experimental data from literature and compare to our theoretically selected subspaces. 
The selected subspaces not only encompass the top-performing chemistries computationally identified in this study, but also account for many known, top-performing half-Heuslers.  For example, well-known $p$-type YNiSb and YPtSb fall into the yttrium-based half-Heusler phases highlighted for low thermal conductivity. Several of the representatives of this family have been reported with promising thermoelectric properties. YPtSb\cite{li2013high, oestreich2003thermoelectrical, ouardi2011transport},  YNiBi\cite{li2015synthesis}, YNiSb\cite{oestreich2003thermoelectrical}, and YPdSb\cite{oestreich2003thermoelectrical} exhibited in experiment low lattice thermal conductivities of 2-4 Wm$^{-1}$K$^{1}$ range, 2.5, 2.3, and 2.4 Wm$^{-1}$K$^{-1}$, respectively. These values are significantly lower than those for other well known half-Heusler thermoelectrics in their pristine form at 300 K, \textit{e.g.}   for ZrNiSn\cite{hohl1997new}, TiNiSn\cite{hohl1997new},  ZrCoBi\cite{zhu2018discovery}, and
VFeSb \cite{young2000thermoelectric}, which exhibit lattice thermal conductivity of 8-13~Wm$^{-1}$K$^{-1}$. 

Reaching out to other properties, our calculations predicted extremely high mobility at 300 K for YPtSb: 2200 cm$^2$/Vs. One of the experimental reports confirms the prediction within the order of magnitude, reporting even higher mobility at room temperature, 4000 cm$^2$/Vs\cite{shekhar2012ultrahigh}.  Here, it is worth noting that other experimental reports for YPtSb indicate lower mobility (164 cm$^2$/Vs\cite{li2013high} and 300 cm$^2$/Vs\cite{ouardi2011transport}), which may stem from large susceptibility of transport properties to disorder, typical for half-Heusler compounds\cite{xie2014intrinsic, xie2012interrelation, gnida2021origin}.
When it comes to overall performance, YPtSb, for which one of the highest $\beta$ is predicted here, was shown in experiment to have a $zT$ of 0.6 at 1000 K in a single-sample study\cite{li2013high}. This, in combination with our identification of yttrium subspaces, suggests that there may be additional promising materials to be identified in these subspaces, given efforts of optimization by doping, alloying and other more intricate techniques.

It is of interest to consider well-known TE materials not present in the selected subspaces, such as $n$-type NbCoSn and VFeSb. 
Although these compounds rank highly in our dataset, their high performance is an exception among typically low-performing chemistries based on the A-element. Conversely, some known compounds like $n$-type TaCoSn and $p$-type HfPtSn appear in the bottom third of our dataset when ranked.
These discrepancies suggest that some compounds cannot be selected based solely on elemental subspaces and may require additional descriptors. Further, our computational analysis predicts idealized performance based on intrinsic properties, assuming optimal carrier type and concentration. Many of the known half-Heusler TEs may have been identified for their performance without targeted doping and optimization.

Our findings imply that higher performance may be achievable with less-explored species, provided carrier type and concentration are optimized. For example, while Zr compounds like ZrCoBi, ZrCoSb, and ZrPtSn are typically reported for p-type behavior, we predict high potential for n-type performance in Zr containing compounds. Specifically, we predict exceptional n-type performance for ZrIrSb, though achieving this will likely require specific synthesis and doping strategies. 
Figure 5(a) shows a large stability region for ZrIrSb, indicated a variety of thermodynamic environments under which the material may be synthesized. Promising synthesis strategies likely lie in cation-rich regions of phase space, near phase boundaries with ZrSb and Zr$_5$Sb$_3$, to suppress V$_{Zr}$ and achieve optimal performance.

\subsubsection{Our experimental studies}

The above experimental considerations focus on prior results for yttrium pnictides with a transition metal from the tenth group of the periodic table (Ni, Pd, Pt). 
Our calculations similarly predict interesting properties for the less known group of rare-earth gold stannides (REAuSn), explored here as a testament to the chemistries encompassed by the subspaces identified in this work. While several Y- and Zr-based compounds are already known as decent thermoelectrics,\textit{ e.g.} ZrNiSn\cite{xie2014intrinsic}, ZrCoBi\cite{zhu2018discovery},  YPtSb\cite{li2013high}, YNiSb\cite{winiarski2019high}, the Au-subspace appears uncharted and further motivates our experimental exploration. Thus, from the Au-based secondary subspace, we anticipated uniquely low lattice thermal conductivity for REAuSn half-Heuslers. 

We first synthesized ScAuSn, which we predicted to be a narrow band gap semiconductor (E$_{g}$ $\approx$ 100 meV, see Table S3) and stable (Figures \ref{fig3:stability-chart} and \ref{fig4:ZrIrSb}b). 
The experimental low electrical resistivity and low Seebeck coefficient (Fig. \ref{fig8:experiment}b, c) are in concert with the predicted band structure. 
The measured room temperature mobility (\textit{ca}. 30 cm$^2$/Vs) is also within a factor of two of the theoretical value for the valence band (17 cm$^2$/Vs). 
The lattice thermal conductivity for ScAuSn drops from 11 to 8 Wm$^{-1}$K$^{-1}$ from RT to 575 K, which is in agreement with our predicted value of 8 Wm$^{-1}$K$^{-1}$ at RT; see Figure \ref{fig8:experiment}d. 

The next natural step would be an attempt to synthesize the heavier counterpart in the series of rare-earth gold stannides: YAuSn, thereby also incorporating Y from the primary subspace. 
For this material we predicted that, consistent with many other yttrium-bearing half-Heuslers, it will show ultra low lattice thermal conductivity and good overall TE performance (ESI Tables S4). As indicated in Figure \ref{fig3:stability-chart}, although computationally predicted to be stable, experimentally this compound is known to crystallize in hexagonal symmetry (space group \textit{P}63\textit{mc}) instead of the half-Heusler structure. \cite{sebastian2006crystal}  
Given that the two phases are energetically competitive, a targeted synthesis approach addressing the possible multi-phase behavior could be explored to stabilize the half-Heusler phase.

Fortunately, the heaviest rare-earth element (Lu) is able to substitute yttrium, leading to LuAuSn crystallizing in a half-Heusler structure.\cite{sebastian2006crystal} 
Hence, we synthesized LuAuSn, and obtained a semimetal with carrier concentration even higher than for ScAuSn; see Table  \ref{tbl:gold-stannides} for values of carrier concentration at RT and Figure  \ref{fig8:experiment}b for resistivity. Accordingly, values of mobility at 300 K for LuAuSn are revealed to be rather low (4 cm$^2$/Vs) compared to the Sc-based counterpart; \textit{c.f.} Table \ref{tbl:gold-stannides}. 
The lattice thermal conductivity showed  a relatively steady value of 9 Wm$^{-1}$K$^{-1}$ between 300 K and 650 K (see Figure \ref{fig8:experiment}d).

In order to optimize the properties measured for each 1:1:1 chemistry and more fully characterize the promising REAuSn space, we took the opportunity of making a heavily strained alloy between the gold stannides: Sc$_{0.5}$Lu$_{0.5}$AuSn. 
The so-obtained material turned out to be single phase (see Figure  \ref{fig8:experiment}a for XRD), with lattice parameter of 6.491~\AA -- close to the average of the parent ternaries.\cite{sebastian2006crystal}
The carrier concentration was also close to the average values of ScAuSn and LuAuSn (Table \ref{tbl:gold-stannides}). Interestingly, despite significant differences in masses and sizes between Sc and Lu, the mobility maintained relatively high values of 19 cm$^2$/Vs at RT (Table \ref{tbl:gold-stannides}). 
Most importantly, we observed strong suppression of phonon transport.  
When we used the standard procedure for calculating the Lorenz number from Seebeck coefficient\cite{kim2015characterization}, the lattice thermal conductivity dropped to 1.1 Wm$^{-1}$K$^{-1}$ at 650 K. 
This is an ultra-low value compared to other half-Heusler thermoelectrics, including many best performing compounds.\cite{li2024half, xia2021half} 

Here, however, it is worth noting that for Sc$_{0.5}$Lu$_{0.5}$AuSn the majority contribution to thermal conductivity is coming from the electronic component. 
In systems of that kind, precise estimation of lattice thermal conductivity may be difficult due to imperfections of the Franz-Wiedemann Law, see e.g. Ref. \cite{shawon2024alloying}. 
The Lorenz number generally is known to be in range 1.5-2.44$\times$10$^{-8} \ $W$\Omega$K$^2$. \cite{kim2015characterization}
The lower values are typical for intrinsic semiconductors. With increasing conductivity, the Lorenz number also increases; the top limit is relevant for metals.

To ensure transparency of our results, we calculated the upper and lower limit of $\kappa_L$, using the smallest and the largest available Lorenz numbers respectively. The range of possible $\kappa_L$ values are visualized by dashed lines on Figure \ref{fig8:experiment}d. At the highest temperature of measurement, $\kappa_L$ calculated as such ranges between 0.9 Wm$^{-1}$K$^{-1}$ and 2.3 Wm$^{-1}$K$^{-1}$.


To get a preliminary insight into the origin of this interesting finding, we measured sound velocities ($v_s$). The results are displayed in Table \ref{tbl:gold-stannides} and reveal that rare-earth gold stannides show relatively low $v_s$, down to \textit{ca.} 2200 m/s for for LuAuSn and Sc$_{0.5}$Lu$_{0.5}$AuSn. This is indicative of soft bonding in the studied group. Here, it is worth mentioning that in recent years $v_s$ values of 2800 m/s for ZrCoBi\cite{zhu2018discovery} and 2500 m/s for \textit{R}NiSb compounds (\textit{R} = Er, Tm, Lu)\cite{ciesielski2020thermoelectric} were considered among the lowest for half-Heusler phases. In summary, we succeeded in obtaining ultra-low thermal conductivity due to point-defect disorder and extremely soft bonding for Sc$_{0.5}$Lu$_{0.5}$AuSn, as guided by the design rules and trends across chemistry identified in this work.


\begin{table}[h]
\small
  \caption{\ Carrier concentration ($n_H$), Hall mobility ($\mu_H$), and mean sound velocity ($v_s$) for the synthesized rare-earth stannides.}
  \label{tbl:gold-stannides}
  \begin{tabular*}{0.48\textwidth}{@{\extracolsep{\fill}}llll}
    \hline
     & $n_H$ [1/cm$^3$] & $\mu_H$ [cm$^2$/Vs]  & $v_s$ [m/s]\\ 
   \hline
    ScAuSn & 2.6$\times$10$^{20}$ & 31 & 2906\\
    Sc$_{0.5}$Lu$_{0.5}$AuSn & 9.5$\times$10$^{20}$ & 19 & 2191\\
    LuAuSn & 1.7$\times$10$^{21}$ & 4 & 2217\\
    \hline
  \end{tabular*}
\end{table}

\section{Conclusions}

In this study, we examine 332 semiconducting ABC half-Heusler compounds, analyzing their electronic structure, stability, and transport properties critical for optimal thermoelectric performance. 
Our modified screening approach emphasizes patterns across families of materials rather than relying on individual material predictions. 
This strategy is especially valuable for transport properties - derived from semi-empirical models at the high-throughput scale - and allows us to extract robust trends and link them to chemical composition for tuning half-Heusler properties.

Such trends yield insights into desirable properties for different polarity semiconductors. For $n$-type materials, we seek half-Heuslers with sharp conduction band edges that yield low band effective mass, which directly contributes to high electron mobility. This characteristic is preferred over a high DOS effective mass due to the limited conduction band edge degeneracy typical of half-Heuslers. In $p$-type materials, uniquely high valence band degeneracy, achievable through various positions of valence band edge extrema, emerges as the most desirable trait. Such degeneracy, when combined with steep valence band edges where possible, is also favored over simple DOS effective mass trends. 
For both carrier types, we seek materials with lower lattice thermal conductivity ($\kappa_L$) to further maximize the thermoelectric quality factor, $\beta$.

Using these insights, we statistically identify promising elemental subspaces. Our analysis highlights A-elements Y and Zr as well as B-element Ir for $n$-type thermoelectric materials as a result of high electron mobility. Both $n$- and $p$-type performance in Y, as well as secondary subspace based on B-element Au, are driven by low lattice thermal conductivity. B-element Ir is notable for its subset of exceptionally high TE quality factors, driven by ultra-high valence band degeneracy in $p$-type.
The design rules provided by this subspace approach led to the synthesis of Sc$_{0.5}$Lu$_{0.5}$AuSn, a compound with uniquely low lattice thermal conductivity attributed to the presence of Lu and Au. Our findings offer valuable insights for the continued exploration of half-Heusler compounds. The subspace approach combines insights from proven high-performance compounds with opportunities to explore new chemistries, identifying drivers of TE performance and guiding element-specific optimization of half-Heusler transport properties.



\section*{Conflicts of interest}
There are no conflicts to declare.

\section*{Acknowledgements}
All authors acknowledge support from NSF Harnessing the Data Revolution program under Grant No. 2118201. 
AP and EE acknowledge support from the NSF DIGI-MAT program, Grant No. 1922758. 
Computational resources were provided by the Advanced Cyberinfrastructure Coordination Ecosystem: Services Support (ACCESS) program through Bridges-2 at the Pittsburgh Supercomputing Center, allocation TG-MAT220011P.



\balance


\bibliography{mybib} 

\providecommand*{\mcitethebibliography}{\thebibliography}
\csname @ifundefined\endcsname{endmcitethebibliography}
{\let\endmcitethebibliography\endthebibliography}{}
\begin{mcitethebibliography}{64}
\providecommand*{\natexlab}[1]{#1}
\providecommand*{\mciteSetBstSublistMode}[1]{}
\providecommand*{\mciteSetBstMaxWidthForm}[2]{}
\providecommand*{\mciteBstWouldAddEndPuncttrue}
  {\def\EndOfBibitem{\unskip.}}
\providecommand*{\mciteBstWouldAddEndPunctfalse}
  {\let\EndOfBibitem\relax}
\providecommand*{\mciteSetBstMidEndSepPunct}[3]{}
\providecommand*{\mciteSetBstSublistLabelBeginEnd}[3]{}
\providecommand*{\EndOfBibitem}{}
\mciteSetBstSublistMode{f}
\mciteSetBstMaxWidthForm{subitem}
{(\emph{\alph{mcitesubitemcount}})}
\mciteSetBstSublistLabelBeginEnd{\mcitemaxwidthsubitemform\space}
{\relax}{\relax}

\bibitem[Graf \emph{et~al.}(2011)Graf, Felser, and Parkin]{graf2011simple}
T.~Graf, C.~Felser and S.~S. Parkin, \emph{Progress in solid state chemistry}, 2011, \textbf{39}, 1--50\relax
\mciteBstWouldAddEndPuncttrue
\mciteSetBstMidEndSepPunct{\mcitedefaultmidpunct}
{\mcitedefaultendpunct}{\mcitedefaultseppunct}\relax
\EndOfBibitem
\bibitem[Lim \emph{et~al.}(2021)Lim, Zhang, Duran, Tan, Tan, Xu, and Suwardi]{lim2021systematic}
W.~Y.~S. Lim, D.~Zhang, S.~S.~F. Duran, X.~Y. Tan, C.~K.~I. Tan, J.~Xu and A.~Suwardi, \emph{Frontiers in Materials}, 2021, \textbf{8}, 745698\relax
\mciteBstWouldAddEndPuncttrue
\mciteSetBstMidEndSepPunct{\mcitedefaultmidpunct}
{\mcitedefaultendpunct}{\mcitedefaultseppunct}\relax
\EndOfBibitem
\bibitem[Nesper(2014)]{nesper2014zintl}
R.~Nesper, \emph{Zeitschrift f{\"u}r anorganische und allgemeine Chemie}, 2014, \textbf{640}, 2639--2648\relax
\mciteBstWouldAddEndPuncttrue
\mciteSetBstMidEndSepPunct{\mcitedefaultmidpunct}
{\mcitedefaultendpunct}{\mcitedefaultseppunct}\relax
\EndOfBibitem
\bibitem[Bergerhoff \emph{et~al.}(1983)Bergerhoff, Hundt, Sievers, and Brown]{bergerhoff1983inorganic}
G.~Bergerhoff, R.~Hundt, R.~Sievers and I.~Brown, \emph{Journal of chemical information and computer sciences}, 1983, \textbf{23}, 66--69\relax
\mciteBstWouldAddEndPuncttrue
\mciteSetBstMidEndSepPunct{\mcitedefaultmidpunct}
{\mcitedefaultendpunct}{\mcitedefaultseppunct}\relax
\EndOfBibitem
\bibitem[Gautier \emph{et~al.}(2015)Gautier, Zhang, Hu, Yu, Lin, Sunde, Chon, Poeppelmeier, and Zunger]{gautier2015prediction}
R.~Gautier, X.~Zhang, L.~Hu, L.~Yu, Y.~Lin, T.~O. Sunde, D.~Chon, K.~R. Poeppelmeier and A.~Zunger, \emph{Nature chemistry}, 2015, \textbf{7}, 308--316\relax
\mciteBstWouldAddEndPuncttrue
\mciteSetBstMidEndSepPunct{\mcitedefaultmidpunct}
{\mcitedefaultendpunct}{\mcitedefaultseppunct}\relax
\EndOfBibitem
\bibitem[Ciesielski \emph{et~al.}(2021)Ciesielski, Wolanska, Synoradzki, Szymanski, and Kaczorowski]{ciesielski2021mobility}
K.~Ciesielski, I.~Wolanska, K.~Synoradzki, D.~Szymanski and D.~Kaczorowski, \emph{Physical Review Applied}, 2021, \textbf{15}, 044047\relax
\mciteBstWouldAddEndPuncttrue
\mciteSetBstMidEndSepPunct{\mcitedefaultmidpunct}
{\mcitedefaultendpunct}{\mcitedefaultseppunct}\relax
\EndOfBibitem
\bibitem[Ciesielski \emph{et~al.}(2020)Ciesielski, Synoradzki, Wolanska, Stachowiak, Kepinski, Jezowski, Tolinski, and Kaczorowski]{ciesielski2020high}
K.~Ciesielski, K.~Synoradzki, I.~Wolanska, P.~Stachowiak, L.~Kepinski, A.~Jezowski, T.~Tolinski and D.~Kaczorowski, \emph{Journal of Alloys and Compounds}, 2020, \textbf{816}, 152596\relax
\mciteBstWouldAddEndPuncttrue
\mciteSetBstMidEndSepPunct{\mcitedefaultmidpunct}
{\mcitedefaultendpunct}{\mcitedefaultseppunct}\relax
\EndOfBibitem
\bibitem[Carrete \emph{et~al.}(2014)Carrete, Mingo, Wang, and Curtarolo]{carrete2014nanograined}
J.~Carrete, N.~Mingo, S.~Wang and S.~Curtarolo, \emph{Advanced Functional Materials}, 2014, \textbf{24}, 7427--7432\relax
\mciteBstWouldAddEndPuncttrue
\mciteSetBstMidEndSepPunct{\mcitedefaultmidpunct}
{\mcitedefaultendpunct}{\mcitedefaultseppunct}\relax
\EndOfBibitem
\bibitem[Chadov \emph{et~al.}(2010)Chadov, Qi, K{\"u}bler, Fecher, Felser, and Zhang]{chadov2010tunable}
S.~Chadov, X.~Qi, J.~K{\"u}bler, G.~H. Fecher, C.~Felser and S.~C. Zhang, \emph{Nature materials}, 2010, \textbf{9}, 541--545\relax
\mciteBstWouldAddEndPuncttrue
\mciteSetBstMidEndSepPunct{\mcitedefaultmidpunct}
{\mcitedefaultendpunct}{\mcitedefaultseppunct}\relax
\EndOfBibitem
\bibitem[Lin \emph{et~al.}(2010)Lin, Wray, Xia, Xu, Jia, Cava, Bansil, and Hasan]{lin2010half}
H.~Lin, L.~A. Wray, Y.~Xia, S.~Xu, S.~Jia, R.~J. Cava, A.~Bansil and M.~Z. Hasan, \emph{Nature materials}, 2010, \textbf{9}, 546--549\relax
\mciteBstWouldAddEndPuncttrue
\mciteSetBstMidEndSepPunct{\mcitedefaultmidpunct}
{\mcitedefaultendpunct}{\mcitedefaultseppunct}\relax
\EndOfBibitem
\bibitem[Chasmar and Stratton(1959)]{chasmar1959thermoelectric}
R.~Chasmar and R.~Stratton, \emph{International journal of electronics}, 1959, \textbf{7}, 52--72\relax
\mciteBstWouldAddEndPuncttrue
\mciteSetBstMidEndSepPunct{\mcitedefaultmidpunct}
{\mcitedefaultendpunct}{\mcitedefaultseppunct}\relax
\EndOfBibitem
\bibitem[Bipasha \emph{et~al.}(2022)Bipasha, Gomes, Qu, and Ertekin]{bipasha2022intrinsic}
F.~A. Bipasha, L.~C. Gomes, J.~Qu and E.~Ertekin, \emph{Frontiers in Electronic Materials}, 2022, \textbf{2}, 1059684\relax
\mciteBstWouldAddEndPuncttrue
\mciteSetBstMidEndSepPunct{\mcitedefaultmidpunct}
{\mcitedefaultendpunct}{\mcitedefaultseppunct}\relax
\EndOfBibitem
\bibitem[Snyder and Toberer(2008)]{snyder2008complex}
G.~J. Snyder and E.~S. Toberer, \emph{Nature materials}, 2008, \textbf{7}, 105--114\relax
\mciteBstWouldAddEndPuncttrue
\mciteSetBstMidEndSepPunct{\mcitedefaultmidpunct}
{\mcitedefaultendpunct}{\mcitedefaultseppunct}\relax
\EndOfBibitem
\bibitem[Qu \emph{et~al.}(2020)Qu, Stevanovi{\'c}, Ertekin, and Gorai]{qu2020doping}
J.~Qu, V.~Stevanovi{\'c}, E.~Ertekin and P.~Gorai, \emph{Journal of Materials Chemistry A}, 2020, \textbf{8}, 25306--25315\relax
\mciteBstWouldAddEndPuncttrue
\mciteSetBstMidEndSepPunct{\mcitedefaultmidpunct}
{\mcitedefaultendpunct}{\mcitedefaultseppunct}\relax
\EndOfBibitem
\bibitem[Gorai \emph{et~al.}(2017)Gorai, Stevanovi{\'c}, and Toberer]{gorai2017computationally}
P.~Gorai, V.~Stevanovi{\'c} and E.~S. Toberer, \emph{Nature Reviews Materials}, 2017, \textbf{2}, 1--16\relax
\mciteBstWouldAddEndPuncttrue
\mciteSetBstMidEndSepPunct{\mcitedefaultmidpunct}
{\mcitedefaultendpunct}{\mcitedefaultseppunct}\relax
\EndOfBibitem
\bibitem[Chen \emph{et~al.}(2016)Chen, P{\"o}hls, Hautier, Broberg, Bajaj, Aydemir, Gibbs, Zhu, Asta, Snyder,\emph{et~al.}]{chen2016understanding}
W.~Chen, J.-H. P{\"o}hls, G.~Hautier, D.~Broberg, S.~Bajaj, U.~Aydemir, Z.~M. Gibbs, H.~Zhu, M.~Asta, G.~J. Snyder \emph{et~al.}, \emph{Journal of Materials Chemistry C}, 2016, \textbf{4}, 4414--4426\relax
\mciteBstWouldAddEndPuncttrue
\mciteSetBstMidEndSepPunct{\mcitedefaultmidpunct}
{\mcitedefaultendpunct}{\mcitedefaultseppunct}\relax
\EndOfBibitem
\bibitem[Gorai \emph{et~al.}(2015)Gorai, Parilla, Toberer, and Stevanovic]{gorai2015computational}
P.~Gorai, P.~Parilla, E.~S. Toberer and V.~Stevanovic, \emph{Chemistry of Materials}, 2015, \textbf{27}, 6213--6221\relax
\mciteBstWouldAddEndPuncttrue
\mciteSetBstMidEndSepPunct{\mcitedefaultmidpunct}
{\mcitedefaultendpunct}{\mcitedefaultseppunct}\relax
\EndOfBibitem
\bibitem[Madsen(2006)]{madsen2006automated}
G.~K. Madsen, \emph{Journal of the American Chemical Society}, 2006, \textbf{128}, 12140--12146\relax
\mciteBstWouldAddEndPuncttrue
\mciteSetBstMidEndSepPunct{\mcitedefaultmidpunct}
{\mcitedefaultendpunct}{\mcitedefaultseppunct}\relax
\EndOfBibitem
\bibitem[Miller \emph{et~al.}(2017)Miller, Gorai, Ortiz, Goyal, Gao, Barnett, Mason, Snyder, Lv, Stevanović,\emph{et~al.}]{miller2017capturing}
S.~A. Miller, P.~Gorai, B.~R. Ortiz, A.~Goyal, D.~Gao, S.~A. Barnett, T.~O. Mason, G.~J. Snyder, Q.~Lv, V.~Stevanović \emph{et~al.}, \emph{Chemistry of Materials}, 2017, \textbf{29}, 2494--2501\relax
\mciteBstWouldAddEndPuncttrue
\mciteSetBstMidEndSepPunct{\mcitedefaultmidpunct}
{\mcitedefaultendpunct}{\mcitedefaultseppunct}\relax
\EndOfBibitem
\bibitem[Yan \emph{et~al.}(2015)Yan, Gorai, Ortiz, Miller, Barnett, Mason, Stevanovi{\'c}, and Toberer]{yan2015material}
J.~Yan, P.~Gorai, B.~Ortiz, S.~Miller, S.~A. Barnett, T.~Mason, V.~Stevanovi{\'c} and E.~S. Toberer, \emph{Energy \& Environmental Science}, 2015, \textbf{8}, 983--994\relax
\mciteBstWouldAddEndPuncttrue
\mciteSetBstMidEndSepPunct{\mcitedefaultmidpunct}
{\mcitedefaultendpunct}{\mcitedefaultseppunct}\relax
\EndOfBibitem
\bibitem[Madsen and Singh(2006)]{madsen2006boltztrap}
G.~K. Madsen and D.~J. Singh, \emph{Computer Physics Communications}, 2006, \textbf{175}, 67--71\relax
\mciteBstWouldAddEndPuncttrue
\mciteSetBstMidEndSepPunct{\mcitedefaultmidpunct}
{\mcitedefaultendpunct}{\mcitedefaultseppunct}\relax
\EndOfBibitem
\bibitem[Ganose \emph{et~al.}(2021)Ganose, Park, Faghaninia, Woods-Robinson, Persson, and Jain]{ganose2021efficient}
A.~M. Ganose, J.~Park, A.~Faghaninia, R.~Woods-Robinson, K.~A. Persson and A.~Jain, \emph{Nature communications}, 2021, \textbf{12}, 2222\relax
\mciteBstWouldAddEndPuncttrue
\mciteSetBstMidEndSepPunct{\mcitedefaultmidpunct}
{\mcitedefaultendpunct}{\mcitedefaultseppunct}\relax
\EndOfBibitem
\bibitem[Qu \emph{et~al.}(2021)Qu, Porter, Gomes, Adamczyk, Toriyama, Ortiz, Toberer, and Ertekin]{qu2021controlling}
J.~Qu, C.~E. Porter, L.~C. Gomes, J.~M. Adamczyk, M.~Y. Toriyama, B.~R. Ortiz, E.~S. Toberer and E.~Ertekin, \emph{Journal of Materials Chemistry A}, 2021, \textbf{9}, 26189--26201\relax
\mciteBstWouldAddEndPuncttrue
\mciteSetBstMidEndSepPunct{\mcitedefaultmidpunct}
{\mcitedefaultendpunct}{\mcitedefaultseppunct}\relax
\EndOfBibitem
\bibitem[Bhattacharya and Madsen(2015)]{bhattacharya2015high}
S.~Bhattacharya and G.~K. Madsen, \emph{Physical Review B}, 2015, \textbf{92}, 085205\relax
\mciteBstWouldAddEndPuncttrue
\mciteSetBstMidEndSepPunct{\mcitedefaultmidpunct}
{\mcitedefaultendpunct}{\mcitedefaultseppunct}\relax
\EndOfBibitem
\bibitem[Jia \emph{et~al.}(2020)Jia, Feng, Guo, Zhang, and Zhang]{jia2020screening}
T.~Jia, Z.~Feng, S.~Guo, X.~Zhang and Y.~Zhang, \emph{ACS applied materials \& interfaces}, 2020, \textbf{12}, 11852--11864\relax
\mciteBstWouldAddEndPuncttrue
\mciteSetBstMidEndSepPunct{\mcitedefaultmidpunct}
{\mcitedefaultendpunct}{\mcitedefaultseppunct}\relax
\EndOfBibitem
\bibitem[Becke and Johnson(2006)]{becke2006simple}
A.~D. Becke and E.~R. Johnson, \emph{The Journal of chemical physics}, 2006, \textbf{124}, year\relax
\mciteBstWouldAddEndPuncttrue
\mciteSetBstMidEndSepPunct{\mcitedefaultmidpunct}
{\mcitedefaultendpunct}{\mcitedefaultseppunct}\relax
\EndOfBibitem
\bibitem[Wang \emph{et~al.}(2011)Wang, Wang, Setyawan, Mingo, and Curtarolo]{wang2011assessing}
S.~Wang, Z.~Wang, W.~Setyawan, N.~Mingo and S.~Curtarolo, \emph{Physical Review X}, 2011, \textbf{1}, 021012\relax
\mciteBstWouldAddEndPuncttrue
\mciteSetBstMidEndSepPunct{\mcitedefaultmidpunct}
{\mcitedefaultendpunct}{\mcitedefaultseppunct}\relax
\EndOfBibitem
\bibitem[Xi \emph{et~al.}(2018)Xi, Pan, Li, Xu, Ni, Sun, Yang, Luo, Xi, Zhu,\emph{et~al.}]{xi2018discovery}
L.~Xi, S.~Pan, X.~Li, Y.~Xu, J.~Ni, X.~Sun, J.~Yang, J.~Luo, J.~Xi, W.~Zhu \emph{et~al.}, \emph{Journal of the American Chemical Society}, 2018, \textbf{140}, 10785--10793\relax
\mciteBstWouldAddEndPuncttrue
\mciteSetBstMidEndSepPunct{\mcitedefaultmidpunct}
{\mcitedefaultendpunct}{\mcitedefaultseppunct}\relax
\EndOfBibitem
\bibitem[Joshi \emph{et~al.}(2011)Joshi, Yan, Wang, Liu, Chen, and Ren]{joshi2011enhancement}
G.~Joshi, X.~Yan, H.~Wang, W.~Liu, G.~Chen and Z.~Ren, \emph{Advanced Energy Materials}, 2011, \textbf{1}, 643--647\relax
\mciteBstWouldAddEndPuncttrue
\mciteSetBstMidEndSepPunct{\mcitedefaultmidpunct}
{\mcitedefaultendpunct}{\mcitedefaultseppunct}\relax
\EndOfBibitem
\bibitem[Li \emph{et~al.}(2019)Li, Zhu, Mao, Feng, Li, Chen, Cao, Liu, Singh, Ren,\emph{et~al.}]{li2019n}
S.~Li, H.~Zhu, J.~Mao, Z.~Feng, X.~Li, C.~Chen, F.~Cao, X.~Liu, D.~J. Singh, Z.~Ren \emph{et~al.}, \emph{ACS applied materials \& interfaces}, 2019, \textbf{11}, 41321--41329\relax
\mciteBstWouldAddEndPuncttrue
\mciteSetBstMidEndSepPunct{\mcitedefaultmidpunct}
{\mcitedefaultendpunct}{\mcitedefaultseppunct}\relax
\EndOfBibitem
\bibitem[Luo \emph{et~al.}(2023)Luo, Lin, Li, Zhang, and Luo]{luo2023tafesb}
P.~Luo, C.~Lin, Z.~Li, J.~Zhang and J.~Luo, \emph{ACS Applied Energy Materials}, 2023, \textbf{6}, 10070--10077\relax
\mciteBstWouldAddEndPuncttrue
\mciteSetBstMidEndSepPunct{\mcitedefaultmidpunct}
{\mcitedefaultendpunct}{\mcitedefaultseppunct}\relax
\EndOfBibitem
\bibitem[Rogl \emph{et~al.}(2017)Rogl, Sauerschnig, Rykavets, Romaka, Heinrich, Hinterleitner, Grytsiv, Bauer, and Rogl]{rogl2017v}
G.~Rogl, P.~Sauerschnig, Z.~Rykavets, V.~Romaka, P.~Heinrich, B.~Hinterleitner, A.~Grytsiv, E.~Bauer and P.~Rogl, \emph{Acta Materialia}, 2017, \textbf{131}, 336--348\relax
\mciteBstWouldAddEndPuncttrue
\mciteSetBstMidEndSepPunct{\mcitedefaultmidpunct}
{\mcitedefaultendpunct}{\mcitedefaultseppunct}\relax
\EndOfBibitem
\bibitem[Zhu \emph{et~al.}(2018)Zhu, He, Mao, Zhu, Li, Sun, Ren, Wang, Liu, Tang,\emph{et~al.}]{zhu2018discovery}
H.~Zhu, R.~He, J.~Mao, Q.~Zhu, C.~Li, J.~Sun, W.~Ren, Y.~Wang, Z.~Liu, Z.~Tang \emph{et~al.}, \emph{Nature Communications}, 2018, \textbf{9}, 2497\relax
\mciteBstWouldAddEndPuncttrue
\mciteSetBstMidEndSepPunct{\mcitedefaultmidpunct}
{\mcitedefaultendpunct}{\mcitedefaultseppunct}\relax
\EndOfBibitem
\bibitem[Sagar \emph{et~al.}(2024)Sagar, Bhardwaj, Lamba, Novitskii, Khovaylo, and Patnaik]{sagar2024substantial}
A.~Sagar, A.~Bhardwaj, M.~Lamba, A.~Novitskii, V.~Khovaylo and S.~Patnaik, \emph{Bulletin of Materials Science}, 2024, \textbf{47}, 146\relax
\mciteBstWouldAddEndPuncttrue
\mciteSetBstMidEndSepPunct{\mcitedefaultmidpunct}
{\mcitedefaultendpunct}{\mcitedefaultseppunct}\relax
\EndOfBibitem
\bibitem[Zeier \emph{et~al.}(2016)Zeier, Schmitt, Hautier, Aydemir, Gibbs, Felser, and Snyder]{zeier2016engineering}
W.~G. Zeier, J.~Schmitt, G.~Hautier, U.~Aydemir, Z.~M. Gibbs, C.~Felser and G.~J. Snyder, \emph{Nature Reviews Materials}, 2016, \textbf{1}, 1--10\relax
\mciteBstWouldAddEndPuncttrue
\mciteSetBstMidEndSepPunct{\mcitedefaultmidpunct}
{\mcitedefaultendpunct}{\mcitedefaultseppunct}\relax
\EndOfBibitem
\bibitem[Kresse and Furthm{\"u}ller(1996)]{kresse1996efficient}
G.~Kresse and J.~Furthm{\"u}ller, \emph{Physical review B}, 1996, \textbf{54}, 11169\relax
\mciteBstWouldAddEndPuncttrue
\mciteSetBstMidEndSepPunct{\mcitedefaultmidpunct}
{\mcitedefaultendpunct}{\mcitedefaultseppunct}\relax
\EndOfBibitem
\bibitem[Bl{\"o}chl(1994)]{blochl1994projector}
P.~E. Bl{\"o}chl, \emph{Physical review B}, 1994, \textbf{50}, 17953\relax
\mciteBstWouldAddEndPuncttrue
\mciteSetBstMidEndSepPunct{\mcitedefaultmidpunct}
{\mcitedefaultendpunct}{\mcitedefaultseppunct}\relax
\EndOfBibitem
\bibitem[Perdew \emph{et~al.}(1996)Perdew, Burke, and Ernzerhof]{perdew1996generalized}
J.~P. Perdew, K.~Burke and M.~Ernzerhof, \emph{Physical review letters}, 1996, \textbf{77}, 3865\relax
\mciteBstWouldAddEndPuncttrue
\mciteSetBstMidEndSepPunct{\mcitedefaultmidpunct}
{\mcitedefaultendpunct}{\mcitedefaultseppunct}\relax
\EndOfBibitem
\bibitem[Birch(1988)]{birch1988elasticity}
F.~Birch, \emph{Elastic Properties and Equations of State}, 1988, \textbf{26}, 31--90\relax
\mciteBstWouldAddEndPuncttrue
\mciteSetBstMidEndSepPunct{\mcitedefaultmidpunct}
{\mcitedefaultendpunct}{\mcitedefaultseppunct}\relax
\EndOfBibitem
\bibitem[Ong \emph{et~al.}(2013)Ong, Richards, Jain, Hautier, Kocher, Cholia, Gunter, Chevrier, Persson, and Ceder]{ong2013python}
S.~P. Ong, W.~D. Richards, A.~Jain, G.~Hautier, M.~Kocher, S.~Cholia, D.~Gunter, V.~L. Chevrier, K.~A. Persson and G.~Ceder, \emph{Computational Materials Science}, 2013, \textbf{68}, 314--319\relax
\mciteBstWouldAddEndPuncttrue
\mciteSetBstMidEndSepPunct{\mcitedefaultmidpunct}
{\mcitedefaultendpunct}{\mcitedefaultseppunct}\relax
\EndOfBibitem
\bibitem[Borup \emph{et~al.}(2012)Borup, Toberer, Zoltan, Nakatsukasa, Errico, Fleurial, Iversen, and Snyder]{borup2012measurement}
K.~A. Borup, E.~S. Toberer, L.~D. Zoltan, G.~Nakatsukasa, M.~Errico, J.-P. Fleurial, B.~B. Iversen and G.~J. Snyder, \emph{Review of Scientific Instruments}, 2012, \textbf{83}, year\relax
\mciteBstWouldAddEndPuncttrue
\mciteSetBstMidEndSepPunct{\mcitedefaultmidpunct}
{\mcitedefaultendpunct}{\mcitedefaultseppunct}\relax
\EndOfBibitem
\bibitem[Iwanaga \emph{et~al.}(2011)Iwanaga, Toberer, LaLonde, and Snyder]{iwanaga2011high}
S.~Iwanaga, E.~S. Toberer, A.~LaLonde and G.~J. Snyder, \emph{Review of Scientific Instruments}, 2011, \textbf{82}, year\relax
\mciteBstWouldAddEndPuncttrue
\mciteSetBstMidEndSepPunct{\mcitedefaultmidpunct}
{\mcitedefaultendpunct}{\mcitedefaultseppunct}\relax
\EndOfBibitem
\bibitem[Kim \emph{et~al.}(2015)Kim, Gibbs, Tang, Wang, and Snyder]{kim2015characterization}
H.-S. Kim, Z.~M. Gibbs, Y.~Tang, H.~Wang and G.~J. Snyder, \emph{APL materials}, 2015, \textbf{3}, year\relax
\mciteBstWouldAddEndPuncttrue
\mciteSetBstMidEndSepPunct{\mcitedefaultmidpunct}
{\mcitedefaultendpunct}{\mcitedefaultseppunct}\relax
\EndOfBibitem
\bibitem[Guo \emph{et~al.}(2022)Guo, Anand, Brod, Zhang, and Snyder]{guo2022conduction}
S.~Guo, S.~Anand, M.~K. Brod, Y.~Zhang and G.~J. Snyder, \emph{Journal of Materials Chemistry A}, 2022, \textbf{10}, 3051--3057\relax
\mciteBstWouldAddEndPuncttrue
\mciteSetBstMidEndSepPunct{\mcitedefaultmidpunct}
{\mcitedefaultendpunct}{\mcitedefaultseppunct}\relax
\EndOfBibitem
\bibitem[Xie \emph{et~al.}(2012)Xie, Weidenkaff, Tang, Zhang, Poon, and Tritt]{xie2012recent}
W.~Xie, A.~Weidenkaff, X.~Tang, Q.~Zhang, J.~Poon and T.~M. Tritt, \emph{Nanomaterials}, 2012, \textbf{2}, 379--412\relax
\mciteBstWouldAddEndPuncttrue
\mciteSetBstMidEndSepPunct{\mcitedefaultmidpunct}
{\mcitedefaultendpunct}{\mcitedefaultseppunct}\relax
\EndOfBibitem
\bibitem[Yan \emph{et~al.}(2011)Yan, Joshi, Liu, Lan, Wang, Lee, Simonson, Poon, Tritt, Chen,\emph{et~al.}]{yan2011enhanced}
X.~Yan, G.~Joshi, W.~Liu, Y.~Lan, H.~Wang, S.~Lee, J.~Simonson, S.~Poon, T.~Tritt, G.~Chen \emph{et~al.}, \emph{Nano letters}, 2011, \textbf{11}, 556--560\relax
\mciteBstWouldAddEndPuncttrue
\mciteSetBstMidEndSepPunct{\mcitedefaultmidpunct}
{\mcitedefaultendpunct}{\mcitedefaultseppunct}\relax
\EndOfBibitem
\bibitem[Sakurada and Shutoh(2005)]{sakurada2005effect}
S.~Sakurada and N.~Shutoh, \emph{Applied Physics Letters}, 2005, \textbf{86}, year\relax
\mciteBstWouldAddEndPuncttrue
\mciteSetBstMidEndSepPunct{\mcitedefaultmidpunct}
{\mcitedefaultendpunct}{\mcitedefaultseppunct}\relax
\EndOfBibitem
\bibitem[Dylla \emph{et~al.}(2020)Dylla, Dunn, Anand, Jain, and Snyder]{dylla2020machine}
M.~T. Dylla, A.~Dunn, S.~Anand, A.~Jain and G.~J. Snyder, \emph{Research}, 2020\relax
\mciteBstWouldAddEndPuncttrue
\mciteSetBstMidEndSepPunct{\mcitedefaultmidpunct}
{\mcitedefaultendpunct}{\mcitedefaultseppunct}\relax
\EndOfBibitem
\bibitem[Li \emph{et~al.}(2013)Li, Kurosaki, Ohishi, Muta, and Yamanaka]{li2013high}
G.~Li, K.~Kurosaki, Y.~Ohishi, H.~Muta and S.~Yamanaka, \emph{Japanese Journal of Applied Physics}, 2013, \textbf{52}, 041804\relax
\mciteBstWouldAddEndPuncttrue
\mciteSetBstMidEndSepPunct{\mcitedefaultmidpunct}
{\mcitedefaultendpunct}{\mcitedefaultseppunct}\relax
\EndOfBibitem
\bibitem[Oestreich \emph{et~al.}(2003)Oestreich, Probst, Richardt, and Bucher]{oestreich2003thermoelectrical}
J.~Oestreich, U.~Probst, F.~Richardt and E.~Bucher, \emph{Journal of Physics: Condensed Matter}, 2003, \textbf{15}, 635\relax
\mciteBstWouldAddEndPuncttrue
\mciteSetBstMidEndSepPunct{\mcitedefaultmidpunct}
{\mcitedefaultendpunct}{\mcitedefaultseppunct}\relax
\EndOfBibitem
\bibitem[Ouardi \emph{et~al.}(2011)Ouardi, Fecher, Felser, Hamrle, Postava, and Pi{\v{s}}tora]{ouardi2011transport}
S.~Ouardi, G.~H. Fecher, C.~Felser, J.~Hamrle, K.~Postava and J.~Pi{\v{s}}tora, \emph{Applied Physics Letters}, 2011, \textbf{99}, year\relax
\mciteBstWouldAddEndPuncttrue
\mciteSetBstMidEndSepPunct{\mcitedefaultmidpunct}
{\mcitedefaultendpunct}{\mcitedefaultseppunct}\relax
\EndOfBibitem
\bibitem[Li \emph{et~al.}(2015)Li, Zhao, Li, Jin, and Gu]{li2015synthesis}
S.~Li, H.~Zhao, D.~Li, S.~Jin and L.~Gu, \emph{Journal of Applied Physics}, 2015, \textbf{117}, year\relax
\mciteBstWouldAddEndPuncttrue
\mciteSetBstMidEndSepPunct{\mcitedefaultmidpunct}
{\mcitedefaultendpunct}{\mcitedefaultseppunct}\relax
\EndOfBibitem
\bibitem[Hohl \emph{et~al.}(1997)Hohl, Ramirez, Kaefer, Fess, Thurner, Kloc, and Bucher]{hohl1997new}
H.~Hohl, A.~Ramirez, W.~Kaefer, K.~Fess, C.~Thurner, C.~Kloc and E.~Bucher, \emph{MRS Online Proceedings Library (OPL)}, 1997, \textbf{478}, 109\relax
\mciteBstWouldAddEndPuncttrue
\mciteSetBstMidEndSepPunct{\mcitedefaultmidpunct}
{\mcitedefaultendpunct}{\mcitedefaultseppunct}\relax
\EndOfBibitem
\bibitem[Young \emph{et~al.}(2000)Young, Khalifah, Cava, and Ramirez]{young2000thermoelectric}
D.~Young, P.~Khalifah, R.~J. Cava and A.~Ramirez, \emph{Journal of Applied Physics}, 2000, \textbf{87}, 317--321\relax
\mciteBstWouldAddEndPuncttrue
\mciteSetBstMidEndSepPunct{\mcitedefaultmidpunct}
{\mcitedefaultendpunct}{\mcitedefaultseppunct}\relax
\EndOfBibitem
\bibitem[Shekhar \emph{et~al.}(2012)Shekhar, Ouardi, Nayak, Fecher, Schnelle, and Felser]{shekhar2012ultrahigh}
C.~Shekhar, S.~Ouardi, A.~K. Nayak, G.~H. Fecher, W.~Schnelle and C.~Felser, \emph{Physical Review B}, 2012, \textbf{86}, 155314\relax
\mciteBstWouldAddEndPuncttrue
\mciteSetBstMidEndSepPunct{\mcitedefaultmidpunct}
{\mcitedefaultendpunct}{\mcitedefaultseppunct}\relax
\EndOfBibitem
\bibitem[Xie \emph{et~al.}(2014)Xie, Wang, Fu, Liu, Snyder, Zhao, and Zhu]{xie2014intrinsic}
H.~Xie, H.~Wang, C.~Fu, Y.~Liu, G.~J. Snyder, X.~Zhao and T.~Zhu, \emph{Scientific reports}, 2014, \textbf{4}, 6888\relax
\mciteBstWouldAddEndPuncttrue
\mciteSetBstMidEndSepPunct{\mcitedefaultmidpunct}
{\mcitedefaultendpunct}{\mcitedefaultseppunct}\relax
\EndOfBibitem
\bibitem[Xie \emph{et~al.}(2012)Xie, Mi, Hu, Lock, Chirstensen, Fu, Iversen, Zhao, and Zhu]{xie2012interrelation}
H.-H. Xie, J.-L. Mi, L.-P. Hu, N.~Lock, M.~Chirstensen, C.-G. Fu, B.~B. Iversen, X.-B. Zhao and T.-J. Zhu, \emph{CrystEngComm}, 2012, \textbf{14}, 4467--4471\relax
\mciteBstWouldAddEndPuncttrue
\mciteSetBstMidEndSepPunct{\mcitedefaultmidpunct}
{\mcitedefaultendpunct}{\mcitedefaultseppunct}\relax
\EndOfBibitem
\bibitem[Gnida \emph{et~al.}(2021)Gnida, Ciesielski, and Kaczorowski]{gnida2021origin}
D.~Gnida, K.~Ciesielski and D.~Kaczorowski, \emph{Physical Review B}, 2021, \textbf{103}, 174206\relax
\mciteBstWouldAddEndPuncttrue
\mciteSetBstMidEndSepPunct{\mcitedefaultmidpunct}
{\mcitedefaultendpunct}{\mcitedefaultseppunct}\relax
\EndOfBibitem
\bibitem[Winiarski and Bili{\'n}ska(2019)]{winiarski2019high}
M.~J. Winiarski and K.~Bili{\'n}ska, \emph{Intermetallics}, 2019, \textbf{108}, 55--60\relax
\mciteBstWouldAddEndPuncttrue
\mciteSetBstMidEndSepPunct{\mcitedefaultmidpunct}
{\mcitedefaultendpunct}{\mcitedefaultseppunct}\relax
\EndOfBibitem
\bibitem[Sebastian \emph{et~al.}(2006)Sebastian, Eckert, Rayaprol, Hoffmann, and P{\"o}ttgen]{sebastian2006crystal}
C.~P. Sebastian, H.~Eckert, S.~Rayaprol, R.-D. Hoffmann and R.~P{\"o}ttgen, \emph{Solid state sciences}, 2006, \textbf{8}, 560--566\relax
\mciteBstWouldAddEndPuncttrue
\mciteSetBstMidEndSepPunct{\mcitedefaultmidpunct}
{\mcitedefaultendpunct}{\mcitedefaultseppunct}\relax
\EndOfBibitem
\bibitem[Li \emph{et~al.}(2024)Li, Ghosh, Liu, and Poudel]{li2024half}
W.~Li, S.~Ghosh, N.~Liu and B.~Poudel, \emph{Joule}, 2024\relax
\mciteBstWouldAddEndPuncttrue
\mciteSetBstMidEndSepPunct{\mcitedefaultmidpunct}
{\mcitedefaultendpunct}{\mcitedefaultseppunct}\relax
\EndOfBibitem
\bibitem[Xia \emph{et~al.}(2021)Xia, Hu, Fu, Zhao, and Zhu]{xia2021half}
K.~Xia, C.~Hu, C.~Fu, X.~Zhao and T.~Zhu, \emph{Applied Physics Letters}, 2021, \textbf{118}, year\relax
\mciteBstWouldAddEndPuncttrue
\mciteSetBstMidEndSepPunct{\mcitedefaultmidpunct}
{\mcitedefaultendpunct}{\mcitedefaultseppunct}\relax
\EndOfBibitem
\bibitem[Shawon \emph{et~al.}(2024)Shawon, Guetari, Ciesielski, Orenstein, Qu, Chanakian, Rahman, Ertekin, Toberer, and Zevalkink]{shawon2024alloying}
A.~A. Shawon, W.~Guetari, K.~Ciesielski, R.~Orenstein, J.~Qu, S.~Chanakian, M.~T. Rahman, E.~Ertekin, E.~Toberer and A.~Zevalkink, \emph{Chemistry of Materials}, 2024\relax
\mciteBstWouldAddEndPuncttrue
\mciteSetBstMidEndSepPunct{\mcitedefaultmidpunct}
{\mcitedefaultendpunct}{\mcitedefaultseppunct}\relax
\EndOfBibitem
\bibitem[Ciesielski \emph{et~al.}(2020)Ciesielski, Synoradzki, Veremchuk, Skokowski, Szyma{\'n}ski, Grin, and Kaczorowski]{ciesielski2020thermoelectric}
K.~Ciesielski, K.~Synoradzki, I.~Veremchuk, P.~Skokowski, D.~Szyma{\'n}ski, Y.~Grin and D.~Kaczorowski, \emph{Physical Review Applied}, 2020, \textbf{14}, 054046\relax
\mciteBstWouldAddEndPuncttrue
\mciteSetBstMidEndSepPunct{\mcitedefaultmidpunct}
{\mcitedefaultendpunct}{\mcitedefaultseppunct}\relax
\EndOfBibitem
\end{mcitethebibliography}
\bibliographystyle{rsc} 



\section*{Supplementary material}

# Electronic Supplementary Information

Angela Pak,[a] Kamil Ciesielski,[b] Maria Wroblewska[b], Eric S. Toberer[b], and Elif Ertekin[a]

January 20, 2025

[a] *University of Illinois at Urbana-Champaign, Urbana, IL 61801, USA; E-mail: ertekin@illinois.edu* [b] *Colorado School of Mines, Golden, CO 80401, USA.*

# 1 Band Structure Comparisons.

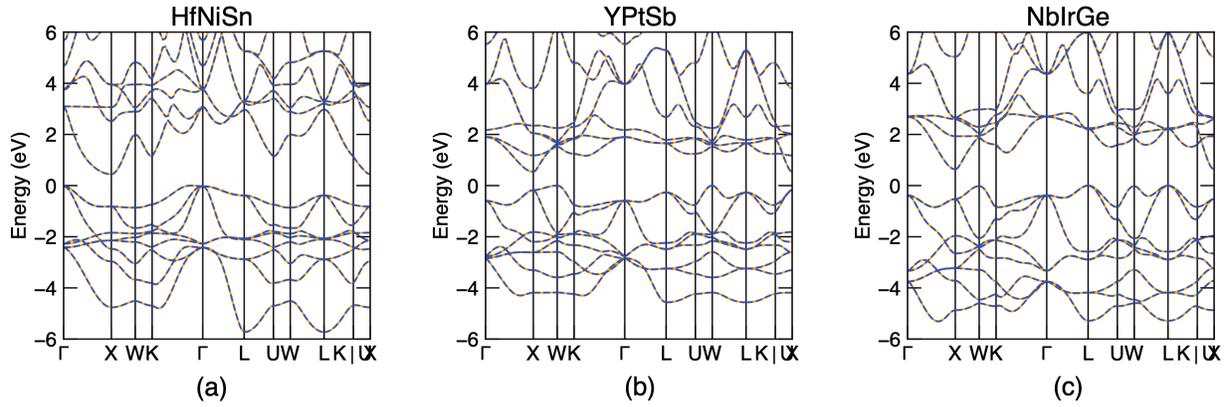

Figure 1: Calculated band structures for a) HfNiSn, a typical half-Heusler with conventional band structure features b) YPtSb, a half-Heusler predicted stable in this work with uniquely high electron moblity as anticipated by the steeper conduction band (CB) edge but still with a CB edge degeneracy of 3 at $X$-point and c) NbIrGe, a half-Heusler with uniquely high degeneracy at the valence band edge as seen by $W$, $\Gamma$, and $X$-points all reaching similar energy eigenvalues.

## 2 Correlations.

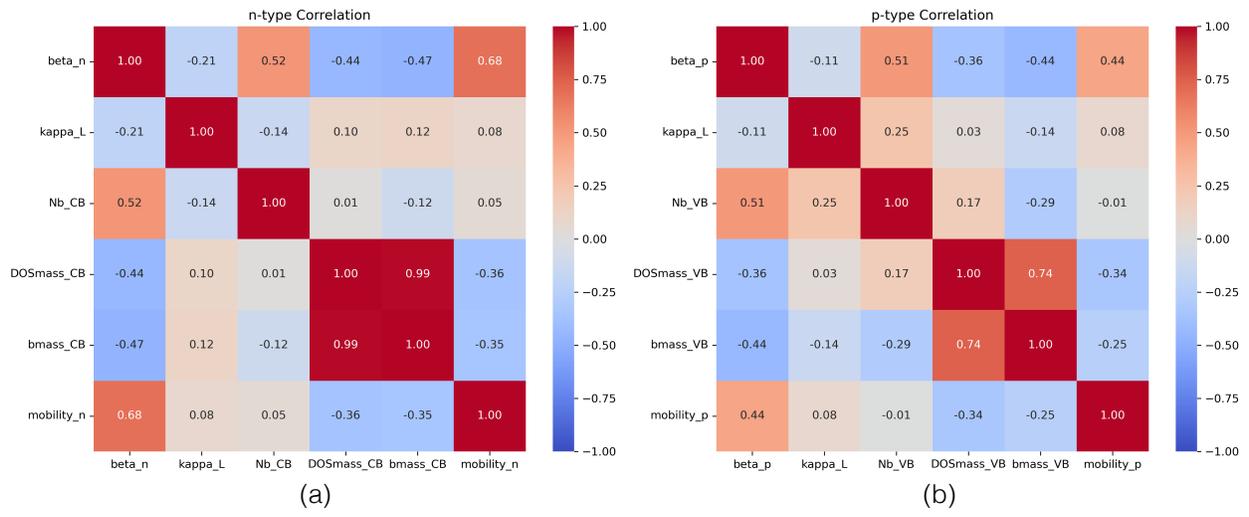

(a)　(b)

Figure 2: Pearson correlation coefficients of TE quality factor $\beta$ and each of its components for $n$- and $p$-type data.

# 3 Group Number Comparisons.

As chemistry-specific trends were pursued to offer design rules in high-performing thermoelectrics, their performance alongside other descriptors were evaluated to ensure that chosen subspaces maintained high $\beta$. A common descriptor with which to analyze half-Heuslers is the group number difference between each of their ABC elements respectively. This is particularly useful in the case of our work as it is focused on 18-electron systems, so the chemistries considered fall into a rather limited range of group number differences. The highest-performing half-Heuslers were found to spread across nearly the entirety of possible group number differences. This indicates that a selection of subspace by group number would not have captured the desirable transport trends evaluated in this work to the extent that chemistry-based subspaces did. Additionally, the highest $\beta_n$ overall for each pair-wise group number comparison was often found in a compound that belonged to a high-performing subspace. The same is true for $\beta_p$ as well.

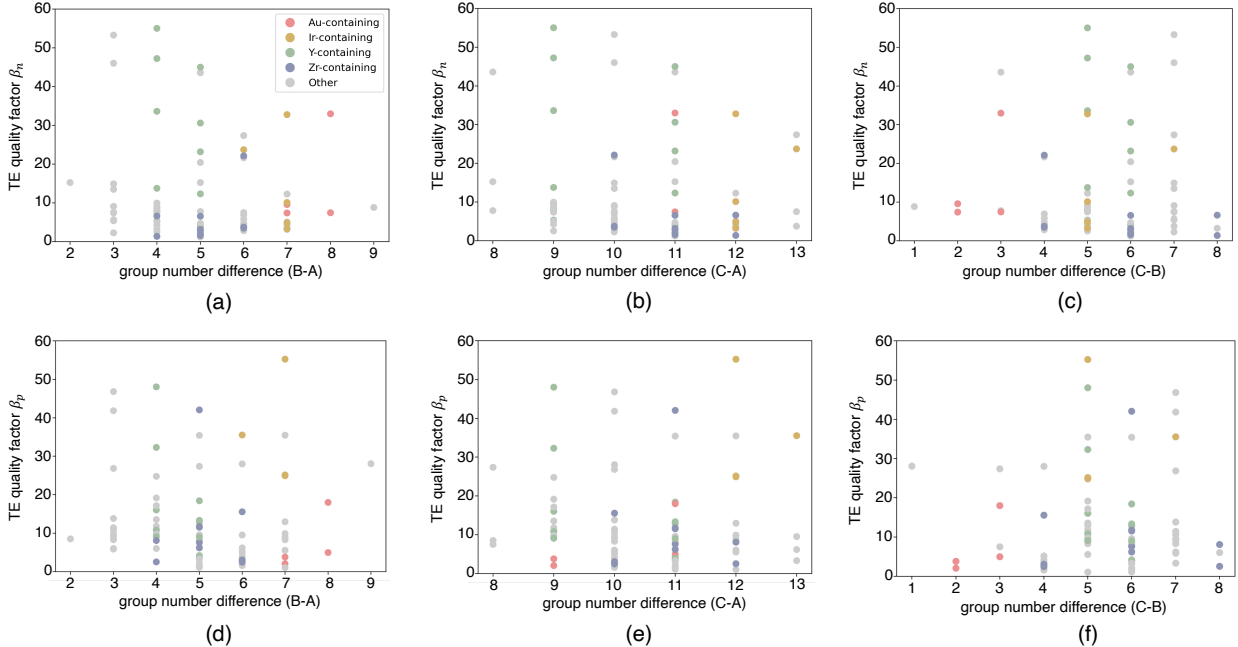

Figure 3: $\beta$ values for n-type performance (a), (b), and (c) as well as for p-type performance (d), (e), and (f), visualized by difference in group numbers between ABC element pairs. Chemistries are colored accordingly if they belonged to a subspace named in this work.

# 4 Additional Semiconductor Data - Transport.

Table 1: Electronic Transport, n-type

| Compound | $m_{DOS}(m_e)$ | $N_b$ | $m_b(m_e)$ | $\mu_n(\frac{cm^2}{Vs})$ | $\kappa(\frac{W}{mK})$ |
|---|---|---|---|---|---|
| AcNiAs | 0.076 | 3 | 0.036 | 1321.419 | 5.876 |
| AcNiP | 2.717 | 3 | 1.306 | 6.633 | 7.25 |
| AcNiSb | 2.387 | 3 | 1.148 | 6.814 | 5.157 |
| CoMoAl | 1.961 | 9 | 0.453 | 55.478 | 22.953 |
| CrCoAl | 2.859 | 3 | 1.374 | 12.112 | 31.889 |
| CrCoB | 0.765 | 3 | 0.368 | 129.21 | 51.699 |
| CrCoGa | 1.706 | 3 | 0.82 | 27.171 | 24.873 |
| CrCoIn | 6.682 | 6 | 2.024 | 5.809 | 17.093 |
| CrFeC | 2.153 | 3 | 1.035 | 30.82 | 61.675 |
| CrFeGe | 2.168 | 3 | 1.042 | 21.423 | 31.988 |
| CrFeSi | 4.218 | 3 | 2.028 | 8.628 | 48.319 |
| CrFeSn | 1.269 | 3 | 0.61 | 39.375 | 21.467 |
| CrIrAl | 3.658 | 3 | 1.759 | 9.217 | 23.511 |
| CrIrB | 4.395 | 3 | 2.113 | 9.271 | 31.7 |
| CrIrGa | 2.505 | 3 | 1.204 | 16.617 | 21.008 |
| CrIrIn | 4.329 | 3 | 2.081 | 6.184 | 16.043 |
| CrIrTl | 2.464 | 3 | 1.185 | 36.501 | 97.764 |
| CrOsGe | 4.827 | 3 | 2.321 | 6.872 | 25.716 |
| CrOsPb | 4.195 | 3 | 2.017 | 6.289 | 12.579 |
| CrOsSi | 7.374 | 3 | 3.545 | 3.923 | 32.723 |
| CrOsSn | 4.933 | 3 | 2.372 | 5.726 | 20.175 |
| CrReAs | 4.709 | 3 | 2.264 | 6.973 | 25.314 |
| CrReP | 3.635 | 3 | 1.748 | 11.373 | 32.444 |
| CrReSb | 6.269 | 3 | 3.014 | 3.995 | 20.774 |
| CrRhAl | 5.215 | 3 | 2.507 | 4.821 | 28.548 |
| CrRhB | 1.995 | 3 | 0.959 | 27.765 | 41.078 |
| CrRhGa | 4.461 | 3 | 2.144 | 6.234 | 23.483 |
| CrRuC | 4.221 | 3 | 2.029 | 10.023 | 47.53 |
| CrRuGe | 3.751 | 3 | 1.803 | 9.217 | 30.488 |
| CrRuPb | 4.043 | 3 | 1.944 | 6.279 | 13.891 |
| CrRuSi | 5.917 | 3 | 2.845 | 5.017 | 42.159 |
| CrRuSn | 4.042 | 3 | 1.943 | 7.065 | 22.551 |
| CrTcAs | 3.799 | 3 | 1.826 | 9.104 | 31.938 |
| CrTcP | 5.317 | 3 | 2.556 | 6.049 | 43.702 |
| CrTcSb | 3.465 | 3 | 1.666 | 9.094 | 24.251 |
| FeMoSi | 6.254 | 3 | 3.007 | 4.057 | 33.98 |
| HfAgAl | 0.795 | 3 | 0.382 | 53.927 | 10.457 |
| HfAgB | 1.119 | 3 | 0.538 | 41.258 | 13.008 |
| HfAgGa | 0.906 | 3 | 0.435 | 44.841 | 9.005 |
| HfAuAl | 1.188 | 3 | 0.571 | 35.27 | 11.248 |
| HfAuB | 1.574 | 3 | 0.757 | 27.838 | 12.876 |
| HfAuGa | 1.266 | 3 | 0.608 | 32.074 | 9.886 |
| HfAuIn | 1.125 | 3 | 0.541 | 34.872 | 8.702 |
| HfAuTl | 1.039 | 3 | 0.499 | 36.495 | 6.512 |
| HfCoAs | 4.295 | 3 | 2.065 | 6.52 | 16.463 |
| HfCoBi | 3.523 | 3 | 1.694 | 6.826 | 9.148 |
| HfCoP | 3.888 | 3 | 1.869 | 8.292 | 20.825 |
| HfCoSb | 3.188 | 3 | 1.533 | 9.086 | 13.888 |
| HfCuB | 0.943 | 3 | 0.453 | 61.423 | 15.441 |
| HfFeTe | 0.742 | 3 | 0.357 | 82.067 | 14 |
| HfIrSb | 0.282 | 3 | 0.136 | 381.509 | 14.884 |

| Compound | | | | | |
|---|---|---|---|---|---|
| HfNiC | 2.696 | 3 | 1.296 | 14.986 | 18.405 |
| HfNiGe | 1.817 | 3 | 0.874 | 21.485 | 14.278 |
| HfNiSi | 1.762 | 3 | 0.847 | 24.24 | 18.593 |
| HfNiSn | 1.592 | 3 | 0.765 | 23.172 | 11.798 |
| HfPdSn | 1.464 | 3 | 0.704 | 25.639 | 11.835 |
| HfPtGe | 2.541 | 3 | 1.221 | 13.727 | 13.747 |
| HfPtPb | 1.914 | 3 | 0.92 | 17.359 | 8.779 |
| HfPtSi | 2.379 | 3 | 1.144 | 16.227 | 16.772 |
| HfPtSn | 1.902 | 3 | 0.914 | 19.731 | 12.551 |
| HfRhAs | 0.257 | 3 | 0.124 | 427.148 | 16.498 |
| HfRhBi | 0.238 | 3 | 0.114 | 399.445 | 10.308 |
| HfRhP | 0.616 | 3 | 0.296 | 124.674 | 20.229 |
| HfRhSb | 3.26 | 3 | 1.567 | 8.907 | 15.157 |
| HfRuTe | 0.207 | 9 | 0.048 | 1693.412 | 15.359 |
| LaAuSi | 1.331 | 6 | 0.403 | 35.932 | 6.657 |
| LaAuSn | 1.28 | 6 | 0.388 | 34.087 | 4.964 |
| LaCoS | 0.604 | 9 | 0.14 | 227.013 | 11.445 |
| LaCoSe | 0.248 | 9 | 0.057 | 792.901 | 8.946 |
| LaCoTe | 4.619 | 9 | 1.068 | 9.137 | 7.761 |
| LaCuC | 1.852 | 3 | 0.89 | 12.665 | 9.033 |
| LaFeF | 13.1 | 9 | 3.028 | 0.801 | 2.357 |
| LaHgAl | 1.07 | 9 | 0.247 | 55.481 | 4.438 |
| LaHgIn | 1 | 3 | 0.481 | 18.794 | 3.311 |
| LaNiAs | 2.906 | 9 | 0.672 | 18.786 | 8.716 |
| LaNiP | 3.062 | 9 | 0.708 | 18.906 | 11.431 |
| LaNiSb | 2.491 | 9 | 0.576 | 21.568 | 7.328 |
| LaPtSb | 0.056 | 9 | 0.013 | 6903.559 | 6.935 |
| LaRhS | 0.2 | 9 | 0.046 | 1162.399 | 11.125 |
| LaRhSe | 0.151 | 9 | 0.035 | 1659.377 | 9.172 |
| LaRhTe | 2.653 | 9 | 0.613 | 21.554 | 8.343 |
| LaRuF | 3.016 | 9 | 0.697 | 17.112 | 6.573 |
| MnNiB | 19.488 | 3 | 9.369 | 0.81 | 34.551 |
| MoIrB | 1.567 | 3 | 0.753 | 44.496 | 34.243 |
| MoIrTl | 1.76 | 3 | 0.846 | 25.461 | 14.783 |
| MoOsPb | 2.861 | 3 | 1.375 | 12.925 | 16.684 |
| MoRhB | 1.143 | 3 | 0.55 | 66.078 | 41.263 |
| NbCoC | 5.046 | 3 | 2.426 | 7.541 | 45.876 |
| NbCoGe | 3.725 | 6 | 1.128 | 18.082 | 28.371 |
| NbCoPb | 3.292 | 3 | 1.582 | 8.505 | 13.353 |
| NbCoSi | 1.942 | 3 | 0.934 | 25.698 | 39.32 |
| NbCoSn | 3.463 | 9 | 0.8 | 26.213 | 21.288 |
| NbFeAs | 0.733 | 3 | 0.352 | 108.219 | 30.924 |
| NbFeBi | 1.133 | 3 | 0.544 | 43.533 | 14.71 |
| NbFeP | 0.662 | 3 | 0.318 | 139.896 | 43.627 |
| NbFeSb | 0.931 | 3 | 0.448 | 66.827 | 24.157 |
| NbIrC | 0.226 | 3 | 0.109 | 727.35 | 28.682 |
| NbIrGe | 0.366 | 3 | 0.176 | 310.733 | 24.695 |
| NbIrPb | 0.93 | 3 | 0.447 | 63.195 | 14.604 |
| NbIrSi | 0.608 | 3 | 0.292 | 155.315 | 30.969 |
| NbIrSn | 0.785 | 3 | 0.377 | 90.355 | 21.168 |
| NbNiAl | 0.416 | 3 | 0.2 | 201.235 | 26.615 |
| NbNiB | 2.127 | 3 | 1.023 | 24.087 | 39.501 |
| NbNiTl | 2.587 | 3 | 1.244 | 11.319 | 11.483 |
| NbOsAs | 0.733 | 3 | 0.352 | 112.42 | 25.856 |
| NbOsBi | 0.912 | 3 | 0.438 | 67.354 | 16.047 |
| NbOsP | 0.553 | 3 | 0.266 | 185.802 | 32.018 |
| NbOsSb | 0.706 | 3 | 0.34 | 111.517 | 23.51 |

| Compound | Value1 | Value2 | Value3 | Value4 | Value5 |
|---|---|---|---|---|---|
| NbPtAl | 0.487 | 3 | 0.234 | 175.48 | 22.028 |
| NbPtB | 0.741 | 3 | 0.356 | 115.026 | 27.043 |
| NbPtGa | 0.453 | 3 | 0.218 | 198.484 | 19.438 |
| NbPtIn | 0.647 | 3 | 0.311 | 103.858 | 16.101 |
| NbPtTl | 4.902 | 3 | 2.357 | 4.722 | 12.012 |
| NbRhC | 2.354 | 3 | 1.131 | 21.15 | 38.361 |
| NbRhGe | 0.677 | 3 | 0.326 | 113.678 | 27.798 |
| NbRhPb | 1.246 | 3 | 0.599 | 36.865 | 14.454 |
| NbRhSi | 0.662 | 3 | 0.318 | 125.215 | 36.656 |
| NbRhSn | 0.929 | 3 | 0.446 | 63.614 | 22.356 |
| NbRuAs | 0.659 | 3 | 0.317 | 124.545 | 30.583 |
| NbRuBi | 0.967 | 3 | 0.465 | 57.232 | 16.689 |
| NbRuP | 0.634 | 3 | 0.305 | 143.345 | 40.237 |
| NbRuSb | 0.768 | 3 | 0.369 | 91.293 | 25.982 |
| ScAgC | 2.384 | 3 | 1.146 | 10.417 | 13.918 |
| ScAuGe | 1.838 | 3 | 0.884 | 14.266 | 8.983 |
| ScAuSi | 1.815 | 3 | 0.872 | 16.161 | 12.259 |
| ScAuSn | 1.596 | 3 | 0.767 | 16.46 | 7.998 |
| ScCdIn | 2.127 | 3 | 1.023 | 6.778 | 5.12 |
| ScCdTl | 2.125 | 3 | 1.022 | 6.047 | 3.198 |
| ScCoS | 1 | 3 | 0.481 | 47.661 | 23.676 |
| ScCoSe | 0.411 | 3 | 0.197 | 165.818 | 16.688 |
| ScCoTe | 4.093 | 3 | 1.967 | 4.919 | 13.946 |
| ScCuC | 1.828 | 3 | 0.879 | 19.205 | 21.654 |
| ScHgAl | 1.155 | 3 | 0.555 | 22.261 | 7.324 |
| ScHgB | 1.625 | 3 | 0.781 | 16.472 | 8.589 |
| ScHgGa | 1.339 | 3 | 0.644 | 16.976 | 5.606 |
| ScHgIn | 1.301 | 3 | 0.626 | 16.335 | 4.911 |
| ScHgTl | 1.384 | 3 | 0.666 | 12.814 | 3.12 |
| ScNiAs | 2.667 | 3 | 1.282 | 9.834 | 17.218 |
| ScNiBi | 2.308 | 3 | 1.11 | 9.315 | 7.512 |
| ScNiP | 2.997 | 3 | 1.441 | 9.201 | 26.032 |
| ScNiSb | 2.222 | 3 | 1.068 | 11.478 | 13.143 |
| ScPtP | 0.119 | 3 | 0.057 | 1146.887 | 15.557 |
| ScPtSb | 2.984 | 3 | 1.434 | 8.11 | 11.6 |
| ScRhTe | 0.987 | 3 | 0.474 | 41.671 | 14.184 |
| TaCoC | 4.935 | 3 | 2.372 | 7.878 | 27.988 |
| TaCoGe | 3.19 | 3 | 1.533 | 12.12 | 21.702 |
| TaCoPb | 3.059 | 3 | 1.471 | 10.089 | 11.659 |
| TaCoSi | 3.444 | 3 | 1.656 | 11.588 | 27.858 |
| TaCoSn | 2.854 | 3 | 1.372 | 12.459 | 17.37 |
| TaFeAs | 0.827 | 3 | 0.397 | 94.991 | 23.159 |
| TaFeBi | 3.754 | 3 | 1.805 | 7.636 | 12.74 |
| TaFeP | 0.711 | 3 | 0.342 | 132.24 | 30.139 |
| TaFeSb | 1.036 | 9 | 0.239 | 182.019 | 19.694 |
| TaIrGe | 0.7 | 3 | 0.337 | 124.639 | 21.433 |
| TaIrSi | 0.629 | 3 | 0.302 | 155.655 | 25.763 |
| TaIrSn | 0.831 | 9 | 0.192 | 264.631 | 18.999 |
| TaNiAl | 1.341 | 3 | 0.645 | 37.487 | 19.322 |
| TaNiB | 1.764 | 3 | 0.848 | 33.493 | 26.023 |
| TaNiGa | 1.54 | 3 | 0.74 | 31.667 | 17.265 |
| TaNiIn | 1.574 | 3 | 0.757 | 26.423 | 13.59 |
| TaNiTl | 1.511 | 3 | 0.726 | 26.885 | 9.943 |
| TaOsBi | 2.411 | 3 | 1.159 | 16.368 | 14.417 |
| TaOsP | 0.183 | 3 | 0.088 | 1005.257 | 25.608 |
| TaOsSb | 0.724 | 3 | 0.348 | 113.447 | 20.732 |
| TaPtAl | 0.647 | 3 | 0.311 | 123.767 | 19.148 |

| | | | | | |
|---|---|---|---|---|---|
| TaPtB | 1.08 | 3 | 0.519 | 69.073 | 22.289 |
| TaPtGa | 0.895 | 3 | 0.43 | 77.138 | 17.454 |
| TaPtIn | 2.308 | 3 | 1.11 | 16.615 | 14.747 |
| TaPtTl | 3.518 | 3 | 1.691 | 8.342 | 11.369 |
| TaRhC | 0.32 | 3 | 0.154 | 425.167 | 25.588 |
| TaRhGe | 0.853 | 3 | 0.41 | 85.417 | 22.374 |
| TaRhPb | 2.86 | 3 | 1.375 | 11.292 | 12.957 |
| TaRhSi | 0.64 | 3 | 0.308 | 139.215 | 27.616 |
| TaRhSn | 1.354 | 3 | 0.651 | 38.52 | 18.934 |
| TaRuAs | 0.457 | 3 | 0.22 | 225.25 | 23.838 |
| TaRuBi | 0.365 | 3 | 0.175 | 260.342 | 14.695 |
| TaRuP | 0.647 | 3 | 0.311 | 144.094 | 29.147 |
| TaRuSb | 0.743 | 3 | 0.357 | 101.495 | 21.745 |
| TiAgAl | 1.323 | 3 | 0.636 | 23.199 | 14.475 |
| TiAgB | 2.208 | 3 | 1.062 | 14.625 | 20.379 |
| TiAgGa | 1.625 | 3 | 0.781 | 17.157 | 11.304 |
| TiAgIn | 1.929 | 3 | 0.927 | 11.668 | 8.723 |
| TiAuAl | 3.163 | 3 | 1.521 | 7.681 | 13.674 |
| TiAuB | 3.85 | 3 | 1.851 | 7.358 | 17.467 |
| TiAuGa | 3.196 | 3 | 1.537 | 7.535 | 11.444 |
| TiAuIn | 2.925 | 3 | 1.406 | 7.593 | 9.318 |
| TiAuTl | 2.888 | 3 | 1.389 | 7.046 | 6.414 |
| TiCoAs | 6.119 | 3 | 2.942 | 3.909 | 27.012 |
| TiCoBi | 5.518 | 3 | 2.653 | 3.439 | 11.586 |
| TiCoP | 5.408 | 3 | 2.6 | 5.305 | 41.292 |
| TiCoSb | 5.346 | 3 | 2.57 | 4.167 | 20.007 |
| TiCuB | 1.508 | 3 | 0.725 | 31.047 | 30.147 |
| TiCuGa | 1.098 | 3 | 0.528 | 35.651 | 14.812 |
| TiFePo | 0.952 | 3 | 0.458 | 48.752 | 12.179 |
| TiFeS | 1.777 | 3 | 0.854 | 26.948 | 37.198 |
| TiFeSe | 1.23 | 3 | 0.591 | 42.092 | 25.25 |
| TiFeTe | 1.952 | 3 | 0.939 | 19.394 | 20.72 |
| TiIrAs | 0.984 | 9 | 0.227 | 189.135 | 20.877 |
| TiIrBi | 2.723 | 3 | 1.309 | 11.076 | 12.211 |
| TiIrP | 1.344 | 3 | 0.646 | 43.503 | 26.911 |
| TiIrSb | 3.139 | 3 | 1.509 | 10.327 | 18.847 |
| TiNiC | 3.714 | 3 | 1.786 | 10.253 | 42.616 |
| TiNiGe | 2.661 | 3 | 1.279 | 12.185 | 22.947 |
| TiNiPb | 2.95 | 3 | 1.418 | 7.873 | 9.439 |
| TiNiSi | 2.49 | 3 | 1.197 | 14.6 | 34.866 |
| TiNiSn | 2.688 | 3 | 1.292 | 10.23 | 16.271 |
| TiOsTe | 0.27 | 3 | 0.13 | 415.51 | 19.158 |
| TiPtC | 0.225 | 3 | 0.108 | 611.196 | 21.407 |
| TiPtGe | 4.282 | 3 | 2.058 | 6.274 | 17.623 |
| TiPtPb | 8.32 | 3 | 4 | 1.819 | 9.516 |
| TiPtSi | 1.559 | 3 | 0.75 | 31.021 | 23.056 |
| TiPtSn | 7.17 | 3 | 3.447 | 2.607 | 14.813 |
| TiRhAs | 1.86 | 3 | 0.894 | 22.306 | 24.728 |
| TiRhBi | 7.145 | 3 | 3.435 | 2.355 | 12.402 |
| TiRhP | 1.585 | 3 | 0.762 | 31.349 | 34.664 |
| TiRhSb | 4.431 | 3 | 2.13 | 5.536 | 20.385 |
| TiRuTe | 2.81 | 3 | 1.351 | 11.325 | 21.373 |
| VCoC | 1.278 | 3 | 0.615 | 62.521 | 54.667 |
| VCoGe | 1.898 | 3 | 0.913 | 24.372 | 28.749 |
| VCoPb | 5.496 | 3 | 2.642 | 3.659 | 11.589 |
| VCoSi | 1.331 | 3 | 0.64 | 45.403 | 43.617 |
| VCoSn | 3.531 | 3 | 1.697 | 8.095 | 20.19 |

| Compound | | | | | |
|---|---|---|---|---|---|
| VFeAs | 1.493 | 3 | 0.718 | 37.493 | 33.002 |
| VFeBi | 4.83 | 3 | 2.322 | 4.697 | 13.387 |
| VFeP | 1.088 | 3 | 0.523 | 68.179 | 50.492 |
| VFeSb | 1.705 | 9 | 0.394 | 78.98 | 23.948 |
| VIrC | 2.387 | 3 | 1.148 | 22.358 | 29.679 |
| VIrGe | 1.271 | 3 | 0.611 | 47.462 | 23.481 |
| VIrSi | 0.995 | 3 | 0.478 | 74.254 | 30.364 |
| VIrSn | 1.547 | 3 | 0.744 | 31.143 | 19.14 |
| VNiAl | 6.03 | 3 | 2.899 | 3.489 | 27.727 |
| VNiB | 4.588 | 3 | 2.206 | 7.666 | 43.682 |
| VNiGa | 4.611 | 3 | 2.217 | 5.411 | 21.417 |
| VNiIn | 8.035 | 3 | 3.863 | 1.956 | 14.656 |
| VOsAs | 1.209 | 3 | 0.581 | 53.809 | 25.884 |
| VOsBi | 3.802 | 3 | 1.828 | 7.531 | 14.302 |
| VOsP | 0.984 | 3 | 0.473 | 80.583 | 32.945 |
| VOsSb | 1.788 | 3 | 0.86 | 26.834 | 21.861 |
| VPtAl | 1.141 | 3 | 0.548 | 46.467 | 19.725 |
| VPtB | 1.221 | 3 | 0.587 | 54.953 | 26.006 |
| VPtGa | 1.835 | 3 | 0.882 | 23.034 | 17.137 |
| VPtTl | 10.568 | 3 | 5.08 | 1.332 | 9.599 |
| VReTe | 1.92 | 3 | 0.923 | 23.038 | 20.366 |
| VRhC | 1.26 | 3 | 0.606 | 55.722 | 41.134 |
| VRhGe | 1.367 | 6 | 0.414 | 76.818 | 26.799 |
| VRhPb | 7.717 | 3 | 3.71 | 2.226 | 12.653 |
| VRhSi | 1.133 | 3 | 0.545 | 55.437 | 38.003 |
| VRhSn | 2.099 | 3 | 1.009 | 17.689 | 20.572 |
| VRuAs | 1.297 | 3 | 0.624 | 45.022 | 31.356 |
| VRuBi | 3.691 | 3 | 1.775 | 7.278 | 15.129 |
| VRuP | 1.034 | 3 | 0.497 | 69.881 | 43.453 |
| VRuSb | 1.482 | 3 | 0.712 | 32.719 | 24.605 |
| VTcTe | 1.23 | 3 | 0.592 | 41.725 | 23.452 |
| WCoAl | 0.584 | 3 | 0.281 | 157.497 | 25.067 |
| WCoB | 0.415 | 3 | 0.199 | 353.811 | 34.205 |
| WCoGa | 0.554 | 3 | 0.266 | 176.202 | 22.46 |
| WCoIn | 2.412 | 9 | 0.558 | 49.424 | 17.223 |
| WFeC | 0.497 | 3 | 0.239 | 280.748 | 34.998 |
| WFeGe | 0.417 | 3 | 0.201 | 291.356 | 26.57 |
| WFeSi | 0.298 | 3 | 0.143 | 521.644 | 34.286 |
| WFeSn | 0.602 | 3 | 0.289 | 145.838 | 21.081 |
| WIrAl | 1.393 | 3 | 0.67 | 46.587 | 24.788 |
| WIrB | 0.342 | 3 | 0.164 | 463.913 | 29.443 |
| WIrGa | 1.057 | 3 | 0.508 | 71.557 | 22.785 |
| WIrIn | 0.521 | 3 | 0.251 | 183.891 | 19.291 |
| WIrTl | 0.819 | 3 | 0.394 | 87.779 | 14.692 |
| WOsGe | 1.606 | 3 | 0.772 | 0.77 | 0.188 |
| WOsPb | 0.58 | 3 | 0.279 | 152.608 | 16.166 |
| WOsSn | 1.434 | 3 | 0.689 | 43.498 | 22.495 |
| WRhAl | 1.762 | 3 | 0.847 | 29.48 | 25.705 |
| WRhB | 0.28 | 3 | 0.135 | 582.309 | 32.079 |
| WRhGa | 0.644 | 3 | 0.31 | 136.165 | 22.988 |
| WRhIn | 0.763 | 3 | 0.367 | 93.28 | 18.786 |
| WRhTl | 0.929 | 3 | 0.446 | 65.967 | 13.97 |
| WRuPb | 0.578 | 3 | 0.278 | 142.292 | 16.132 |
| YAgC | 0.862 | 3 | 0.414 | 39.584 | 9.571 |
| YAgSi | 1.265 | 3 | 0.608 | 20.419 | 10.09 |
| YAuGe | 0.655 | 6 | 0.198 | 113.613 | 6.864 |
| YAuSi | 1.525 | 6 | 0.462 | 34.961 | 8.958 |

| Compound | $m_{DOS}(m_e)$ | $N_b$ | $m_b(m_e)$ | $\mu_p(\frac{cm^2}{Vs})$ | $\kappa(\frac{W}{mK})$ |
|---|---|---|---|---|---|
| YAuSn | 1.082 | 6 | 0.328 | 51.097 | 6.347 |
| YCoS | 0.303 | 3 | 0.146 | 235.286 | 14.88 |
| YCoTe | 4.193 | 3 | 2.016 | 4.039 | 10.156 |
| YCuC | 1.403 | 3 | 0.674 | 22.965 | 13.348 |
| YCuGe | 1.327 | 3 | 0.638 | 19.454 | 8.906 |
| YCuSi | 1.173 | 3 | 0.564 | 25.146 | 12.422 |
| YHgAl | 0.926 | 6 | 0.28 | 54.494 | 5.878 |
| YHgB | 1.374 | 3 | 0.66 | 17.45 | 6.348 |
| YNiAs | 2.58 | 3 | 1.24 | 8.653 | 11.81 |
| YNiBi | 2.033 | 3 | 0.977 | 9.807 | 5.961 |
| YNiP | 2.557 | 3 | 1.229 | 9.573 | 16.19 |
| YNiSb | 2.102 | 3 | 1.01 | 10.664 | 9.611 |
| YPtAs | 0.663 | 3 | 0.319 | 67.794 | 9.285 |
| YPtBi | 0.774 | 3 | 0.372 | 46.457 | 6.182 |
| YPtP | 0.236 | 3 | 0.114 | 338.009 | 11.022 |
| YPtSb | 0.064 | 3 | 0.031 | 2203.969 | 8.753 |
| YRhTe | 0.332 | 3 | 0.159 | 182.777 | 10.473 |
| ZrAgB | 1.136 | 6 | 0.344 | 75.632 | 17.937 |
| ZrAuAl | 1.212 | 6 | 0.367 | 63.459 | 13.318 |
| ZrAuB | 1.729 | 6 | 0.524 | 45.178 | 15.809 |
| ZrAuGa | 1.362 | 6 | 0.413 | 53.135 | 11.294 |
| ZrAuTl | 1.425 | 3 | 0.685 | 21.3 | 7.128 |
| ZrCoAs | 3.977 | 3 | 1.912 | 6.873 | 22.023 |
| ZrCoBi | 3.405 | 3 | 1.637 | 6.802 | 10.823 |
| ZrCoP | 4.289 | 3 | 2.062 | 6.764 | 30.713 |
| ZrCoSb | 3.401 | 3 | 1.635 | 7.75 | 17.458 |
| ZrCuB | 0.817 | 6 | 0.247 | 143.089 | 23.408 |
| ZrFePo | 0.335 | 3 | 0.161 | 221.121 | 11.063 |
| ZrFeS | 0.468 | 3 | 0.225 | 178.142 | 27.06 |
| ZrFeSe | 0.313 | 3 | 0.151 | 299.739 | 20.319 |
| ZrFeTe | 7.255 | 3 | 3.488 | 2.518 | 17.543 |
| ZrIrAs | 0.118 | 3 | 0.057 | 1398.387 | 18.216 |
| ZrIrBi | 0.112 | 6 | 0.034 | 2560.732 | 11.814 |
| ZrIrP | 0.191 | 9 | 0.044 | 2190.353 | 22.455 |
| ZrIrSb | 4.586 | 6 | 1.389 | 11.057 | 17.285 |
| ZrNiC | 2.625 | 3 | 1.262 | 15.109 | 30.211 |
| ZrNiGe | 1.774 | 3 | 0.853 | 20.922 | 19.182 |
| ZrNiPb | 1.698 | 9 | 0.393 | 53.479 | 9.374 |
| ZrNiSi | 1.942 | 3 | 0.934 | 19.491 | 26.606 |
| ZrNiSn | 1.781 | 3 | 0.856 | 18.318 | 14.748 |
| ZrPtPb | 2.039 | 3 | 0.98 | 14.95 | 9.775 |
| ZrPtSi | 2.737 | 6 | 0.829 | 24.653 | 20.274 |
| ZrPtSn | 2.059 | 3 | 0.99 | 16.519 | 14.412 |
| ZrRhAs | 3.305 | 3 | 1.589 | 8.892 | 21.726 |
| ZrRhBi | 4.045 | 6 | 1.225 | 10.915 | 12.261 |
| ZrRhP | 4.439 | 6 | 1.344 | 12.348 | 28.405 |
| ZrRhSb | 3.712 | 3 | 1.785 | 6.933 | 18.575 |
| ZrRuTe | 0.728 | 3 | 0.35 | 81.82 | 19.157 |

Table 2: Electronic Transport, p-type

| Compound | $m_{DOS}(m_e)$ | $N_b$ | $m_b(m_e)$ | $\mu_p(\frac{cm^2}{Vs})$ | $\kappa(\frac{W}{mK})$ |
|---|---|---|---|---|---|
| AcNiAs | 9.703 | 14 | 1.67 | 4.236 | 5.876 |
| AcNiP | 9.032 | 8 | 2.258 | 2.918 | 7.25 |
| AcNiSb | 1.684 | 14 | 0.29 | 53.657 | 5.157 |
| CoMoAl | 1.08 | 14 | 0.186 | 211.221 | 22.953 |

| | | | | | |
|---|---|---|---|---|---|
| CrCoAl | 2.786 | 14 | 0.48 | 58.757 | 31.889 |
| CrCoB | 3.145 | 8 | 0.786 | 41.312 | 51.699 |
| CrCoGa | 2.492 | 14 | 0.429 | 71.815 | 24.873 |
| CrCoIn | 2.853 | 14 | 0.491 | 48.575 | 17.093 |
| CrFeC | 4.379 | 8 | 1.095 | 28.327 | 61.675 |
| CrFeGe | 9.713 | 8 | 2.428 | 6.026 | 31.988 |
| CrFeSi | 6.914 | 8 | 1.729 | 10.964 | 48.319 |
| CrFeSn | 9.504 | 14 | 1.636 | 8.967 | 21.467 |
| CrIrAl | 4.885 | 14 | 0.841 | 27.872 | 23.511 |
| CrIrB | 3.366 | 8 | 0.841 | 36.896 | 31.7 |
| CrIrGa | 4.707 | 14 | 0.81 | 30.109 | 21.008 |
| CrIrIn | 4.323 | 14 | 0.744 | 28.927 | 16.043 |
| CrIrTl | 2.488 | 17 | 0.376 | 203.945 | 97.764 |
| CrOsGe | 4.815 | 8 | 1.204 | 18.398 | 25.716 |
| CrOsPb | 5.266 | 14 | 0.907 | 20.864 | 12.579 |
| CrOsSi | 4.83 | 8 | 1.207 | 19.737 | 32.723 |
| CrOsSn | 4.83 | 14 | 0.832 | 27.579 | 20.175 |
| CrReAs | 5.28 | 8 | 1.32 | 15.663 | 25.314 |
| CrReP | 4.878 | 8 | 1.219 | 19.511 | 32.444 |
| CrReSb | 4.122 | 8 | 1.03 | 19.979 | 20.774 |
| CrRhAl | 6.964 | 14 | 1.199 | 14.577 | 28.548 |
| CrRhB | 1.911 | 8 | 0.478 | 78.964 | 41.078 |
| CrRhGa | 4.352 | 14 | 0.749 | 30.187 | 23.483 |
| CrRuC | 3.567 | 8 | 0.892 | 34.408 | 47.53 |
| CrRuGe | 4.299 | 14 | 0.74 | 35.058 | 30.488 |
| CrRuPb | 3.934 | 14 | 0.677 | 30.53 | 13.891 |
| CrRuSi | 3.889 | 14 | 0.669 | 43.945 | 42.159 |
| CrRuSn | 3.846 | 14 | 0.662 | 35.524 | 22.551 |
| CrTcAs | 3.461 | 8 | 0.865 | 27.919 | 31.938 |
| CrTcP | 5.834 | 8 | 1.458 | 14.037 | 43.702 |
| CrTcSb | 4.227 | 8 | 1.057 | 17.999 | 24.251 |
| FeMoSi | 3.714 | 8 | 0.929 | 23.638 | 33.98 |
| HfAgAl | 6.052 | 3 | 2.909 | 2.569 | 10.457 |
| HfAgB | 2.139 | 3 | 1.028 | 15.615 | 13.008 |
| HfAgGa | 3.658 | 3 | 1.759 | 5.523 | 9.005 |
| HfAuAl | 8.699 | 3 | 4.182 | 1.781 | 11.248 |
| HfAuB | 2.319 | 3 | 1.115 | 15.557 | 12.876 |
| HfAuGa | 4.244 | 3 | 2.04 | 5.223 | 9.886 |
| HfAuIn | 3.379 | 3 | 1.624 | 6.701 | 8.702 |
| HfAuTl | 2.486 | 3 | 1.195 | 9.859 | 6.512 |
| HfCoAs | 8.412 | 8 | 2.103 | 6.344 | 16.463 |
| HfCoBi | 6.507 | 11 | 1.316 | 9.972 | 9.148 |
| HfCoP | 8.331 | 8 | 2.083 | 7.05 | 20.825 |
| HfCoSb | 9.607 | 8 | 2.402 | 4.632 | 13.888 |
| HfCuB | 2.575 | 3 | 1.238 | 13.613 | 15.441 |
| HfFeTe | 5.758 | 8 | 1.44 | 10.113 | 14 |
| HfIrSb | 3.024 | 3 | 1.454 | 10.878 | 14.884 |
| HfNiC | 1.449 | 3 | 0.697 | 38.044 | 18.405 |
| HfNiGe | 2.783 | 3 | 1.338 | 11.334 | 14.278 |
| HfNiSi | 4.536 | 3 | 2.181 | 5.868 | 18.593 |
| HfNiSn | 2.761 | 3 | 1.328 | 10.141 | 11.798 |
| HfPdSn | 2.186 | 3 | 1.051 | 14.06 | 11.835 |
| HfPtGe | 2.285 | 3 | 1.098 | 16.099 | 13.747 |
| HfPtPb | 2.037 | 3 | 0.979 | 15.807 | 8.779 |
| HfPtSi | 3.227 | 3 | 1.552 | 10.272 | 16.772 |
| HfPtSn | 2.443 | 3 | 1.175 | 13.55 | 12.551 |
| HfRhAs | 1.778 | 3 | 0.855 | 23.533 | 16.498 |

| | | | | |
|---|---|---|---|---|
| HfRhBi | 1.811 | 3 | 0.871 | 18.968 | 10.308 |
| HfRhP | 1.884 | 3 | 0.906 | 23.287 | 20.229 |
| HfRhSb | 2.129 | 3 | 1.023 | 16.877 | 15.157 |
| HfRuTe | 4.851 | 3 | 2.332 | 4.974 | 15.359 |
| LaAuSi | 1.121 | 6 | 0.339 | 46.514 | 6.657 |
| LaAuSn | 1.417 | 14 | 0.244 | 68.231 | 4.964 |
| LaCoS | 10.086 | 10 | 2.173 | 3.697 | 11.445 |
| LaCoSe | 10.47 | 10 | 2.256 | 3.214 | 8.946 |
| LaCoTe | 9.49 | 1 | 9.49 | 0.345 | 7.761 |
| LaCuC | 4.779 | 6 | 1.447 | 6.111 | 9.033 |
| LaFeF | 11.622 | 5 | 3.975 | 0.533 | 2.357 |
| LaHgAl | 4.102 | 17 | 0.62 | 13.964 | 4.438 |
| LaHgIn | 4.6 | 17 | 0.696 | 10.792 | 3.311 |
| LaNiAs | 7.573 | 1 | 7.573 | 0.496 | 8.716 |
| LaNiP | 8.187 | 10 | 1.764 | 4.804 | 11.431 |
| LaNiSb | 1.674 | 1 | 1.674 | 4.352 | 7.328 |
| LaPtSb | 1.176 | 14 | 0.202 | 112.472 | 6.935 |
| LaRhS | 4.893 | 5 | 1.673 | 5.334 | 11.125 |
| LaRhSe | 4.274 | 2 | 2.692 | 2.459 | 9.172 |
| LaRhTe | 3.247 | 2 | 2.046 | 3.537 | 8.343 |
| LaRuF | 8.297 | 1 | 8.297 | 0.417 | 6.573 |
| MnNiB | 21.477 | 6 | 6.504 | 1.401 | 34.551 |
| MoIrB | 2.326 | 8 | 0.582 | 65.625 | 34.243 |
| MoIrTl | 1.31 | 6 | 0.397 | 79.26 | 14.783 |
| MoOsPb | 3.887 | 14 | 0.669 | 38.087 | 16.684 |
| MoRhB | 1.303 | 14 | 0.224 | 253.598 | 41.263 |
| NbCoC | 4.881 | 8 | 1.22 | 21.14 | 45.876 |
| NbCoGe | 7.198 | 14 | 1.239 | 15.705 | 28.371 |
| NbCoPb | 7.896 | 14 | 1.359 | 10.682 | 13.353 |
| NbCoSi | 7.016 | 6 | 2.125 | 7.486 | 39.32 |
| NbCoSn | 9.107 | 14 | 1.568 | 9.564 | 21.288 |
| NbFeAs | 4.522 | 8 | 1.131 | 18.827 | 30.924 |
| NbFeBi | 4.49 | 8 | 1.122 | 14.706 | 14.71 |
| NbFeP | 4.125 | 8 | 1.031 | 23.978 | 43.627 |
| NbFeSb | 4.652 | 10 | 1.002 | 19.955 | 24.157 |
| NbIrC | 2.772 | 8 | 0.693 | 45.077 | 28.682 |
| NbIrGe | 5.902 | 14 | 1.016 | 22.42 | 24.695 |
| NbIrPb | 5.466 | 14 | 0.941 | 20.693 | 14.604 |
| NbIrSi | 4.626 | 14 | 0.796 | 34.512 | 30.969 |
| NbIrSn | 4.541 | 14 | 0.782 | 30.29 | 21.168 |
| NbNiAl | 0.446 | 6 | 0.135 | 362.004 | 26.615 |
| NbNiB | 2.848 | 6 | 0.862 | 31.099 | 39.501 |
| NbNiTl | 1.248 | 9 | 0.289 | 101.285 | 11.483 |
| NbOsAs | 2.099 | 8 | 0.525 | 61.845 | 25.856 |
| NbOsBi | 2.562 | 8 | 0.641 | 38.144 | 16.047 |
| NbOsP | 2.14 | 8 | 0.535 | 65.079 | 32.018 |
| NbOsSb | 2.564 | 8 | 0.641 | 42.995 | 23.51 |
| NbPtAl | 2.059 | 6 | 0.624 | 40.375 | 22.028 |
| NbPtB | 2.481 | 6 | 0.751 | 37.532 | 27.043 |
| NbPtGa | 2.41 | 6 | 0.73 | 32.336 | 19.438 |
| NbPtIn | 1.538 | 6 | 0.466 | 56.706 | 16.101 |
| NbPtTl | 1.375 | 6 | 0.416 | 63.572 | 12.012 |
| NbRhC | 3.714 | 8 | 0.928 | 28.454 | 38.361 |
| NbRhGe | 2.821 | 6 | 0.854 | 26.74 | 27.798 |
| NbRhPb | 6.426 | 6 | 1.946 | 6.297 | 14.454 |
| NbRhSi | 2.086 | 6 | 0.632 | 44.738 | 36.656 |
| NbRhSn | 5.948 | 6 | 1.801 | 7.85 | 22.356 |

| Compound | Col2 | Col3 | Col4 | Col5 | Col6 |
|---|---|---|---|---|---|
| NbRuAs | 2.604 | 8 | 0.651 | 42.265 | 30.583 |
| NbRuBi | 3.185 | 8 | 0.796 | 25.529 | 16.689 |
| NbRuP | 2.579 | 8 | 0.645 | 46.566 | 40.237 |
| NbRuSb | 2.952 | 8 | 0.738 | 32.326 | 25.982 |
| ScAgC | 0.956 | 3 | 0.459 | 41.05 | 13.918 |
| ScAuGe | 1.572 | 3 | 0.756 | 18.035 | 8.983 |
| ScAuSi | 1.783 | 3 | 0.857 | 16.592 | 12.259 |
| ScAuSn | 1.752 | 3 | 0.842 | 14.306 | 7.998 |
| ScCdIn | 2.674 | 3 | 1.286 | 4.808 | 5.12 |
| ScCdTl | 2.42 | 3 | 1.163 | 4.976 | 3.198 |
| ScCoS | 5.325 | 2 | 3.354 | 2.585 | 23.676 |
| ScCoSe | 4.677 | 3 | 2.249 | 4.312 | 16.688 |
| ScCoTe | 3.192 | 3 | 1.535 | 7.14 | 13.946 |
| ScCuC | 0.896 | 3 | 0.431 | 55.983 | 21.654 |
| ScHgAl | 3.62 | 3 | 1.741 | 4.014 | 7.324 |
| ScHgB | 1.802 | 3 | 0.866 | 14.112 | 8.589 |
| ScHgGa | 2.791 | 3 | 1.342 | 5.642 | 5.606 |
| ScHgIn | 2.662 | 3 | 1.28 | 5.583 | 4.911 |
| ScHgTl | 2.371 | 3 | 1.14 | 5.716 | 3.12 |
| ScNiAs | 1.418 | 3 | 0.682 | 25.378 | 17.218 |
| ScNiBi | 1.312 | 3 | 0.631 | 21.747 | 7.512 |
| ScNiP | 1.607 | 3 | 0.773 | 23.428 | 26.032 |
| ScNiSb | 1.551 | 3 | 0.746 | 19.676 | 13.143 |
| ScPtP | 1.09 | 2 | 0.687 | 27.739 | 15.557 |
| ScPtSb | 1.257 | 3 | 0.604 | 29.667 | 11.6 |
| ScRhTe | 1.18 | 3 | 0.567 | 31.86 | 14.184 |
| TaCoC | 5.02 | 8 | 1.255 | 20.475 | 27.988 |
| TaCoGe | 7.87 | 14 | 1.355 | 14.595 | 21.702 |
| TaCoPb | 9.4 | 17 | 1.422 | 10.612 | 11.659 |
| TaCoSi | 8.629 | 8 | 2.157 | 7.792 | 27.858 |
| TaCoSn | 9.229 | 14 | 1.589 | 9.997 | 17.37 |
| TaFeAs | 4.144 | 8 | 1.036 | 22.576 | 23.159 |
| TaFeBi | 4.754 | 8 | 1.188 | 14.288 | 12.74 |
| TaFeP | 4.079 | 8 | 1.02 | 25.664 | 30.139 |
| TaFeSb | 4.777 | 8 | 1.194 | 16.347 | 19.694 |
| TaIrGe | 4.481 | 8 | 1.12 | 20.525 | 21.433 |
| TaIrSi | 6.045 | 8 | 1.511 | 13.927 | 25.763 |
| TaIrSn | 6.442 | 8 | 1.611 | 10.906 | 18.999 |
| TaNiAl | 0.988 | 14 | 0.17 | 276.409 | 19.322 |
| TaNiB | 4.138 | 11 | 0.837 | 34.179 | 26.023 |
| TaNiGa | 1.29 | 14 | 0.222 | 192.613 | 17.265 |
| TaNiIn | 1.463 | 11 | 0.296 | 108.21 | 13.59 |
| TaNiTl | 2.564 | 3 | 1.233 | 12.164 | 9.943 |
| TaOsBi | 2.528 | 8 | 0.632 | 40.667 | 14.417 |
| TaOsP | 2.372 | 8 | 0.593 | 57.653 | 25.608 |
| TaOsSb | 2.516 | 8 | 0.629 | 46.648 | 20.732 |
| TaPtAl | 2.776 | 8 | 0.694 | 37.112 | 19.148 |
| TaPtB | 6.746 | 3 | 3.243 | 4.423 | 22.289 |
| TaPtGa | 4.401 | 11 | 0.89 | 25.941 | 17.454 |
| TaPtIn | 4.024 | 3 | 1.935 | 7.219 | 14.747 |
| TaPtTl | 3.945 | 3 | 1.896 | 7.025 | 11.369 |
| TaRhC | 5.181 | 11 | 1.047 | 23.982 | 25.588 |
| TaRhGe | 9.161 | 11 | 1.852 | 8.893 | 22.374 |
| TaRhPb | 6.398 | 3 | 3.076 | 3.375 | 12.957 |
| TaRhSi | 7.415 | 8 | 1.854 | 9.41 | 27.616 |
| TaRhSn | 8.942 | 11 | 1.808 | 8.321 | 18.934 |
| TaRuAs | 2.816 | 8 | 0.704 | 39.233 | 23.838 |

| | | | | | |
|---|---|---|---|---|---|
| TaRuBi | 3.106 | 8 | 0.777 | 27.956 | 14.695 |
| TaRuP | 2.795 | 8 | 0.699 | 42.766 | 29.147 |
| TaRuSb | 3.144 | 8 | 0.786 | 31.117 | 21.745 |
| TiAgAl | 8.202 | 3 | 3.943 | 1.503 | 14.475 |
| TiAgB | 3.813 | 3 | 1.833 | 6.445 | 20.379 |
| TiAgGa | 6.771 | 3 | 3.255 | 2.017 | 11.304 |
| TiAgIn | 6.134 | 3 | 2.949 | 2.057 | 8.723 |
| TiAuAl | 12.11 | 3 | 5.822 | 1.025 | 13.674 |
| TiAuB | 4.128 | 3 | 1.984 | 6.628 | 17.467 |
| TiAuGa | 8.274 | 3 | 3.978 | 1.809 | 11.444 |
| TiAuIn | 7.126 | 3 | 3.426 | 1.997 | 9.318 |
| TiAuTl | 6.068 | 3 | 2.917 | 2.314 | 6.414 |
| TiCoAs | 9.759 | 3 | 4.692 | 1.941 | 27.012 |
| TiCoBi | 7.742 | 3 | 3.722 | 2.07 | 11.586 |
| TiCoP | 10.337 | 11 | 2.09 | 7.361 | 41.292 |
| TiCoSb | 10.583 | 11 | 2.14 | 5.487 | 20.007 |
| TiCuB | 4.091 | 3 | 1.967 | 6.947 | 30.147 |
| TiCuGa | 6.622 | 3 | 3.183 | 2.406 | 14.812 |
| TiFePo | 7.996 | 11 | 1.617 | 7.34 | 12.179 |
| TiFeS | 7.086 | 8 | 1.772 | 9.027 | 37.198 |
| TiFeSe | 6.03 | 8 | 1.508 | 10.334 | 25.25 |
| TiFeTe | 6.729 | 8 | 1.682 | 8.081 | 20.72 |
| TiIrAs | 4.69 | 3 | 2.255 | 6.057 | 20.877 |
| TiIrBi | 4.564 | 3 | 2.194 | 5.104 | 12.211 |
| TiIrP | 4.86 | 3 | 2.337 | 6.329 | 26.911 |
| TiIrSb | 4.927 | 3 | 2.369 | 5.251 | 18.847 |
| TiNiC | 2.717 | 3 | 1.306 | 16.385 | 42.616 |
| TiNiGe | 5.285 | 3 | 2.541 | 4.354 | 22.947 |
| TiNiPb | 3.865 | 3 | 1.858 | 5.249 | 9.439 |
| TiNiSi | 7.874 | 3 | 3.785 | 2.597 | 34.866 |
| TiNiSn | 5.026 | 3 | 2.416 | 4.002 | 16.271 |
| TiOsTe | 6.4 | 3 | 3.077 | 3.598 | 19.158 |
| TiPtC | 2.395 | 2 | 1.509 | 11.756 | 21.407 |
| TiPtGe | 4.539 | 3 | 2.182 | 5.748 | 17.623 |
| TiPtPb | 4.069 | 3 | 1.956 | 5.317 | 9.516 |
| TiPtSi | 6.095 | 3 | 2.93 | 4.014 | 23.056 |
| TiPtSn | 5.232 | 3 | 2.515 | 4.183 | 14.813 |
| TiRhAs | 4.254 | 3 | 2.045 | 6.447 | 24.728 |
| TiRhBi | 4.25 | 3 | 2.043 | 5.134 | 12.402 |
| TiRhP | 3.384 | 3 | 1.627 | 10.053 | 34.664 |
| TiRhSb | 4.6 | 3 | 2.211 | 5.234 | 20.385 |
| TiRuTe | 5.747 | 3 | 2.763 | 3.872 | 21.373 |
| VCoC | 5.644 | 11 | 1.141 | 24.714 | 54.667 |
| VCoGe | 11.765 | 8 | 2.941 | 4.213 | 28.749 |
| VCoPb | 11.841 | 3 | 5.692 | 1.157 | 11.589 |
| VCoSi | 9.743 | 8 | 2.436 | 6.11 | 43.617 |
| VCoSn | 13.44 | 11 | 2.717 | 3.997 | 20.19 |
| VFeAs | 5.431 | 8 | 1.358 | 14.41 | 33.002 |
| VFeBi | 7.428 | 8 | 1.857 | 6.567 | 13.387 |
| VFeP | 4.795 | 8 | 1.199 | 19.654 | 50.492 |
| VFeSb | 6.419 | 8 | 1.605 | 9.608 | 23.948 |
| VIrC | 3.309 | 8 | 0.827 | 36.538 | 29.679 |
| VIrGe | 8.13 | 11 | 1.644 | 10.757 | 23.481 |
| VIrSi | 7.112 | 11 | 1.438 | 14.245 | 30.364 |
| VIrSn | 9.536 | 11 | 1.928 | 7.459 | 19.14 |
| VNiAl | 1.411 | 8 | 0.353 | 82.188 | 27.727 |
| VNiB | 7.435 | 3 | 3.575 | 3.715 | 43.682 |

| | | | | | |
|---|---|---|---|---|---|
| VNiGa | 2.466 | 11 | 0.499 | 50.713 | 21.417 |
| VNiIn | 2.816 | 3 | 1.354 | 9.425 | 14.656 |
| VOsAs | 3.18 | 3 | 1.529 | 12.615 | 25.884 |
| VOsBi | 3.644 | 3 | 1.752 | 8.027 | 14.302 |
| VOsP | 3.072 | 3 | 1.477 | 14.617 | 32.945 |
| VOsSb | 3.483 | 11 | 0.704 | 36.195 | 21.861 |
| VPtAl | 3.013 | 11 | 0.609 | 39.683 | 19.725 |
| VPtB | 6.763 | 3 | 3.252 | 4.216 | 26.006 |
| VPtGa | 4.723 | 3 | 2.271 | 5.58 | 17.137 |
| VPtTl | 4.764 | 3 | 2.29 | 4.403 | 9.599 |
| VReTe | 2.753 | 11 | 0.557 | 49.183 | 20.366 |
| VRhC | 5.386 | 3 | 2.589 | 6.306 | 41.134 |
| VRhGe | 8.75 | 3 | 4.206 | 2.371 | 26.799 |
| VRhPb | 12.27 | 3 | 5.899 | 1.11 | 12.653 |
| VRhSi | 8.628 | 3 | 4.148 | 2.637 | 38.003 |
| VRhSn | 10.81 | 3 | 5.197 | 1.513 | 20.572 |
| VRuAs | 3.868 | 11 | 0.782 | 32.066 | 31.356 |
| VRuBi | 6.888 | 11 | 1.393 | 10.469 | 15.129 |
| VRuP | 3.658 | 3 | 1.759 | 10.506 | 43.453 |
| VRuSb | 5.215 | 11 | 1.054 | 18.175 | 24.605 |
| VTcTe | 2.884 | 11 | 0.583 | 42.645 | 23.452 |
| WCoAl | 1.818 | 6 | 0.551 | 57.275 | 25.067 |
| WCoB | 2.901 | 8 | 0.725 | 51.013 | 34.205 |
| WCoGa | 2.39 | 6 | 0.724 | 39.291 | 22.46 |
| WCoIn | 2.179 | 6 | 0.66 | 38.372 | 17.223 |
| WFeC | 3.373 | 8 | 0.843 | 42.346 | 34.998 |
| WFeGe | 4.867 | 8 | 1.217 | 19.508 | 26.57 |
| WFeSi | 5.317 | 8 | 1.329 | 18.425 | 34.286 |
| WFeSn | 7.217 | 14 | 1.242 | 16.378 | 21.081 |
| WIrAl | 1.382 | 6 | 0.419 | 94.25 | 24.788 |
| WIrB | 2.766 | 8 | 0.692 | 53.712 | 29.443 |
| WIrGa | 1.073 | 6 | 0.325 | 139.898 | 22.785 |
| WIrIn | 1.28 | 6 | 0.388 | 95.602 | 19.291 |
| WIrTl | 1.083 | 6 | 0.328 | 115.43 | 14.692 |
| WOsGe | 2.414 | 8 | 0.603 | 1.115 | 0.188 |
| WOsPb | 3.1 | 14 | 0.534 | 57.688 | 16.166 |
| WOsSn | 3.207 | 14 | 0.552 | 60.673 | 22.495 |
| WRhAl | 1.729 | 6 | 0.524 | 60.648 | 25.705 |
| WRhB | 1.445 | 8 | 0.361 | 132.414 | 32.079 |
| WRhGa | 1.631 | 6 | 0.494 | 67.589 | 22.988 |
| WRhIn | 1.659 | 6 | 0.503 | 58.215 | 18.786 |
| WRhTl | 1.379 | 6 | 0.418 | 72.938 | 13.97 |
| WRuPb | 3.185 | 14 | 0.548 | 51.364 | 16.132 |
| YAgC | 0.755 | 9 | 0.174 | 144.993 | 9.571 |
| YAgSi | 1.227 | 6 | 0.372 | 42.74 | 10.09 |
| YAuGe | 1.168 | 6 | 0.354 | 47.681 | 6.864 |
| YAuSi | 1.318 | 6 | 0.399 | 43.52 | 8.958 |
| YAuSn | 1.213 | 6 | 0.367 | 43.038 | 6.347 |
| YCoS | 10.938 | 4 | 4.341 | 1.445 | 14.88 |
| YCoTe | 8.574 | 8 | 2.143 | 3.684 | 10.156 |
| YCuC | 0.709 | 8 | 0.177 | 170.292 | 13.348 |
| YCuGe | 1.292 | 6 | 0.391 | 40.505 | 8.906 |
| YCuSi | 1.334 | 6 | 0.404 | 41.466 | 12.422 |
| YHgAl | 2.773 | 6 | 0.84 | 10.506 | 5.878 |
| YHgB | 1.524 | 6 | 0.462 | 29.864 | 6.348 |
| YNiAs | 0.974 | 8 | 0.243 | 99.5 | 11.81 |
| YNiBi | 0.837 | 8 | 0.209 | 98.956 | 5.961 |

| | | | | | |
|---|---|---|---|---|---|
| YNiP | 1.161 | 8 | 0.29 | 83.496 | 16.19 |
| YNiSb | 0.979 | 8 | 0.245 | 89.458 | 9.611 |
| YPtAs | 0.763 | 14 | 0.131 | 256.551 | 9.285 |
| YPtBi | 0.871 | 6 | 0.264 | 77.731 | 6.182 |
| YPtP | 0.968 | 8 | 0.242 | 108.632 | 11.022 |
| YPtSb | 0.888 | 6 | 0.269 | 84.551 | 8.753 |
| YRhTe | 1.139 | 8 | 0.285 | 76.543 | 10.473 |
| ZrAgB | 2.825 | 6 | 0.855 | 19.278 | 17.937 |
| ZrAuAl | 2.28 | 6 | 0.69 | 24.605 | 13.318 |
| ZrAuB | 2.752 | 6 | 0.834 | 22.486 | 15.809 |
| ZrAuGa | 6.404 | 6 | 1.939 | 5.212 | 11.294 |
| ZrAuTl | 4.329 | 6 | 1.311 | 8.045 | 7.128 |
| ZrCoAs | 6.374 | 8 | 1.594 | 9.034 | 22.023 |
| ZrCoBi | 9.252 | 8 | 2.313 | 4.05 | 10.823 |
| ZrCoP | 8.081 | 8 | 2.02 | 6.976 | 30.713 |
| ZrCoSb | 8.241 | 8 | 2.06 | 5.48 | 17.458 |
| ZrCuB | 2.611 | 6 | 0.791 | 25.041 | 23.408 |
| ZrFePo | 4.951 | 4 | 1.965 | 5.187 | 11.063 |
| ZrFeS | 4.147 | 4 | 1.646 | 8.998 | 27.06 |
| ZrFeSe | 4.882 | 4 | 1.938 | 6.496 | 20.319 |
| ZrFeTe | 4.843 | 4 | 1.922 | 6.157 | 17.543 |
| ZrIrAs | 4.459 | 8 | 1.115 | 15.971 | 18.216 |
| ZrIrBi | 2.838 | 8 | 0.71 | 26.811 | 11.814 |
| ZrIrP | 4.777 | 8 | 1.194 | 15.536 | 22.455 |
| ZrIrSb | 4.72 | 8 | 1.18 | 14.12 | 17.285 |
| ZrNiC | 1.824 | 8 | 0.456 | 69.576 | 30.211 |
| ZrNiGe | 3.453 | 8 | 0.863 | 20.554 | 19.182 |
| ZrNiPb | 2.766 | 8 | 0.691 | 22.874 | 9.374 |
| ZrNiSi | 11.496 | 8 | 2.874 | 3.61 | 26.606 |
| ZrNiSn | 3.499 | 3 | 1.682 | 6.653 | 14.748 |
| ZrPtPb | 2.19 | 3 | 1.053 | 13.438 | 9.775 |
| ZrPtSi | 3.719 | 6 | 1.126 | 15.559 | 20.274 |
| ZrPtSn | 3.084 | 3 | 1.483 | 9.011 | 14.412 |
| ZrRhAs | 2.898 | 8 | 0.724 | 28.884 | 21.726 |
| ZrRhBi | 1.911 | 8 | 0.478 | 44.83 | 12.261 |
| ZrRhP | 4.472 | 8 | 1.118 | 16.286 | 28.405 |
| ZrRhSb | 2.587 | 8 | 0.647 | 31.78 | 18.575 |
| ZrRuTe | 4.87 | 8 | 1.218 | 12.603 | 19.157 |

# 5 Additional Semiconductor Data - Structure.

Table 3: Half-Heusler Structure

| Compound | Lattice Parameter (Å) | $E_g$ (eV) | $E_g$ Type | Bulk Modulus (GPa) |
|---|---|---|---|---|
| AcNiAs | 6.583 | 0.145 | PBE | 76.206 |
| AcNiP | 6.462 | 0.615 | PBE | 82.504 |
| AcNiSb | 6.826 | 0.391 | PBE | 69.803 |
| CoMoAl | 5.772 | 0.306 | MBJ | 141.056 |
| CrCoAl | 5.461 | 0.545 | MBJ | 162.610 |
| CrCoB | 4.984 | 0.84 | MBJ | 239.975 |
| CrCoGa | 5.457 | 0.622 | MBJ | 168.191 |
| CrCoIn | 5.752 | 0.699 | MBJ | 139.333 |
| CrFeC | 4.895 | 0.563 | MBJ | 270.387 |
| CrFeGe | 5.450 | 0.423 | MBJ | 190.022 |
| CrFeSi | 5.339 | 0.368 | MBJ | 207.641 |
| CrFeSn | 5.743 | 0.466 | MBJ | 156.398 |
| CrIrAl | 5.760 | 0.351 | MBJ | 179.152 |
| CrIrB | 5.376 | 0.528 | MBJ | 237.306 |
| CrIrGa | 5.775 | 0.361 | MBJ | 182.999 |
| CrIrIn | 6.042 | 0.476 | MBJ | 154.748 |
| CrIrTl | 6.125 | 0.584 | MBJ | 392.268 |
| CrOsGe | 5.767 | 0.354 | MBJ | 202.465 |
| CrOsPb | 6.135 | 0.338 | MBJ | 150.077 |
| CrOsSi | 5.662 | 0.31 | MBJ | 218.221 |
| CrOsSn | 6.026 | 0.419 | MBJ | 174.263 |
| CrReAs | 5.794 | 0.152 | MBJ | 197.964 |
| CrReP | 5.641 | 0.123 | MBJ | 218.949 |
| CrReSb | 6.057 | 0.241 | MBJ | 174.160 |
| CrRhAl | 5.729 | 0.444 | MBJ | 159.459 |
| CrRhB | 5.315 | 0.63 | MBJ | 217.260 |
| CrRhGa | 5.733 | 0.44 | MBJ | 163.138 |
| CrRuC | 5.232 | 0.36 | MBJ | 241.483 |
| CrRuGe | 5.722 | 0.203 | MBJ | 185.993 |
| CrRuPb | 6.090 | 0.333 | MBJ | 141.803 |
| CrRuSi | 5.620 | 0.147 | MBJ | 200.581 |
| CrRuSn | 5.987 | 0.287 | MBJ | 159.508 |
| CrTcAs | 5.752 | 0.162 | MBJ | 187.237 |
| CrTcP | 5.597 | 0.097 | MBJ | 206.015 |
| CrTcSb | 6.018 | 0.234 | MBJ | 162.928 |
| FeMoSi | 5.674 | 0.183 | MBJ | 176.267 |
| HfAgAl | 6.268 | 0.068 | PBE | 106.243 |
| HfAgB | 5.862 | 0.147 | PBE | 135.707 |
| HfAgGa | 6.237 | 0.090 | MBJ | 107.349 |
| HfAuAl | 6.263 | 0.641 | PBE | 126.918 |
| HfAuB | 5.919 | 0.72 | PBE | 152.648 |
| HfAuGa | 6.248 | 0.562 | PBE | 126.841 |
| HfAuIn | 6.461 | 0.369 | PBE | 115.610 |
| HfAuTl | 6.526 | 0.254 | PBE | 107.310 |
| HfCoAs | 5.792 | 1.279 | PBE | 161.215 |
| HfCoBi | 6.186 | 0.969 | PBE | 125.393 |
| HfCoP | 5.649 | 1.368 | PBE | 176.580 |
| HfCoSb | 6.054 | 1.127 | PBE | 143.683 |
| HfCuB | 5.528 | 0.085 | MBJ | 156.225 |
| HfFeTe | 6.042 | 0.824 | PBE | 145.573 |
| HfIrSb | 6.320 | 0.881 | PBE | 158.922 |
| HfNiC | 5.336 | 0.908 | PBE | 184.298 |

| Compound | a (Å) | gap (eV) | Functional | Value |
|---|---|---|---|---|
| HfNiGe | 5.849 | 0.563 | PBE | 146.182 |
| HfNiSi | 5.771 | 0.685 | PBE | 157.476 |
| HfNiSn | 6.105 | 0.376 | PBE | 129.269 |
| HfPdSn | 6.343 | 0.394 | PBE | 126.208 |
| HfPtGe | 6.161 | 1.138 | PBE | 154.420 |
| HfPtPb | 6.462 | 0.753 | PBE | 127.647 |
| HfPtSi | 6.082 | 1.293 | PBE | 165.434 |
| HfPtSn | 6.366 | 0.906 | PBE | 143.738 |
| HfRhAs | 6.061 | 0.295 | PBE | 154.922 |
| HfRhBi | 6.404 | 0.145 | PBE | 128.435 |
| HfRhP | 5.936 | 0.855 | PBE | 167.249 |
| HfRhSb | 6.282 | 1.126 | PBE | 145.615 |
| HfRuTe | 6.277 | 0.066 | PBE | 147.603 |
| LaAuSi | 6.751 | 0.241 | PBE | 76.672 |
| LaAuSn | 6.995 | 0.143 | PBE | 68.537 |
| LaCoS | 6.130 | 0.32 | PBE | 98.687 |
| LaCoSe | 6.295 | 0.066 | PBE | 90.744 |
| LaCoTe | 6.545 | 0.632 | PBE | 83.991 |
| LaCuC | 6.009 | 0.333 | PBE | 88.674 |
| LaFeF | 6.494 | 0.056 | MBJ | 35.166 |
| LaHgAl | 7.012 | 0.066 | PBE | 56.874 |
| LaHgIn | 7.165 | 0.121 | PBE | 52.197 |
| LaNiAs | 6.370 | 0.626 | PBE | 86.178 |
| LaNiP | 6.247 | 0.698 | PBE | 93.785 |
| LaNiSb | 6.619 | 0.453 | PBE | 78.517 |
| LaPtSb | 6.859 | 0.05 | PBE | 85.383 |
| LaRhS | 6.404 | 0.366 | PBE | 96.228 |
| LaRhSe | 6.545 | 0.128 | PBE | 90.539 |
| LaRhTe | 6.760 | 0.749 | PBE | 86.242 |
| LaRuF | 6.037 | 0.056 | MBJ | 82.975 |
| MnNiB | 5.057 | 0.060 | PBE | 193.676 |
| MoIrB | 5.599 | 0.325 | MBJ | 242.510 |
| MoIrTl | 6.217 | 0.169 | MBJ | 165.141 |
| MoOsPb | 6.237 | 0.116 | MBJ | 173.706 |
| MoRhB | 5.529 | 0.497 | MBJ | 224.441 |
| NbCoC | 5.224 | 1.446 | PBE | 237.448 |
| NbCoGe | 5.706 | 1.071 | PBE | 180.535 |
| NbCoPb | 6.051 | 0.942 | PBE | 141.093 |
| NbCoSi | 5.623 | 0.862 | PBE | 193.214 |
| NbCoSn | 5.962 | 0.982 | PBE | 156.446 |
| NbFeAs | 5.694 | 0.581 | PBE | 188.598 |
| NbFeBi | 6.081 | 0.596 | PBE | 145.740 |
| NbFeP | 5.552 | 0.507 | PBE | 209.284 |
| NbFeSb | 5.951 | 0.507 | PBE | 166.863 |
| NbIrC | 5.656 | 0.283 | PBE | 216.764 |
| NbIrGe | 6.019 | 0.634 | PBE | 191.362 |
| NbIrPb | 6.321 | 0.715 | PBE | 157.385 |
| NbIrSi | 5.941 | 0.485 | PBE | 204.370 |
| NbIrSn | 6.230 | 0.631 | PBE | 174.466 |
| NbNiAl | 5.771 | 0 | PBE | 149.911 |
| NbNiB | 5.349 | 0.58 | PBE | 207.562 |
| NbNiTl | 6.045 | 0.091 | PBE | 130.811 |
| NbOsAs | 6.015 | 0.242 | PBE | 195.936 |
| NbOsBi | 6.346 | 0.318 | PBE | 162.956 |
| NbOsP | 5.893 | 0.16 | PBE | 212.157 |
| NbOsSb | 6.226 | 0.264 | PBE | 183.893 |
| NbPtAl | 6.065 | 0.143 | PBE | 165.732 |

| Compound | a (Å) | Gap (eV) | Method | B (GPa) |
|---|---|---|---|---|
| NbPtB | 5.741 | 0.808 | PBE | 203.722 |
| NbPtGa | 6.058 | 0.457 | PBE | 168.077 |
| NbPtIn | 6.274 | 0.449 | PBE | 150.271 |
| NbPtTl | 6.337 | 0.592 | PBE | 142.360 |
| NbRhC | 5.567 | 1.048 | PBE | 212.120 |
| NbRhGe | 5.972 | 0.649 | PBE | 176.010 |
| NbRhPb | 6.279 | 0.763 | PBE | 142.467 |
| NbRhSi | 5.896 | 0.473 | PBE | 187.189 |
| NbRhSn | 6.193 | 0.659 | PBE | 158.157 |
| NbRuAs | 5.962 | 0.332 | PBE | 185.055 |
| NbRuBi | 6.303 | 0.421 | PBE | 151.150 |
| NbRuP | 5.839 | 0.245 | PBE | 200.906 |
| NbRuSb | 6.184 | 0.356 | PBE | 170.766 |
| ScAgC | 5.736 | 0.228 | PBE | 106.520 |
| ScAuGe | 6.268 | 0.297 | PBE | 98.743 |
| ScAuSi | 6.174 | 0.386 | PBE | 109.741 |
| ScAuSn | 6.493 | 0.146 | PBE | 92.178 |
| ScCdIn | 6.676 | 0.068 | MBJ | 58.410 |
| ScCdTl | 6.748 | 0.113 | MBJ | 52.036 |
| ScCoS | 5.585 | 0.672 | PBE | 132.323 |
| ScCoSe | 5.776 | 0.292 | PBE | 121.174 |
| ScCoTe | 6.043 | 0.889 | PBE | 113.128 |
| ScCuC | 5.352 | 0.118 | PBE | 131.888 |
| ScHgAl | 6.441 | 0.103 | PBE | 76.804 |
| ScHgB | 6.029 | 0.17 | PBE | 94.817 |
| ScHgGa | 6.431 | 0.077 | PBE | 73.059 |
| ScHgIn | 6.665 | 0.378 | MBJ | 67.349 |
| ScHgTl | 6.760 | 0.153 | MBJ | 57.973 |
| ScNiAs | 5.822 | 0.467 | PBE | 119.007 |
| ScNiBi | 6.247 | 0.179 | PBE | 90.751 |
| ScNiP | 5.669 | 0.58 | PBE | 132.603 |
| ScNiSb | 6.103 | 0.263 | PBE | 105.623 |
| ScPtP | 6.019 | 0.085 | PBE | 131.587 |
| ScPtSb | 6.375 | 0.659 | PBE | 116.107 |
| ScRhTe | 6.294 | 0.463 | PBE | 113.466 |
| TaCoC | 5.216 | 1.536 | PBE | 239.881 |
| TaCoGe | 5.698 | 1.206 | PBE | 191.789 |
| TaCoPb | 6.041 | 0.973 | PBE | 149.931 |
| TaCoSi | 5.614 | 1.238 | PBE | 205.737 |
| TaCoSn | 5.952 | 1.039 | PBE | 166.828 |
| TaFeAs | 5.688 | 0.9205 | PBE | 198.364 |
| TaFeBi | 6.072 | 0.92 | PBE | 154.273 |
| TaFeP | 5.545 | 0.837 | PBE | 220.237 |
| TaFeSb | 5.943 | 0.856 | PBE | 177.775 |
| TaIrGe | 6.013 | 0.932 | PBE | 202.793 |
| TaIrSi | 5.935 | 0.795 | PBE | 215.602 |
| TaIrSn | 6.219 | 0.942 | PBE | 185.752 |
| TaNiAl | 5.760 | 0.585 | MBJ | 161.643 |
| TaNiB | 5.338 | 0.778 | PBE | 217.954 |
| TaNiGa | 5.737 | 0.243 | PBE | 168.074 |
| TaNiIn | 5.989 | 0.181 | PBE | 145.008 |
| TaNiTl | 6.032 | 0.427 | MBJ | 138.713 |
| TaOsBi | 6.338 | 0.132 | MBJ | 170.245 |
| TaOsP | 5.890 | 0.049 | PBE | 219.375 |
| TaOsSb | 6.217 | 0.551 | PBE | 193.959 |
| TaPtAl | 6.052 | 0.572 | PBE | 178.819 |
| TaPtB | 5.733 | 1.327 | PBE | 215.282 |

| Compound | a (Å) | gap (eV) | Method | B (GPa) |
|---|---|---|---|---|
| TaPtGa | 6.045 | 0.94 | PBE | 181.458 |
| TaPtIn | 6.259 | 0.923 | PBE | 161.870 |
| TaPtTl | 6.322 | 0.952 | PBE | 152.878 |
| TaRhC | 5.557 | 0.645 | PBE | 214.239 |
| TaRhGe | 5.961 | 1.07 | PBE | 186.818 |
| TaRhPb | 6.266 | 1.049 | PBE | 151.718 |
| TaRhSi | 5.885 | 0.868 | PBE | 197.920 |
| TaRhSn | 6.180 | 1.047 | PBE | 168.556 |
| TaRuAs | 5.955 | 0.452 | PBE | 193.138 |
| TaRuBi | 6.293 | 0.455 | PBE | 159.409 |
| TaRuP | 5.831 | 0.533 | PBE | 208.181 |
| TaRuSb | 6.175 | 0.631 | PBE | 180.683 |
| TiAgAl | 6.112 | 0.151 | PBE | 98.093 |
| TiAgB | 5.660 | 0.337 | PBE | 133.312 |
| TiAgGa | 6.088 | 0.177 | PBE | 98.730 |
| TiAgIn | 6.342 | 0.072 | PBE | 86.833 |
| TiAuAl | 6.094 | 0.695 | PBE | 120.029 |
| TiAuB | 5.700 | 0.864 | PBE | 154.407 |
| TiAuGa | 6.090 | 0.622 | PBE | 119.601 |
| TiAuIn | 6.337 | 0.424 | PBE | 105.515 |
| TiAuTl | 6.419 | 0.326 | PBE | 96.077 |
| TiCoAs | 5.601 | 1.3 | PBE | 164.368 |
| TiCoBi | 6.026 | 0.896 | PBE | 123.841 |
| TiCoP | 5.440 | 1.375 | PBE | 185.328 |
| TiCoSb | 5.881 | 1.053 | PBE | 143.102 |
| TiCuB | 5.300 | 0.216 | PBE | 159.657 |
| TiCuGa | 5.773 | 0.065 | PBE | 113.889 |
| TiFePo | 6.035 | 0.69 | PBE | 125.729 |
| TiFeS | 5.412 | 1.101 | PBE | 177.363 |
| TiFeSe | 5.604 | 0.901 | PBE | 159.409 |
| TiFeTe | 5.874 | 0.957 | PBE | 146.945 |
| TiIrAs | 5.931 | 0.758 | PBE | 170.904 |
| TiIrBi | 6.302 | 0.617 | PBE | 138.221 |
| TiIrP | 5.791 | 0.896 | PBE | 188.372 |
| TiIrSb | 6.165 | 0.807 | PBE | 159.534 |
| TiNiC | 5.099 | 0.97 | PBE | 203.885 |
| TiNiGe | 5.661 | 0.625 | PBE | 146.947 |
| TiNiPb | 6.042 | 0.339 | PBE | 110.792 |
| TiNiSi | 5.568 | 0.743 | PBE | 159.361 |
| TiNiSn | 5.942 | 0.442 | PBE | 125.264 |
| TiOsTe | 6.163 | 0.478 | PBE | 161.820 |
| TiPtC | 5.553 | 0.418 | PBE | 181.595 |
| TiPtGe | 5.983 | 0.845 | PBE | 154.395 |
| TiPtPb | 6.330 | 0.712 | PBE | 121.255 |
| TiPtSi | 5.889 | 0.863 | PBE | 167.809 |
| TiPtSn | 6.218 | 0.764 | PBE | 139.058 |
| TiRhAs | 5.882 | 0.795 | PBE | 157.132 |
| TiRhBi | 6.264 | 0.69 | PBE | 124.931 |
| TiRhP | 5.740 | 0.799 | PBE | 173.820 |
| TiRhSb | 6.130 | 0.719 | PBE | 143.445 |
| TiRuTe | 6.126 | 0.745 | PBE | 148.165 |
| VCoC | 4.948 | 0.936 | PBE | 251.033 |
| VCoGe | 5.507 | 0.66 | PBE | 177.083 |
| VCoPb | 5.899 | 0.638 | PBE | 130.946 |
| VCoSi | 5.404 | 0.526 | PBE | 193.569 |
| VCoSn | 5.795 | 0.614 | PBE | 149.172 |
| VFeAs | 5.491 | 0.342 | PBE | 189.975 |

| | | | | |
|---|---|---|---|---|
| VFeBi | 5.927 | 0.332 | PBE | 138.494 |
| VFeP | 5.325 | 0.302 | PBE | 214.973 |
| VFeSb | 5.782 | 0.321 | PBE | 162.745 |
| VIrC | 5.386 | 0.305 | PBE | 229.091 |
| VIrGe | 5.831 | 0.295 | PBE | 188.927 |
| VIrSi | 5.732 | 0.184 | PBE | 204.679 |
| VIrSn | 6.080 | 0.293 | PBE | 166.402 |
| VNiAl | 5.578 | 0.12 | PBE | 143.521 |
| VNiB | 5.088 | 0.904 | PBE | 209.252 |
| VNiGa | 5.560 | 0.319 | PBE | 148.818 |
| VNiIn | 5.847 | 0.223 | PBE | 123.720 |
| VOsAs | 5.822 | 0.079 | PBE | 198.697 |
| VOsBi | 6.208 | 0.119 | PBE | 155.111 |
| VOsP | 5.674 | 1.099 | MBJ | 218.599 |
| VOsSb | 6.070 | 0.109 | PBE | 178.244 |
| VPtAl | 5.876 | 1.015 | MBJ | 157.257 |
| VPtB | 5.488 | 0.369 | PBE | 205.983 |
| VPtGa | 5.881 | 0.167 | PBE | 159.105 |
| VPtTl | 6.222 | 0.29 | PBE | 127.158 |
| VReTe | 6.107 | 0.244 | MBJ | 170.237 |
| VRhC | 5.301 | 0.452 | PBE | 218.964 |
| VRhGe | 5.787 | 0.319 | PBE | 170.445 |
| VRhPb | 6.147 | 0.405 | PBE | 132.571 |
| VRhSi | 5.690 | 0.187 | PBE | 185.625 |
| VRhSn | 6.042 | 0.346 | PBE | 149.417 |
| VRuAs | 5.773 | 0.14 | PBE | 184.823 |
| VRuBi | 6.166 | 0.199 | PBE | 143.372 |
| VRuP | 5.625 | 0.074 | PBE | 204.176 |
| VRuSb | 6.029 | 0.176 | PBE | 163.953 |
| VTcTe | 6.074 | 0.727 | MBJ | 158.193 |
| WCoAl | 5.630 | 0.455 | MBJ | 195.013 |
| WCoB | 5.232 | 0.753 | PBE | 262.595 |
| WCoGa | 5.622 | 0.346 | PBE | 201.667 |
| WCoIn | 5.874 | 0.377 | PBE | 171.453 |
| WFeC | 5.164 | 0.518 | PBE | 273.211 |
| WFeGe | 5.623 | 0.201 | PBE | 218.169 |
| WFeSi | 5.532 | 0.048 | PBE | 235.323 |
| WFeSn | 5.875 | 0.235 | PBE | 188.992 |
| WIrAl | 5.929 | 0.143 | MBJ | 212.737 |
| WIrB | 5.627 | 0.591 | MBJ | 257.422 |
| WIrGa | 5.933 | 0.233 | MBJ | 215.895 |
| WIrIn | 6.149 | 0.363 | MBJ | 192.332 |
| WIrTl | 6.214 | 0.577 | MBJ | 180.593 |
| WOsGe | 5.937 | 0.104 | MBJ | 4.356 |
| WOsPb | 6.238 | 0.347 | MBJ | 187.378 |
| WOsSn | 6.147 | 0.244 | MBJ | 207.395 |
| WRhAl | 5.889 | 0.206 | MBJ | 191.566 |
| WRhB | 5.553 | 0.764 | MBJ | 239.525 |
| WRhGa | 5.884 | 0.266 | MBJ | 195.501 |
| WRhIn | 6.107 | 0.23 | MBJ | 172.840 |
| WRhTl | 6.163 | 0.078 | PBE | 163.997 |
| WRuPb | 6.192 | 0.159 | MBJ | 173.822 |
| YAgC | 6.061 | 0.851 | MBJ | 88.025 |
| YAgSi | 6.450 | 0.113 | MBJ | 80.654 |
| YAuGe | 6.539 | 0.324 | MBJ | 83.594 |
| YAuSi | 6.458 | 0.239 | PBE | 91.420 |
| YAuSn | 6.736 | 0.077 | MBJ | 79.890 |

| Compound | a (Å) | Gap (eV) | Functional | B (GPa) |
|---|---|---|---|---|
| YCoS | 5.874 | 0.061 | PBE | 108.870 |
| YCoTe | 6.296 | 0.872 | PBE | 96.335 |
| YCuC | 5.698 | 0.085 | PBE | 105.963 |
| YCuGe | 6.233 | 0.299 | MBJ | 82.585 |
| YCuSi | 6.158 | 0.559 | MBJ | 88.692 |
| YHgAl | 6.715 | 0.084 | PBE | 67.389 |
| YHgB | 6.355 | 0.096 | PBE | 78.041 |
| YNiAs | 6.101 | 0.492 | PBE | 99.593 |
| YNiBi | 6.493 | 0.193 | PBE | 78.934 |
| YNiP | 5.964 | 0.594 | PBE | 108.756 |
| YNiSb | 6.359 | 0.285 | PBE | 90.264 |
| YPtAs | 6.423 | 0.826 | MBJ | 101.692 |
| YPtBi | 6.746 | 0.424 | MBJ | 87.827 |
| YPtP | 6.305 | 0.917 | MBJ | 107.722 |
| YPtSb | 6.615 | 0.094 | PBE | 98.290 |
| YRhTe | 6.530 | 0.158 | PBE | 96.961 |
| ZrAgB | 5.925 | 0.24 | PBE | 127.111 |
| ZrAuAl | 6.316 | 0.472 | PBE | 117.633 |
| ZrAuB | 5.976 | 0.79 | PBE | 142.609 |
| ZrAuGa | 6.300 | 0.594 | PBE | 117.321 |
| ZrAuTl | 6.572 | 0.344 | PBE | 100.641 |
| ZrCoAs | 5.836 | 1.208 | PBE | 151.444 |
| ZrCoBi | 6.223 | 0.985 | PBE | 118.732 |
| ZrCoP | 5.699 | 1.284 | PBE | 166.917 |
| ZrCoSb | 6.094 | 1.057 | PBE | 135.048 |
| ZrCuB | 5.587 | 0.084 | PBE | 146.730 |
| ZrFePo | 6.232 | 0.096 | PBE | 119.041 |
| ZrFeS | 5.669 | 0.326 | PBE | 158.293 |
| ZrFeSe | 5.838 | 0.112 | PBE | 145.997 |
| ZrFeTe | 6.084 | 1.156 | PBE | 136.714 |
| ZrIrAs | 6.157 | 0.265 | PBE | 156.638 |
| ZrIrBi | 6.480 | 0.204 | PBE | 133.524 |
| ZrIrP | 6.038 | 0.782 | PBE | 168.954 |
| ZrIrSb | 6.358 | 1.407 | PBE | 150.824 |
| ZrNiC | 5.397 | 0.953 | PBE | 178.536 |
| ZrNiGe | 5.895 | 0.665 | PBE | 137.365 |
| ZrNiPb | 6.239 | 0.365 | PBE | 109.603 |
| ZrNiSi | 5.818 | 0.786 | PBE | 146.561 |
| ZrNiSn | 6.148 | 0.486 | PBE | 120.938 |
| ZrPtPb | 6.502 | 0.826 | PBE | 120.947 |
| ZrPtSi | 6.129 | 1.337 | PBE | 155.002 |
| ZrPtSn | 6.407 | 0.974 | PBE | 135.589 |
| ZrRhAs | 6.104 | 1.204 | PBE | 148.399 |
| ZrRhBi | 6.443 | 1.023 | PBE | 123.348 |
| ZrRhP | 5.984 | 1.427 | PBE | 160.409 |
| ZrRhSb | 6.324 | 1.18 | PBE | 137.754 |
| ZrRuTe | 6.318 | 0.888 | PBE | 141.107 |

# 6 Additional Semiconductor Data - TE Quality Factor, $\beta$.

Table 4: $\beta$ Predictions

| Compound | $\beta_n$ | $\beta_p$ |
|---|---|---|
| AcNiAs | 34.134 | 16.016 |
| AcNiP | 3.49 | 6.702 |
| AcNiSb | 4.487 | 47.799 |
| CoMoAl | 10.671 | 28.335 |
| CrCoAl | 1.517 | 13.314 |
| CrCoB | 3.046 | 5.148 |
| CrCoGa | 2.742 | 18.875 |
| CrCoIn | 3.845 | 20.981 |
| CrFeC | 1.546 | 3.986 |
| CrFeGe | 2.086 | 3.349 |
| CrFeSi | 1.012 | 2.971 |
| CrFeSn | 3.528 | 9.109 |
| CrIrAl | 1.955 | 14.202 |
| CrIrB | 1.72 | 7.971 |
| CrIrGa | 2.805 | 16.603 |
| CrIrIn | 2.237 | 19.349 |
| CrIrTl | 1.305 | 82.541 |
| CrOsGe | 1.71 | 6.763 |
| CrOsPb | 2.82 | 21.259 |
| CrOsSi | 1.123 | 5.717 |
| CrOsSn | 1.852 | 16.21 |
| CrReAs | 1.724 | 6.355 |
| CrReP | 1.738 | 5.752 |
| CrReSb | 1.557 | 7.904 |
| CrRhAl | 1.158 | 8.416 |
| CrRhB | 1.953 | 7.91 |
| CrRhGa | 1.582 | 13.879 |
| CrRuC | 1.196 | 5.224 |
| CrRuGe | 1.542 | 12.278 |
| CrRuPb | 2.466 | 21.667 |
| CrRuSi | 0.915 | 10.169 |
| CrRuSn | 1.709 | 15.217 |
| CrTcAs | 1.47 | 30.813 |
| CrTcP | 0.966 | 29.387 |
| CrTcSb | 1.781 | 30.895 |
| FeMoSi | 0.965 | 5.206 |
| HfAgAl | 6.512 | 1.927 |
| HfAgB | 5.447 | 3.693 |
| HfAgGa | 7.068 | 3.059 |
| HfAuAl | 5.684 | 1.722 |
| HfAuB | 5.046 | 3.998 |
| HfAuGa | 6.223 | 3.011 |
| HfAuIn | 6.915 | 3.575 |
| HfAuTl | 8.999 | 23.766 |
| HfCoAs | 2.282 | 6.018 |
| HfCoBi | 3.597 | 15.348 |
| HfCoP | 2.097 | 5.241 |
| HfCoSb | 2.883 | 5.871 |
| HfCuB | 5.855 | 3.205 |
| HfFeTe | 6.95 | 8.022 |
| HfIrSb | 12.742 | 3.071 |

| | | |
|---|---:|---:|
| HfNiC | 3.085 | 4.478 |
| HfNiGe | 3.997 | 3.095 |
| HfNiSi | 3.368 | 1.91 |
| HfNiSn | 4.631 | 3.328 |
| HfPdSn | 4.739 | 3.727 |
| HfPtGe | 3.586 | 3.823 |
| HfPtPb | 5.503 | 5.3 |
| HfPtSi | 3.276 | 2.728 |
| HfPtSn | 4.351 | 3.744 |
| HfRhAs | 11.844 | 3.715 |
| HfRhBi | 16.492 | 4.874 |
| HfRhP | 6.18 | 3.159 |
| HfRhSb | 2.641 | 3.411 |
| HfRuTe | 64.32 | 2.082 |
| LaAuSi | 14.3 | 15.856 |
| LaAuSn | 17.556 | 54.069 |
| LaCoS | 30.347 | 6.495 |
| LaCoSe | 60.886 | 7.472 |
| LaCoTe | 11.238 | 0.337 |
| LaCuC | 3.789 | 5.661 |
| LaFeF | 8.288 | 3.911 |
| LaHgAl | 32.005 | 34.815 |
| LaHgIn | 8.808 | 28.036 |
| LaNiAs | 13.558 | 0.352 |
| LaNiP | 10.904 | 7.004 |
| LaNiSb | 16.116 | 0.944 |
| LaPtSb | 179.94 | 53.926 |
| LaRhS | 59.1 | 3.811 |
| LaRhSe | 79.688 | 1.308 |
| LaRhTe | 14.972 | 1.615 |
| LaRuF | 16.932 | 0.426 |
| MnNiB | 0.527 | 1.312 |
| MoIrB | 3.022 | 9.412 |
| MoIrTl | 4.446 | 36.728 |
| MoOsPb | 3.096 | 22.26 |
| MoRhB | 2.804 | 22.408 |
| NbCoC | 1.095 | 4.41 |
| NbCoGe | 4.262 | 9.4 |
| NbCoPb | 2.888 | 14.765 |
| NbCoSi | 1.843 | 2.251 |
| NbCoSn | 9.071 | 9.427 |
| NbFeAs | 4.106 | 5.439 |
| NbFeBi | 5.137 | 8.875 |
| NbFeP | 3.433 | 4.521 |
| NbFeSb | 4.027 | 8.278 |
| NbIrC | 10.311 | 9.04 |
| NbIrGe | 7.908 | 48.052 |
| NbIrPb | 6.29 | 18.779 |
| NbIrSi | 4.971 | 12.71 |
| NbIrSn | 5.326 | 16.052 |
| NbNiAl | 5.327 | 21.862 |
| NbNiB | 1.866 | 4.135 |
| NbNiTl | 3.598 | 33.139 |
| NbOsAs | 5.101 | 10.711 |
| NbOsBi | 5.996 | 12.736 |
| NbOsP | 5.283 | 9.259 |
| NbOsSb | 5.383 | 9.805 |

| | | |
|---:|---:|---:|
| NbPtAl | 6.471 | 7.191 |
| NbPtB | 5.039 | 6.439 |
| NbPtGa | 7.769 | 7.52 |
| NbPtIn | 6.768 | 10.626 |
| NbPtTl | 2.551 | 33.771 |
| NbRhC | 1.848 | 5.55 |
| NbRhGe | 4.469 | 5.01 |
| NbRhPb | 4.825 | 4.759 |
| NbRhSi | 3.655 | 4.843 |
| NbRhSn | 4.132 | 3.578 |
| NbRuAs | 4.342 | 7.514 |
| NbRuBi | 5.163 | 9.968 |
| NbRuP | 3.667 | 6.237 |
| NbRuSb | 4.301 | 7.572 |
| ScAgC | 2.539 | 4.394 |
| ScAuGe | 4.262 | 4.681 |
| ScAuSi | 3.498 | 3.535 |
| ScAuSn | 4.864 | 4.599 |
| ScCdIn | 4.052 | 3.532 |
| ScCdTl | 5.783 | 12.857 |
| ScCoS | 3.123 | 0.649 |
| ScCoSe | 6.92 | 1.608 |
| ScCoTe | 1.946 | 2.258 |
| ScCuC | 2.369 | 3.634 |
| ScHgAl | 5.372 | 2.707 |
| ScHgB | 4.608 | 4.331 |
| ScHgGa | 6.111 | 3.933 |
| ScHgIn | 6.541 | 4.258 |
| ScHgTl | 8.541 | 15.121 |
| ScNiAs | 2.143 | 3.132 |
| ScNiBi | 4.086 | 5.735 |
| ScNiP | 1.473 | 2.141 |
| ScNiSb | 2.781 | 3.45 |
| ScPtP | 16.906 | 2.543 |
| ScPtSb | 2.902 | 4.875 |
| ScRhTe | 4.505 | 4.046 |
| TaCoC | 1.837 | 7.18 |
| TaCoGe | 2.462 | 12.374 |
| TaCoPb | 3.673 | 21.24 |
| TaCoSi | 1.965 | 4.47 |
| TaCoSn | 2.86 | 12.222 |
| TaFeAs | 5.364 | 8.05 |
| TaFeBi | 3.059 | 10.481 |
| TaFeP | 5.009 | 6.933 |
| TaFeSb | 22.982 | 7.791 |
| TaIrGe | 6.547 | 8.485 |
| TaIrSi | 6.176 | 6.271 |
| TaIrSn | 28.408 | 7.052 |
| TaNiAl | 3.92 | 40.677 |
| TaNiB | 3.329 | 12.304 |
| TaNiGa | 4.198 | 40.332 |
| TaNiIn | 4.54 | 29.256 |
| TaNiTl | 6.084 | 24.585 |
| TaOsBi | 3.89 | 14.93 |
| TaOsP | 13.24 | 11.253 |
| TaOsSb | 6.346 | 11.861 |
| TaPtAl | 6.776 | 11.162 |

| | | |
|---|---|---|
| TaPtB | 5.153 | 1.716 |
| TaPtGa | 6.208 | 14.719 |
| TaPtIn | 3.712 | 2.659 |
| TaPtTl | 3.532 | 28.83 |
| TaRhC | 9.259 | 10.749 |
| TaRhGe | 5.133 | 7.614 |
| TaRhPb | 3.482 | 2.148 |
| TaRhSi | 5.234 | 4.751 |
| TaRhSn | 4.147 | 8.237 |
| TaRuAs | 7.243 | 9.601 |
| TaRuBi | 11.097 | 12.121 |
| TaRuP | 5.183 | 8.502 |
| TaRuSb | 5.546 | 9.217 |
| TiAgAl | 3.2 | 1.071 |
| TiAgB | 2.272 | 1.637 |
| TiAgGa | 3.646 | 1.549 |
| TiAgIn | 3.749 | 1.873 |
| TiAuAl | 2.457 | 1.098 |
| TiAuB | 2.2 | 2.11 |
| TiAuGa | 2.908 | 1.643 |
| TiAuIn | 3.323 | 1.947 |
| TiAuTl | 4.428 | 19.956 |
| TiCoAs | 1.147 | 0.867 |
| TiCoBi | 2.143 | 1.749 |
| TiCoP | 0.911 | 3.807 |
| TiCoSb | 1.461 | 5.982 |
| TiCuB | 2.313 | 1.271 |
| TiCuGa | 4.062 | 1.382 |
| TiFePo | 5.941 | 10.215 |
| TiFeS | 1.886 | 3.248 |
| TiFeSe | 3.116 | 4.737 |
| TiFeTe | 2.652 | 4.983 |
| TiIrAs | 21.5 | 1.809 |
| TiIrBi | 3.467 | 2.543 |
| TiIrP | 3.274 | 1.514 |
| TiIrSb | 2.381 | 1.816 |
| TiNiC | 1.216 | 1.467 |
| TiNiGe | 1.989 | 1.318 |
| TiNiPb | 3.427 | 2.914 |
| TiNiSi | 1.477 | 0.74 |
| TiNiSn | 2.376 | 1.632 |
| TiOsTe | 10.355 | 1.549 |
| TiPtC | 11.588 | 1.591 |
| TiPtGe | 2.045 | 1.975 |
| TiPtPb | 1.997 | 3.067 |
| TiPtSi | 3.115 | 1.375 |
| TiPtSn | 1.608 | 1.943 |
| TiRhAs | 2.447 | 1.489 |
| TiRhBi | 1.73 | 2.362 |
| TiRhP | 2.125 | 1.348 |
| TiRhSb | 1.609 | 1.573 |
| TiRuTe | 2.084 | 1.356 |
| VCoC | 2.214 | 5.6 |
| VCoGe | 2.342 | 3.095 |
| VCoPb | 2.271 | 1.433 |
| VCoSi | 2.089 | 2.497 |
| VCoSn | 1.936 | 5.354 |

| | | |
|---|---|---|
| VFeAs | 2.529 | 4.6 |
| VFeBi | 2.246 | 6.85 |
| VFeP | 2.261 | 26.847 |
| VFeSb | 12.835 | 4.912 |
| VIrC | 2.558 | 8.303 |
| VIrGe | 3.892 | 7.882 |
| VIrSi | 3.778 | 7.156 |
| VIrSn | 3.739 | 7.74 |
| VNiAl | 0.984 | 9.285 |
| VNiB | 1.073 | 0.803 |
| VNiGa | 1.552 | 13.924 |
| VNiIn | 1.351 | 2.534 |
| VOsAs | 3.827 | 2.142 |
| VOsBi | 2.719 | 2.789 |
| VOsP | 3.742 | 1.89 |
| VOsSb | 3.214 | 13.283 |
| VPtAl | 4.116 | 14.168 |
| VPtB | 3.925 | 1.405 |
| VPtGa | 3.603 | 2.043 |
| VPtTl | 1.798 | 23.346 |
| VReTe | 3.157 | 15.68 |
| VRhC | 2.589 | 1.083 |
| VRhGe | 7.775 | 0.967 |
| VRhPb | 1.718 | 21.472 |
| VRhSi | 2.532 | 0.749 |
| VRhSn | 2.601 | 0.973 |
| VRuAs | 2.817 | 9.017 |
| VRuBi | 2.418 | 10.255 |
| VRuP | 2.572 | 1.206 |
| VRuSb | 2.94 | 8.521 |
| VTcTe | 3.328 | 34.722 |
| WCoAl | 6.005 | 8.013 |
| WCoB | 7.271 | 8.936 |
| WCoGa | 7.152 | 7.848 |
| WCoIn | 15.265 | 9.197 |
| WFeC | 6.635 | 8.302 |
| WFeGe | 7.75 | 7.008 |
| WFeSi | 7.933 | 5.555 |
| WFeSn | 6.794 | 13.223 |
| WIrAl | 3.93 | 10.419 |
| WIrB | 9.303 | 10.472 |
| WIrGa | 5.121 | 13.393 |
| WIrIn | 8.232 | 12.674 |
| WIrTl | 7.743 | 41.701 |
| WOsGe | 9.744 | 30.129 |
| WOsPb | 8.976 | 28.385 |
| WOsSn | 4.15 | 22.122 |
| WRhAl | 2.964 | 7.91 |
| WRhB | 8.955 | 13.206 |
| WRhGa | 6.186 | 9.35 |
| WRhIn | 6.044 | 10.01 |
| WRhTl | 6.856 | 36.704 |
| WRuPb | 8.362 | 25.959 |
| YAgC | 5.616 | 28.319 |
| YAgSi | 3.879 | 10.425 |
| YAuGe | 23.151 | 16.358 |
| YAuSi | 11.682 | 12.752 |

| | | |
|---|---:|---:|
| YAuSn | 17.698 | 16.524 |
| YCoS | 8.372 | 1.456 |
| YCoTe | 2.242 | 5.763 |
| YCuC | 3.62 | 21.515 |
| YCuGe | 4.372 | 11.725 |
| YCuSi | 3.626 | 16.372 |
| YHgAl | 17.707 | 9.166 |
| YHgB | 5.676 | 14.075 |
| YNiAs | 2.668 | 18.898 |
| YNiBi | 4.834 | 32.497 |
| YNiP | 2.136 | 13.548 |
| YNiSb | 3.36 | 20.98 |
| YPtAs | 7.829 | 33.973 |
| YPtBi | 9.258 | 22.748 |
| YPtP | 12.982 | 21.985 |
| YPtSb | 32.755 | 24.938 |
| YRhTe | 10.029 | 18.882 |
| ZrAgB | 9.681 | 5.603 |
| ZrAuAl | 11.602 | 7.943 |
| ZrAuB | 9.576 | 7.244 |
| ZrAuGa | 12.723 | 5.027 |
| ZrAuTl | 6.378 | 27.894 |
| ZrCoAs | 1.678 | 4.991 |
| ZrCoBi | 2.938 | 6.367 |
| ZrCoP | 1.267 | 3.422 |
| ZrCoSb | 2.073 | 4.813 |
| ZrCuB | 10.434 | 5.196 |
| ZrFePo | 11.588 | 3.444 |
| ZrFeS | 5.156 | 2.082 |
| ZrFeSe | 8.054 | 2.319 |
| ZrFeTe | 1.326 | 2.528 |
| ZrIrAs | 17.353 | 7.734 |
| ZrIrBi | 61.934 | 13.331 |
| ZrIrP | 52.883 | 6.494 |
| ZrIrSb | 5.158 | 7.585 |
| ZrNiC | 1.85 | 9.088 |
| ZrNiGe | 2.836 | 7.509 |
| ZrNiPb | 22.131 | 14.006 |
| ZrNiSi | 2.066 | 2.807 |
| ZrNiSn | 3.24 | 2.161 |
| ZrPtPb | 4.507 | 4.319 |
| ZrPtSi | 6.161 | 5.125 |
| ZrPtSn | 3.407 | 2.674 |
| ZrRhAs | 1.863 | 7.957 |
| ZrRhBi | 6.412 | 15.044 |
| ZrRhP | 3.404 | 5.071 |
| ZrRhSb | 1.886 | 9.247 |
| ZrRuTe | 4.98 | 6.284 |


\end{document}